\documentclass[a4paper]{aa}
\usepackage{graphicx}
\usepackage{longtable,lscape}
\usepackage{txfonts}  

\def\atlas9{{\sc ATLAS9}}

\begin{document}

 \title{Long-term variability of the optical spectra of NGC 4151:  I. Light
curves and flux correlations.}

\author{A.I. Shapovalova \inst{1}
\and L.\v C. Popovi\'c\inst{2,3} \and  S. Collin\inst{4} \and  A.N.
Burenkov\inst{1}
\and
V.H. Chavushyan\inst{5} \and  N.G. Bochkarev\inst{6}
 \and E. Ben\'{\i}tez\inst{7} \and D.
Dultzin-Hacyan\inst{7} \and A. Kova\v cevi\'c\inst{8} \and N.
Borisov\inst{1} \and L. Carrasco\inst{5} \and J.
Le\'on-Tavares\inst{5,9} \and A. Mercado\inst{10} \and J.R.
Valdes\inst{5} \and V.V. Vlasuyk\inst{1} \and V.E.
Zhdanova\inst{1} }

\titlerunning{Variability of NGC 4151}
\authorrunning{A.I. Shapovalova et al.}
\offprints{A. I. Shapovalova, \\ \email{ashap@sao.ru}\\ Tables 2-5 and
7-8 are given only in electronic form}

\institute{Special Astrophysical Observatory of the Russian AS,
Nizhnij Arkhyz, Karachaevo-Cherkesia 369167, Russia \and
Astronomical Observatory, Volgina 7, 11160 Belgrade 74, Serbia
\and Alexander von Humboldt Fellow, presently at Max Planck Institute for
Radioastronomy, Bonn, Germany
\and
LUTH,
Observatoire de Paris, CNRS, Universit\'e Paris Diderot; 5 Place
Jules Janssen, 92190 Meudon, France \and
Instituto Nacional de Astrof\'{\i}sica, \'{O}ptica y
Electr\'onica, Apartado Postal 51, CP 72000, Puebla, Pue. M\'exico
\and Sternberg Astronomical Institute, Moscow, Russia
 \and Instituto de
Astronom\'{\i}a, UNAM, Apartado Postal 70-264, CP 04510, M\'exico
\and Department of Astronomy, Faculty of Mathematics, University
of Belgrade, Studentski trg 16, 11000 Belgrade, Serbia \and
Max-Planck Institute f\"{u}r Radioastronomie, Auf dem H\"{u}gel
69, 53121 Bonn, Germany \and Universidad Polit\'ecnica de Baja
California, Av. de la Industria \# 291, CP 21010, Mexicali, B.C.,
M\'exico}
\date{Received  / Accepted }

\abstract
{}
{Results of a long-term spectral monitoring  of the active galactic
nucleus of NGC 4151 are presented (11 years, from 1996 to 2006).}
{High quality spectra (S/N$>50$ in the continuum near H$\alpha$ and H$\beta$)
were obtained in the spectral range $\sim$ 4000 to 7500 \AA, with a resolution
between 5 and 15 \AA, using the 6-m and the 1-m SAO's telescopes (Russia),
the GHAO's 2.1-m telescope (Cananea, M\'exico), and the OAN-SPM's 2.1-m
telescope (San-Pedro, M\'exico). The observed fluxes of the H$\alpha$, H$\beta$, H$\gamma$ and
HeII$\lambda$4686 emission lines and of the continuum at the
observed wavelength 5117\AA, were corrected for the position
angle, the seeing and the aperture effects.}
{ We found that the
continuum and line fluxes varied strongly (up to a factor 6)
during  the monitoring period. The emission was maximum in
1996-1998, and there were two minima, in 2001 and in 2005. As a
consequence, the spectral type of the nucleus changed from a
Sy1.5 in the maximum activity state to a Sy1.8 in the minimum
state. The H$\alpha$, H$\gamma$ and He$\lambda$4686 fluxes were
well correlated with the H$\beta$ flux. The line profiles were
strongly variable, showing changes of the blue and red asymmetry.
The flux ratios of the blue/red wings and of the blue (or red)
wing/core  of  H$\alpha$ and H$\beta$ varied differently. We
considered three characteristic periods during which the H$\beta$
and H$\alpha$ profiles were similar: 1996-1999, 2000-2001 and
2002-2006.  The line to continuum flux ratios
were different; in particular during the first  period
 (1996-2001), the lines were not correlated with the continuum and
saturated at high fluxes. In the third  period
 (2002-2006), the H$\alpha$ and H$\beta$ fluxes were well correlated to
the continuum flux, meaning that the ionizing
continuum was a good extrapolation of the optical continuum. We
thus consider that the values of the time lags
\textbf{ -- line lagging continuum} (0.81$^{+1.55}_{-0.81}$ \textbf{days}
for H$\alpha$ and 0.81$^{+2.19}_{-0.81}$ \textbf{days} for H$\beta$) for
the
third
period give a more realistic estimation of the dimension of the
BLR  than during the other periods. \textbf{Moreover, the time lags
obtained by binning intervals of three years within the whole monitoring
period indicate  the permanent presence of a small  component of the BLR (0.3-0.7 light days)}
  }
 {We discuss the different responses of H$\beta$ and H$\alpha$  to the
continuum during the monitoring period.}


\maketitle

\section{Introduction}

 The brightest Seyfert 1.5-type galaxy NGC 4151 has been studied in  detail
at all  wavelengths (e.g.  Peterson 1988 and  Ulrich  2000) The nucleus of
this galaxy  shows  flux variability in a wide wavelength range, with
time-scales  from a few hours (in the hard X-ray, e.g. Yaqoob et al.
1993) to several months (in the infrared, e.g.  Oknyanskij et al. 1999).

In the optical range, the Active Galactic Nucleus (AGN) of this galaxy is also known to display flux
variations of the continuum and of the lines up to a factor ten or more  (e.g.  Peterson  1988, Clavel et al. 1990, Maoz et al. 1991, Shapovalova
et al. 1996, Ulrich \& Horne 1996, Sergeev et al. 2001, Lyuty 2005). These
variations occur in time scales of several days  (Maoz et al. 1991).

NGC 4151  has been the subject of echo-mapping observational
campaigns.  The main aim of AGN monitoring campaigns was to
determine the size of the Broad Line Region (BLR) by measuring the
time delay between the emission line fluxes, in response to the variations of the
continuum flux (see Peterson 1993 for a review). It is interesting
that different authors found different time lags:  Antonucci \&
Cohen (1983) observed NGC 4151 at least once a month from 1980
May through 1981 July, and found that the BLR radius
was less than ~30 lt-days. These observations were also used
by Peterson \& Cota (1988) in combination with  their own
observations  (performed from 1985 to 1986), and they found a  BLR radius of $6\pm4$ lt-days. Gaskell \& Sparke (1986)
applied the cross-correlation method to the ultraviolet data
of Ulrich et al. (1984) and to the optical data of Antonucci
\& Cohen (1983), and they found a BLR size between 2 lt-days (for
HeII$\lambda$4686) and 20 lt-days (for H$\alpha$).

On the other hand, Maoz et al. (1991) analyzed the observations (optical
continuum,  H$\beta$ and H$\alpha$)  from 67 nights during a 216-day period
between December 15, 1987, and July 18, 1988, and they found a BLR size of $9\pm2$
lt-days.  They also  showed that in a time-scale of $0\pm5$ days, the wings of
H$\alpha$ and H$\beta$ varied in phase, and they ruled out a pure
unidirectional radial motion in the BLR (either inflow or outflow).

A 10--years long monitoring campaign (1988-1998) of the NGC 4151
nucleus was  performed with a sampling of 1-2 observations per month, using
the CCD spectrograph of the 2.6-m CrAO telescope which covers the
H$\alpha$ and H$\beta$ spectral range (Malkov et al. 1997,
Sergeev et al. 2001). The time delays between the broad lines and the continuum at 5100 \AA\ were 1.5-10 days for the Balmer lines, and 0.0 -2.6 days for the HeII $\lambda$4686 line
(Sergeev et al. 2001). Recently, Bentz et al. (2006) analyzed the
observations performed between  February
 27 and  April 10, 2005. They obtained a time lag  for the H$\beta$ line
of $6.6^{+1.1}_{-0.8}$ days.

NGC 4151 was also monitored in the UV. Clavel et al. (1990)
analyzed the spectra obtained with the IUE satellite during two months from November
29, 1988, to January 30, 1989 (with a $\sim$4 days sampling time) and they found
a time lag  of  $4\pm3$ days between the continuum and the lines CIV$\lambda$1549 and Mg
II$\lambda$2798. In addition, NGC
4151 was observed with IUE from November 9 to
December 15, 1991 (35 days), with a one day sampling time (see Ulrich \&
Horne 1996). Ulrich \& Horne { (1996)} found that the time delay in the continuum at
3000 \AA\ with respect  to that at 1320 \AA\ was smaller than one day (i.e. the UV
continuum is emitted by a region  with a dimension less than one light day). The broad UV emission lines showed  large variations that
closely followed the continuum ones. For CIV$\lambda$1549,
the time lag relative to the continuum was 2.4-3.8 days.
Ulrich \& Horne  (1996) observed also deep
blue-shifted absorption lines. They are produced by a low-velocity gas which covers the major part of the rapidly varying
continuum source and the emission-line regions. This material is moving
outwards along the line of sight, and may be located anywhere
beyond 15 lt-days. Kaspi et al. (1996)  observed  NGC 4151
during two months in 1993, with a time resolution of about one day. They
found no evidence for any time lag between the optical and UV
continuum, and a time lag of 0-2 days for
H$\alpha$, and 0-3 days for H$\beta$. Recently, Metzroth et al.
(2006) re-analyzed the IUE spectra of NGC 4151 obtained in 1988
(Clavel et al. 1990) and in 1991 (Ulrich \& Horne 1996), using the
New Spectral Image Processing System (NEWSIPS). This allowed to
improve the photometric precision and to increase the S/N ratio by 10\%-50\%.
They found that the time lags of the revised responses to changes of the continuum were $\sim$ 3-7 days for
 CIV$\lambda$1549, HeII$\lambda$1640,
CIII]$\lambda$1909, and MgII$\lambda$2798.

The results mentioned above indicate that the dimension of the BLR
varies among the emission lines
(radial stratification) and is changing with time.
Contradictory results have been obtained, even for the
same species, after using the modified processing system and
re-analyzing old data (for instance, Antonucci \& Cohen 1983,
Gaskell \& Sparke 1986, Peterson \& Cota 1988, Ulrich \& Horne
1996, Metzroth et al. 2006). Inconsistent time lags
from different monitoring campaigns might be caused by the short
duration of the campaigns, but might also indicate real changes in the
BLR size and geometry. E.g., Lyuty (2005) analyzed  photometric
observations performed during more than 30 years, and  concluded
that  the NGC 4151 nucleus goes through different levels of activity. They could be related to a total destruction of the
accretion disk (AD) that took place in cycle A (from 1968 to 1984), and
the formation of a new AD in cycle B (from 1989 to 1996).
Nevertheless, most UV and optical monitoring campaigns
confirmed that the BLR is small and is
radially stratified.

Unfortunately, spectral optical monitoring of NGC 4151 were
mostly  carried out during short periods (less than one year),
which are insufficient to trace real changes in the BLR
structure. In order to study the evolution of the BLR, more than
10 years of spectral monitoring is needed. Such observations have
been {\bf made} since 1986 in CrAO (Malkov et al. 1997; Sergeev et
al. 2001) and SAO RAS (Bochkarev et al. 1988, 1991; Shapovalova et al.
1996; Nazarova et al. 1998).

 In this paper  we present the analysis of the spectral monitoring of NGC
4151 that covers a 11-years period from 1996 to 2006.
The paper is organized as follows: in \S2 the observations, data
reduction and calibration are explained. In \S3 we study the correlations between the
continuum, the Balmer (H$\alpha$, H$\beta$, H$\gamma$) and the
HeII$\lambda$4686 fluxes, for both the whole lines
and the line wings. In \S4 we discuss different possible
interpretations. The results are summarized in \S5.

\begin{table*}[t]  
\begin{center}
\caption[]{Sources of spectroscopic observations:  1 -  source
(Observatory); 2 -  code assigned to each combination of telescope+equipment and
used throughout this paper;  3 -  telescope aperture  and the spectrograph;
4 -  projected spectrograph entrance apertures (the first number is the
slit width, and the second is the slit length); 5 -  focus of the
telescope. } \label{tab1}
\begin{tabular}{|l|l|l|l|l|}
\hline
Source                 &Code  &Tel.and Equip. & Aperture        &Focus\\
                       &      &               & (arcsec)        &     \\
\hline
1&2&3&4&5\\
\hline
 SAO(Russia)           & L(N) &  6m+Long slit &  2.0$\times$6.0        &  Nasmith\\
 SAO(Russia)           & L(U) &  6m+UAGS      &  2.0$\times$6.0        &  Prime  \\
 SAO(Russia)           & L(S) &  6m+Scorpio   &  1.0$\times$6.0        &  Prime    \\
 Gullermo Haro(M\'exico) & GH   &  2.1m+B$\&$C     &  2.5$\times$6.0        &  Cassegrain\\
 San--Pedro(M\'exico)     & S-P  &  2.1m+B$\&$C     &  2.5$\times$6.0        &  Cassegrain\\
 SAO(Russia)           & L1(G)&  1m+GAD       &  4.2(8.0)$\times$19.8  &  Cassegrain\\
 SAO(Russia)           & L1(U)&  1m+UAGS+CCD2K&  4.0$\times$9.45&  Cassegrain\\
\hline
\end{tabular}
\end{center}
\end{table*}

\section{Observations and data reduction}
\subsection{ Optical Spectroscopy} \label{sec2.1}

{ Spectroscopic observations of NGC 4151 were carried out between
January 11, 1996 (Julian Date = JD 2450094) and April 20, 2006  (JD
2453846), thus covering  a period of more than 10 years. In total 180 blue and
137 red spectra were taken during 220 nights, with the 6-m and 1-m
telescopes of SAO, Russia (1996--2006),
with the 2.1-m telescope of the Guillermo Haro Astrophysical Observatory (GHAO)
at Cananea, Sonora, M\'exico (1998--2006), and with the 2.1-m telescope of
the Observatorio Astron\'omico Nacinal at San Pedro Martir (OAN-SMP), Baja
California, M\'exico (2005--2006).
The spectra were obtained with a long--slit spectrograph equipped with CCDs.
The typical wavelength range was 4000 - 7500 \AA , the spectral
resolution was R=5--15 \AA , and the S/N ratio was $>$ 50 in the
continuum near  H$\alpha$ and H$\beta$. Note that from 2004 to 2006,
the spectral observations with the GHAO's 2.1-m telescope
were carried out with two variants of the equipment:
1) with a grism of 150 l/mm (a low dispersion of R=15 \AA,
like the observations of 1998--2003);
2) with a grism of 300 l/mm (a moderate dispersion of R=7.5\AA).
As a rule, the observations were performed with the moderate dispersion
in the blue or red bands during the first night of each set;
usually during the next night we used the low dispersion
in the whole range 4000-7500 \AA;
the moderate dispersion was used in the following night.
Since the shape of the continuum of active galaxies practically
does not change during adjacent nights, it was easy to link together
the blue and red bands obtained with the moderate dispersion, using
the data obtained for the continuum with the low-dispersion
in the whole wavelength range.
The photometric accuracy is thus considerably improved with respect
to a link obtained by overlapping the extremities of the continuum
(3-5\% instead of 5-10\%).

Spectrophotometric standard stars were observed every night.
Informations on the source of spectroscopic observations are listed
in Table~\ref{tab1}.} Log of the spectroscopic observations is
given in Table~\ref{tab2} (available only in electronic form).

\addtocounter{table}{2}

{The spectrophotometric data reduction was carried out either with the
software developed at the SAO RAS by Vlasyuk (1993), or with IRAF for the
spectra observed in M\'exico. The image reduction process included bias
subtraction, flat-field corrections, cosmic ray removal, 2D wavelength
linearization, sky spectrum subtraction, addition of the spectra for every
night, and {\bf relative flux  calibration} based on standard star
observations.}

\subsection{Absolute calibration and measurements of the
spectra} \label{sec2.2}

{ The standard technique of flux calibration spectra (i.e. comparison with
stars of known spectral energy distribution) is not precise enough for the study
of AGN variability, since even under good photometric conditions, the
accuracy of spectrophotometry is not better than $10\%$. Therefore we used standard
stars only for a relative calibration.

For the absolute calibration, the fluxes of the narrow emission lines are
adopted  for scaling the AGN spectra, because they are known to remain constant on time scales of tens of
years (Peterson 1993).
We thus assume that the flux of the [\ion{O}{iii}]$\lambda$\,5007 line was
constant during the monitoring period. One can indeed check that
it did not change between 1980 (Antonucci \& Cohen 1983) and 1992 (Malkov et al. 1997).
 { This is due to the fact that the forbidden line emitting  region is
very extensive
 (more than a hundred light-years)}.  All blue spectra were thus scaled to
the constant flux $
 F$([\ion{O}{iii}]$\lambda\,5007)= 1.14\times
10^{-11}$\,erg\,s$^{-1}$\,cm$^{-2}$  determined by Malkov et al.
 (1997), and corrected for the position angle (PA), seeing
and aperture effects, as described in \S 2.3. The scaling of the
blue spectra was performed by using the method of Van Groningen
\& Wanders  (1992)
modified by  Shapovalova et al. (2004)\footnote{see Appendix A in Shapovalova et al. (2004)}.
 This method allowed us to obtain a homogeneous set
of spectra with the same wavelength calibration and the same
[OIII]$\lambda$5007 flux.

The spectra obtained using the GHAO 2.1--m telescope (M\'exico) with
a resolution of 15\,\AA\, contain  both the H$\alpha$ and
H$\beta$ regions. These spectra were scaled using the
[\ion{O}{iii}]$\lambda$\,5007 line. In this case the red region
was automatically scaled  also by the [\ion{O}{iii}]$\lambda$\,5007 flux.
However, the
accuracy  of such a scaling depends strongly on  the correct
determination of the continuum slope within the whole wavelength
range (4000-7500), i.e. on a correct correction for the spectral
sensitivity of the equipment, which is determined by a comparison star.
If the night of the observation had not good
photometric conditions (clouds, mist, etc.), the reduction can
give a wrong {\bf spectral} slope and, consequently, the errors in
scaling the H$\alpha$ wavelength band can be large. Most of the
spectra from the 1-m and 6-m SAO telescopes were obtained
separately in the blue  (H$\beta$) and in the red (H$\alpha$) bands, with
a resolution of 8 -- 9\,\AA. Usually, the red
edge of the blue spectra and the blue edge of the red spectra
overlapped within an interval of $\sim 300$\,\AA. Therefore, as
a zero approximation (the first stage) the majority of red
spectra was scaled using the overlapping continuum region with
the blue spectra, which were scaled with the
[\ion{O}{iii}]$\lambda$\,5007 line. In this case the scaling
uncertainty is about  5\%-10\%. However, for some red spectra, this method
could not be used because some spectra
obtained with a higher resolution ($\sim5$\AA) did not overlap
with blue spectra; or some spectrum ends were distorted by
the reduction procedures of the instrumental set-up; or the blue and red
spectra were not taken during the same night. Therefore, to
increase  the precision of the H$\alpha$ spectral region (the second
stage), all red  spectra were \textbf{once more} scaled to a constant flux
value of the
narrow emission line [OI]$\lambda 6300$, using the modified method of Van
Groningen \& Wanders (\cite{van}, see also Shapovalova et al.
2004). As a reference, we used a red spectrum
obtained with the GHAO 2.1-m telescope during a good photometric
night, and well-scaled by the  [OIII]$\lambda$5007 { line}. \textbf{After
scaling all red  spectra using the [OI]$\lambda 6300$ \AA\ line, we
were able to estimate the quality of each spectrum by comparing the [OI] and the [OIII]  scalings, and we eliminated the low quality spectra in the
 further
analysis.}  The
uncertainty of the scaling of red spectra by the line
[OI]$\lambda$6300 (i.e. actually by the flux of the [OIII]$\lambda$5007
{ line})
was then about (2-3)\%.

Then from the scaled spectra we determined an average flux in the
 continuum at the observed wavelength $\sim 5117$ \AA\, (i.e. at
$\sim 5100$ \AA\, in the rest frame of NGC 4151, z=0.0033), by
averaging the fluxes in the band 5092 -- 5142\,\AA.
To determine  the observed H$\beta$ and H$\alpha$ fluxes, it is
necessary to subtract the continuum. \textbf{The continuum was estimated in 30\,\AA\ windows, and
was fitted by a straight
line between two  windows
centred respectively at 4590\,\AA\, and 5125\,\AA\, for H$\beta$, and
at  6200\,\AA\, and 6830\,\AA\, for  H$\alpha$}. After subtracting the
continuum, we measured the observed
fluxes in the lines, in the following wavelength intervals: 4780 --
4950\,\AA\, for the H$\beta$, and 6415 -- 6716\,\AA\, for
 H$\alpha$.

To measure   the fluxes of H$\gamma$ and HeII$\lambda$4686, we used only
115 blue spectra
from a total of 180. The
\textbf{remaining} 65 blue spectra were not suitable, because
they  begin at  $\lambda >$ 4300 \AA, or had a bad correction for
spectral {\bf sensitivity} at the edge of the blue region (for example,
because of bad weather).
The underlying continuum for
 H$\gamma$ and HeII\,$\lambda$4686 was fitted by a straight
line using estimates of the continuum in a   30\,\AA\ window
centered respectively at 4230\,\AA\, and 5125\,\AA\,.
 After continuum
subtraction, the H$\gamma$ and  HeII$\lambda$4686 fluxes were
measured in the
 following  wavelength intervals: 4268 -- 4450\,\AA\,
for  H$\gamma$, and 4607 -- 4783\,\AA\, for
 HeII\,$\lambda$4686. }

\subsection{Correction for the position angle (PA), seeing and
aperture
 effects}
\label{sec2.3}

\textbf{In order to investigate the long term spectral variability of NGC
4151, it is necessary to have  a consistent  set of  spectra.
Since  NGC 4151 was observed
 with different telescopes, in
different position angles, and with different apertures,
first we had to perform
 corrections for the position angle (PA), seeing and aperture
effects. A detailed discussion on the necessity for these corrections is
given in Peterson et al. (1995), and will not be repeated here.
}

 \subsubsection{Correction for the position angle (PA) effect}
\label{sec2.3.1}

\textbf{The position angle corresponds to the position of the slit of the spectrograph
on the sky (from the North to the East direction). Usually, the
observations were performed with PA=90$^\circ$, but  sometimes it was not
possible, e.g. at the 6m - Nesmith focus, etc. Note that the atmospheric
dispersion was very small, since the object was  always observed close to the meridian ($<$ 2h and $z<30^\circ$).}
{To make the  correction for the position angle, 80 spectra of NGC 4151
were taken with the 1-m and 6-m SAO
telescopes on May 8 and 9,
2003, under photometric conditions and a good seeing
(1.2"-1.5"), in different position angles (PA=0, 45, 90, 135
degrees), and with different spectrograph entrance slits (1", 1.5",
2", 4", 8"). Data sets with PA=90 degrees for different slits
were used as standard, since most of  NGC 4151 spectra in our
monitoring campaign were obtained in this  position angle. Then we
determined corrections for the PA effect $k(pa)$,
as:

$$k(pa)=F(90)/F(i),$$
 where $F(i)$ is the observed
flux at PA=$i$ degrees, and $F(90)$ is the flux obtained at PA=90
degrees. In Table 3  { (available only in electronic form)}, we list the
PA corrections $k(pa)$ for the
H$\beta$ and continuum fluxes, obtained for PA=0, 45, and 135 degrees, with
apertures 2.0"$\times$6.0" and 4"$\times$20.25". As  can be
seen from Table 3, the
variations of $k(pa)$ between the H$\beta$ and continuum fluxes in the same PA
is small ($<$1\%), except for PA=45 degrees where they are
$<$3\%. A maximum PA correction for the continuum flux,
$kp(cnt)$=1.1, was obtained with PA=45 degrees and an aperture 2"$\times$6.0". This PA nearly corresponds to the axis of the
ionized cone (PA$\sim$50 degrees). The line and
continuum fluxes were determined for PA=90 degrees, using a linear
interpolation of the $k(pa)$ values of Table 3. }

\onltab{3}{
\begin{table*}

\caption{ Corrections for the position angle effect.
Columns: 1 - position angle (PA) in degrees; 2 -  projected
spectrograph entrance apertures in arcsec; 3 -  slit position
angle (PA) corrections for emission lines (kp); 4 -
estimated PA correction error  for emission line, e(kp); 5 -
position angle (PA) corrections for continuum flux, kp(cnt); 6 -
the estimated PA correction error  for continuum flux, e(kp(cnt));
}
\label{tab3}
\centering
\begin{tabular}{llllll}
\hline \hline
 PA & Apertura  &kp(H$\beta$)&$\pm$e   &  kp(cnt)&$\pm$e \\
deg. &arcsec    &      &    &  &         \\
\hline
 1     & 2      & 3    & 4   &  5   & 6 \\
\hline
 90 & 2.0$\times$6.0  & 1     &        &1     &     \\
 90 & 4.0$\times$20.4 & 1     &        &1     &     \\
  0 & 2.0$\times$6.0  & 0.969 &0.035   &0.968 &0.017\\
  0 & 4.0$\times$20.4 & 1.011 &0.018   &1.019 &0.009\\
 45 & 2.0$\times$6.0  & 1.062 &0.006   &1.098 &0.024\\
 45 &   4$\times$20.4 & 1.048 &0.003   &1.071 &0.006\\
135 & 2.0$\times$6.0  & 0.954 &0.018   &0.943 &0.011\\
135 & 4.0$\times$20.4 & 0.974 &0.004   &0.981 &0.005\\
\hline
\end{tabular}
\end{table*}
}

 \subsubsection{Correction for the seeing effect}
\label{sec2.3.2}

The Narrow Line Region (NLR) of NGC 4151 has an extended bi-conical
structure spreading up to $>$2" from the nucleus (Evans et al.
1993), while the BLR and non-stellar (AGN) continuum are
effectively point-like sources ($\ll 1''$). Consequently, the
measured NLR flux depends on the size of the spectrograph
entrance aperture (see Peterson et al. 1995 for a detailed
discussion). Also, since we observed with different telescopes
and apertures, for each aperture the measured ratio of the
BLR flux (a point-like source) to the NLR flux (a spatially
extended region) depends on the seeing. Therefore, in each
aperture we must find corrections for images and reduce all flux
data to some accepted standard image. The method
suggested by Peterson et al. (1995) has been used for this purpose. For the seeing
in each aperture we can write:

$$             F(s)=k(s) \cdot F(pa)-G(s)$$
 where $F(pa)$ is the
observed flux at  PA=90 degrees, $F(s)$ is the seeing corrected
flux, $k(s)$ is a point-source correction factor, and $G(s)$
is an extended source correction taking into account the
host galaxy light. Obviously, $G(s)=0$ for the broad-line flux
(point-like source).

We have divided the whole range of seeing values into several
intervals, for two different apertures. For the first aperture,
2.0"(2.5")$\times$6.0", which corresponds to the observations
with the 6-m and 2.1-m telescopes (M\'exico), we considered the
following intervals: 1"-1.5"; 1.5"-2.5"; 2.5"-3.5"; 3.5"-4.5";
and $>$4.5". The data set for the interval 1.5"-2.5" was adopted
as a standard one since the average seeing in the period of
observations with this aperture was about 2". A value $k(s)=$1
and $G(s)$=0 was { accepted for this standard data set.} We
obtained the seeing correction $k(s)$ and the extended source
correction $G(s)$ for the seeing intervals mentioned above using
spectra observed with different seeings within a time interval
shorter than 3 days (i.e. $k(s)=F(1.5"-2.5")/F(pa)_i$, where
$F(pa)_i$ is the observed flux in PA=90 for the $i$-seeing, and
$F(1.5"-2.5")$ is the flux for the seeing interval 1.5"-2.5";
these data are separated by 3 days or less). In Table 4
(available only in electronic form) we listed the $k(s)$ and
$G(s)$ corrections and the data obtained from Peterson et al.
(1995, their Figs. 5 and  6) for the aperture 2"$\times$10". From
this table, one can see that our seeing corrections $k(s)$ for an
aperture of 2"$\times$6" practically coincide (within 1\%) with
those of Peterson et al. (1995) for an aperture of 2"$\times$10".
\textbf{The correction ($k(s)$ in Table 4) for the emission line
fluxes is the same for a slit length of $6^{\prime\prime}$ and of
$10^{\prime\prime}$, meaning that the lines are emitted by a
region smaller than $6^{\prime\prime}$, but there are significant
differences in the host galaxy contribution, as expected.}

\addtocounter{table}{4}
\onltab{4}{
\onltab{4}{
\begin{table*}

\caption{Correction for
seeing effects for apertures (2"$\times$6") and (2.5"$\times$6").
Columns: 1 - Interval seeings in arcsec; 2 - mean seeing (arcsec);
3- our seeing correction for the emission lines fluxes, ks(our);
4 -  estimated seeing correction error, e(ks); 5- Peterson's
seeing correction  for the emission lines fluxes from
Peterson et al (1995), ks(Pet); 6 - our host galaxy seeing
correction for the continuum fluxes, in units of $10^{-14}$\,
erg\,s$^{-1}$\,cm$^{-2}$\,\AA$^{-1}$, Gs(our); 7 -  estimated
host galaxy seeing correction error, e(Gs); 8- Peterson's host
galaxy seeing correction Gs(Pet) for the continuum fluxes from
Peterson et al (1995); }

\label{tab4}
\centering
\begin{tabular}{llllllll}
\hline \hline
  interval &  Mean    &ks(our)&$\pm$e     &ks(Pet)  &Gs(our)&$\pm$e   &Gs(Pet)  \\
  seeings  & seeing   &(2"$\times$6")&        &(2"$\times$10") &(2"$\times$6")&      &(2"$\times$10")         \\
  arcsec   & arcsec   &H$\beta$ wings &      & H$\beta$      &10$^{-14}$      & &10$^{-14}$   \\
\hline
    1      &   2      & 3    & 4        &5     & 6     & 7       &8             \\
\hline}
  1"-1.5"  &  1.25"   &0.965 &0.046     &       &-0.160 &0.197  &-0.070 (1.3")  \\
           &  1.5"    &0.977*&          &0.973  &       &       &               \\
           &          &      &          &       &       &       &               \\
  1.5"-2.5"&  2.0"    &1.000 &          &1.000  & 0     &       & 0             \\
           &          &      &          &       &       &       &               \\
  2.5"-3.5"&  3.0"    &1.042 &0.030     &1.052  &-0.039 &       & 0.130         \\
           &          &      &          &       &       &       &               \\
  3.5"-4.5"&  4.0"    &1.069 &0.063     &1.086  &       &       &               \\
  3.5"-5.2"&          &      &          &       & 0.329 &0.203  &               \\
           &  5"      &1.096 &          &1.105  &       &       & 0.31          \\
\hline
\end{tabular}
\end{table*}
}

For the second aperture, 4.2"$\times$19.8", which corresponds to our
observations with the 1-m Zeiss telescope (SAO), we used the seeing
intervals 2"-4", 4"-6", 6"-8".
They are large because with this telescope it is
impossible to determine the seeing quality with a good precision,
owing to the small { scale along the spectrograph slit (2.2"/px).}
The data set for the interval 2"- 4" was used as a standard. The seeing
corrections $k(s)$ and $G(s)$ for the aperture 4.2"$\times$19.8"  were
obtained with the same procedure described above for the
aperture 2"$\times$6". The results are given in  Table 5
(available only in electronic form), together with those of Peterson et al. (1995)  for the aperture
5"$\times$7.5" (their Figs. 5 and 6).

\addtocounter{table}{5}
\onltab{5}{
\begin{table*}
\caption{Corrections for the seeing effects for aperture
(4.2"$\times$19.8") Columns: 1 - Interval seeings in arcsec; 2 -
mean seeing (arcsec); 3 - our seeing correction for the emission
lines fluxes, ks(our); 4 - the estimated seeing correction error,
e(ks); 5 - Peterson's seeing correction ks(Pet) for the emission
lines fluxes from Peterson et al (1995), ks(Pet); 6 - our host
galaxy seeing correction for the continuum fluxes, in units of
$10^{-14}$\, erg\,s$^{-1}$\,cm$^{-2}$\,\AA$^{-1}$), Gs(our); 7 -
the estimated host galaxy seeing correction error, e(Gs); 8-
Peterson's host galaxy seeing correction Gs(Pet) for the
continuum fluxes from Peterson et al (1995); } \label{tab5}
\centering \begin{tabular}{llllllll} \hline \hline
 interval  & Mean    &ks(our)&$\pm$e     &ks(Pet)   &Gs(our)&$\pm$e    &Gs(Pet)\\
 seeings   &seeing   &H$\beta$&   &H$\beta$        &10$^{-14}$ &&10$^{-14}$    \\
 arcsec    &arcsec   &(4.2"$\times$19.8")&   &(5"$\times$7.5") &(4.2"$\times$19.8") & &(2"$\times$10")  \\
\hline
   1        &2       &  3     &4      & 5       &  6    & 7      &  8        \\
\hline
 2"-4"      &3"      &1.000  &        &1.000    &0.000  &         &0.000     \\
            &        &       &        &         &       &         &          \\
 4"-6"      &5"      &1.063  &$\pm$0.022   &1.004    &0.031  &$\pm$0.232    &0.123     \\
            &        &       &        &         &       &         &          \\
 4"-8"      &6"      &1.080  &$\pm$0.035   &1.015    &0.188  &$\pm$0.292    &          \\
            &        &       &        &         &       &         &          \\
 6"-8"      &7"      &1.116  &        &         &       &         &          \\
            &        &       &        &         &0.503  &         &0.245     \\
\hline
\end{tabular}
\end{table*}
}

Emission line and continuum fluxes were scaled to the mean seeing
2" for the apertures 2"$\times$6.0"(2.5"$\times$6.0"), and to the
mean seeing 3" for the aperture 4.2"$\times$19.8", using the
seeing corrections from Tables 4 and 5. \textbf{After that, we scaled all spectra
to the 2"$\times$6.0" aperture (cf. below).}

 \subsubsection{Correction for the aperture effect}
\label{sec2.3.3}

To correct the observed fluxes for aperture effects, we determined a
point-source correction factor $\varphi$ using  the equation (see
Peterson et al. 1995 for a detailed discussion):

$$   F(Hb)true= \varphi\cdot F(Hb)obs,$$
where $F(Hb)obs$ is the observed H$\beta$ flux after correction
for the PA and seeing effects, as described in \S 2.3.1 and \S
2.3.2; $F(Hb)true$ is the H$\beta$ flux corrected for the aperture
effect.

The contribution of the host galaxy to the continuum flux depends
also on the aperture size. The continuum fluxes $ F_\lambda$(5117)
were corrected for different amounts of host-galaxy contamination,
according to the following expression (see Peterson et al. 1995):

$$   F(5117\AA)true=\varphi\cdot F(5117\AA)obs - G(g),$$
where $F(5117A)obs$ is the observed continuum flux after correction
for the PA and seeing effects, as described in \S 2.3.1 and \S
2.3.2;
 $G(g)$ is an aperture -- dependent correction factor to
account for the host galaxy contribution. The cases L (Table~\ref{tab1}),
which correspond to the aperture ($2''\times 6''$)  of the 6-m
telescope, was taken as a  standard (i.e. $\varphi=1.0$, $G(g)$=0
by definition). The corrections $\varphi$ and $G(g)$ were defined
for each aperture via the comparison of a pair of observations
separated by 0 to 2 days. It means that the
variability on shorter times ($<2$ days) was suppressed by the
procedure of data re-calibration. The point-source correction factors
$\varphi$ and $G(g)$ values for various samples are given in
Table~\ref{tab6}. Using these factors, we re-calibrated the
observed fluxes of H$\alpha$, H$\beta$, H$\gamma$ and
HeII$\lambda$4686 and of the continuum to a common scale
corresponding to the aperture  $2''\times
6''$ (Table~\ref{tab7} { -available only in
electronic form}).

\begin{table} 
\begin{center}
\begin{minipage}[t]{\columnwidth}
\renewcommand{\footnoterule}{}
\caption[]{Flux scale factors for optical spectra}\label{tab6}
\begin{tabular}{|l|l|l|l|l|}
\hline
Sample      & Years     & Aperture     &Point-Source     &Extended Source \\
            &           & (arcsec)     &Scale factor     &Correction      \\
            &           &              & ($\varphi$)           & G(g)\footnote{in units $10^{-14}\rm (erg\,s^{-1}\,cm^{-2}\,\AA^{-1})$}  \\
\hline
  L(U,N)    &  1996-2005&   2.0$\times$6.0    &   1.000         &    0.000       \\
  GH,S-P    &  1996-2005&   2.5$\times$6.0    &   1.000         &    0.000       \\
  L(S)      &  2004-2006&   1.0$\times$6.0    &   0.950$\pm$0.000  &   -0.391       \\
  L1(G)     &  1996-2003&   4.2$\times$19.8   &   1.035$\pm$0.021  &    1.133$\pm$0.037\\
  L1(G)     &  1996-2003&   8.0$\times$19.8   &   1.112$\pm$0.005  &    1.623$\pm$0.080\\
  L1(U)     &  2004-2006&   4.0$\times$9.45   &   0.962$\pm$0.044  &    0.750$\pm$0.150\\
\hline
\end{tabular}
\end{minipage}
\end{center}
\end{table}

\addtocounter{table}{7}

The fluxes listed in Table~\ref{tab7}  were not corrected for contamination by the
narrow-line emission components of H$\gamma$, HeII$\lambda 4686$,
H$\beta$, H$\alpha$,  and $[\ion{N}{ii}]\lambda\lambda$\,6548,
6584. These contributions are expected to be constant and to have
no influence on the broad line variability.

The mean error (uncertainty) in our flux determinations for
H$\alpha$ and H$\beta$  and  for the continuum is $<$3\%,
while it is $\sim 5$\% for H$\gamma$ and $\sim 8$\% for
HeII\,$\lambda$4686. These quantities were estimated by comparing
the results from spectra obtained within time intervals
shorter than 2 days. The estimated mean errors for every year and
for the total period of monitoring are given in
Table~\ref{tab8} (available only in electronic form).

\addtocounter{table}{8}
\onltab{8}{
\begin{table*}
\caption{ The mean error (uncertainty) of our flux determinations
for continuum, H$\alpha$, H$\beta$, H$\gamma$ and HeII$\,4686$
emission lines in different years, and in the whole (11 years)
observation period. Columns: 1 - year;
2 - $\varepsilon_{c}$, the estimated mean continuum flux error in
\%; 3 - sigma of the estimated  mean continuum flux  error in \%;
4 - $\varepsilon_{H\alpha}$, the estimated mean H$\alpha$ flux
error in \%; 5 - sigma of the estimated  mean H$\alpha$ flux
error in \%; 6 - $\varepsilon_{H\beta}$, the estimated mean
H$\beta$ flux error in \%; 7 - sigma of the estimated  mean H$\beta$
flux error in \%; 8 - $\varepsilon_{H\gamma}$, the estimated mean
H$\gamma$ flux error in \%; 9 - sigma of the estimated  mean
H$\gamma$ flux error in \%; 10 - $\varepsilon_{HeII}$, the
estimated mean  flux error in \%; 11 - sigma of the estimated  mean
HeII$\,4686$ flux error in \%; Bottom - mean error (uncertainty)
in the whole (11 years) observation period.} \label{tab8} \centering
\begin{tabular}{lllllllllll}
\hline  \hline
 Year   &$\varepsilon_{c}$&$\pm\sigma$&$\varepsilon_{H\alpha}$&$\pm\sigma$
 &$\varepsilon_{H\beta}$&$\pm\sigma$ &$\varepsilon_{H\gamma}$&$\pm\sigma$
 &$\varepsilon_{HeII}$&$\pm\sigma$\\
\hline
  1     &2    &3     &4     &5    &6      &7     &8     &9     &10     &11\\
\hline
 1996  &3.9   &2.27  &1.9   &1.2   &3.90  &1.56  &7.5   &      &10.4   &4.20\\
 1997  &5.17  &2.98  &2.19  &1.63  &1.63  &1.81  &6.0   &1.62  &10.22  &4.20\\
 1998  &2.18  &1.37  &3.1   &2.4   &1.90  &1.42  &5.75  &1.62  &10.18  &4.70\\
 1999  &3.24  &2.22  &1.75  &1.28  &2.29  &1.68  &5.65  &4.95  & 6.26  &3.00\\
 2000  &2.3   &3.65  &3.62  &2.31  &2.12  &0.82  &2.88  &2.53  & 4.65  &4.47\\
 2001  &3.77  &3.16  &2.9   &1.13  &1.63  &2.25  &4.2   &1.7   &12.90  &4.67\\
 2002  &2.05  &1.38  &2.7   &1.66  &3.91  &2.44  &5.37  &5.9   & 5.60  &1.56\\
 2003  &3.34  &2.37  &2.8   &1.49  &2.86  &2.44  &4.82  &2.67  & 7.62  &4.76\\
 2004  &2.65  &2.26  &2.6   &0.8   &2.25  &3.18  &3.85  &4.45  & 7.28  &4.57\\
 2005  &2.37  &2.14  &2.72  &2.63  &1.96  &2.22  &3.3   &1.31  & 7.37  &5.86\\
 2006  &2.42  &1.76  &2.12  &1.68  &1.62  &1.34  &6.58  &1.97  & 8.62  &4.71\\
\hline
 mean  &2.84  &1.23  &2.46  &0.75  &2.37  &0.84  &4.99  &1.46   &7.78  &2.51\\
(1996-2006)&&&&&&&&&&                                                        \\
\hline
\end{tabular}
\end{table*}
}

\subsection{ The narrow emission line contribution}
\label{sec2.3.4}

 In order to estimate the narrow line contributions
to the broad line fluxes, we constructed a spectral
template for the narrow lines. To this end, we used the blue and
red spectra in the minimum activity state (May 12,
2005), obtained with a spectral resolution of $\sim 8\,\AA$. In
these spectra, the broad H$\beta$ component was very weak, and
the broad components from the higher Balmer line series were
absent.

  Both the broad and the
narrow components of  H$\gamma$, HeII$\lambda 4686$,  H$\beta$ and
 H$\alpha$, were fitted by Gaussians. The template spectrum contains the
following lines:  for  H$\gamma$:  the narrow component of
H$\gamma$ and [\ion{O}{iii}]$\lambda$\,4363;  for  H$\beta$:
the narrow component of  H$\beta$ and
[\ion{O}{iii}]$\lambda\lambda$ 4959, 5007; for  H$\alpha$: the
narrow component of H$\alpha$,
[\ion{N}{ii}]$\lambda\lambda$\,6548, 6584,
{ [\ion{O}{i}]$\lambda\lambda$\,6300, 6364,) }
[\ion{S}{ii}]$\lambda\lambda$\,6717, 6731 and HeI
$\lambda$\,6678. In Table~\ref{tab9} are listed the narrow line
contributions obtained from the template spectrum in the same
wavelength integration intervals as for integral fluxes. Our
results are in good  agreement with those given by Sergeev et
al. (2001).

\begin{table} 
\begin{center}
\begin{minipage}[t]{0.8\columnwidth}
\renewcommand{\footnoterule}{}
\caption[]{The wavelength integration intervals and the corresponding NLR contributions
}\label{tab9}
\begin{tabular}{|l|l|l|}
\hline
 Components    &integration    &NLR flux\footnote{in $10^{-12}$ erg cm$^{-2}$ sec$^{-1}$}\\
               &intervals (\AA)  &    \\
\hline
 H$\alpha$+[NII]      &6415-6716      &7.93\\
 H$\beta$-nar.        &4780-4950      &1.00\\
 H$\gamma$+[OIII]$\lambda$4363 &4268-4450      &1.04\\
 HeII+[ArIV]+  &4607-4783      &0.36\\
 $\rm [FeIII]\,$ 4658 & &                 \\
\hline
\end{tabular}
\end{minipage}
\end{center}
\end{table}

\section{Data analysis}
\label{sec3}

\subsection{ Variability of the emission lines and of the optical continuum}
\label{sec3.1}

\begin{figure}
\centering
\includegraphics[width=8.5cm]{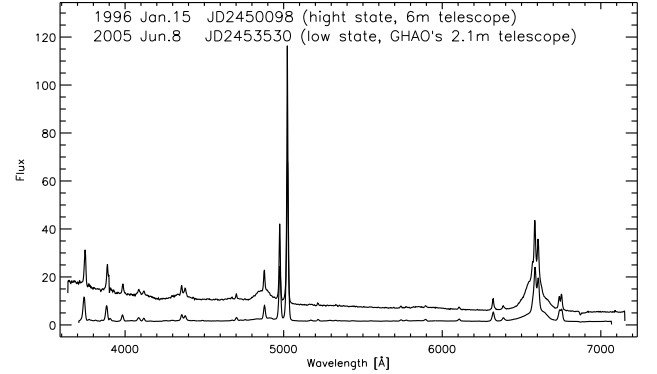}
\caption{The spectra of NGC 4151 corresponding to the high
activity state
  (top) and to the low activity state (bottom).
   The observed wavelength
  (we recall that z=0.0033) is displayed on the X-axis, and the flux (in units of
   $10^{-14}$\,erg\,cm$^{-2}$\,s$^{-1}$\,\AA$^{-1}$) is displayed
on the Y-axis.}\label{fig1}
\end{figure}

\begin{figure}
\includegraphics[width=9.5cm]{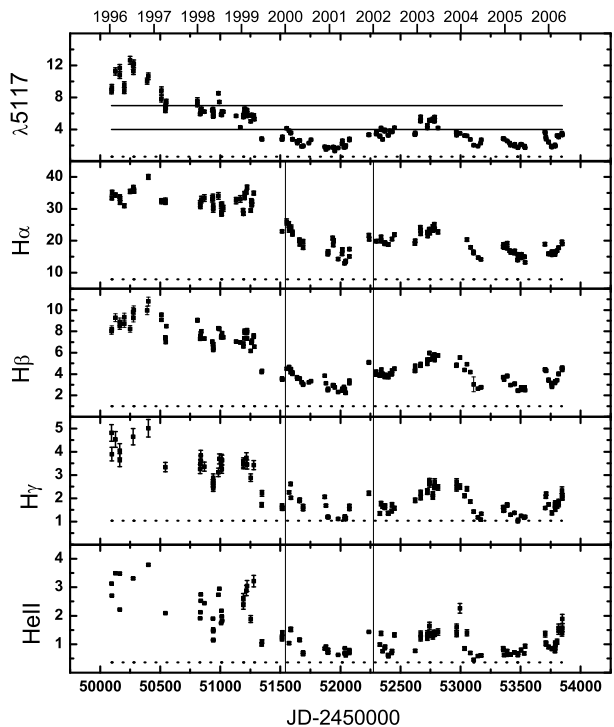}
\caption{The light curves of  H$\alpha$, H$\beta$, H$\gamma$,
HeII$\lambda$4686 and of the continuum at the observed wavelength
$\lambda=$5117 \AA, in the period 1996-2006. {Constant
contributions from the narrow lines (Table~\ref{tab9}) and from
of the host galaxy (see 3.1) are shown by horizontal dashed
lines.} The line fluxes  are given in units of 10$^{-12} \rm erg\
cm^{-2} s^{-1}$, and  the continuum flux in units of 10$^{-14}\rm
erg\ cm^{-2} s^{-1} \AA^{-1}$. \textbf{The horizontal lines in the first
panel
correspond to the division of the continuum flux into three intervals  (see \S
3.4.2), and the vertical lines correspond to the division into time intervals based on
the similarity of the line shapes  (see \S
3.3).}} \label{fig2}
\end{figure}

{ The spectra for the high- and
low-activity states, obtained respectively on January 15, 1996 (6m
SAO's telescope) and on June 8, 2005 (2.1 m GHAO's telescope) are
presented in Fig.~\ref{fig1}. As can be seen, the continuum flux decreased
by a large factor ($\sim 5.6$ times) in the
low-activity state, and the slope of the
continuum in the blue was significantly flatter than  in the
high-activity state. Besides, the
wings of  H$\beta$ and H$\alpha$ became
extremely weak in the minimum state, and those of
  H$\gamma$ and of the higher Balmer line series
 could not be detected at all. These profiles correspond to a
Sy 1.8 type and not to a Sy1-Sy1.5, as this AGN could be classified
in the maximum state. So, the spectral type of the object is
changing with time. This was noted earlier. In 1984-1989,  the
nucleus of NGC 4151 went through a very deep minimum. At that time, the
brightness of the source fell down to the level of the
host galaxy for an aperture of 27" in the V-band, the broad wings of hydrogen
lines became much weaker (they almost completely vanished in April
1984) and the spectrum of the nucleus was identified as a Sy 2
(Penston \& Perez 1984).

In Fig.~\ref{fig2} are presented the light curves obtained from
Table~\ref{tab7}, for the H$\alpha$, H$\beta$, H$\gamma$ and
HeII$\lambda$4686 integrated line fluxes and for the
continuum at the observed wavelength 5117\AA\,. { The
fluxes  of
H$\gamma$, HeII$\lambda$ 4686, H$\beta$, and H$\alpha$  were not corrected
for contamination by the constant contributions of the narrow lines.
The contributions of the narrow lines given in Table~\ref{tab9} are shown
in Fig.~\ref{fig2} by horizontal dashed lines.
It is clearly  seen that  if the fluxes of the narrow lines
are subtracted from  H$\gamma$ and HeII$\lambda$ 4686 during the
minimum of activity, these lines disappear.}

 The continuum flux
presented in Table~\ref{tab7} and in Fig.~\ref{fig2} contains also
a constant contribution from the starlight  of the  host galaxy, which is estimated as
$F{\rm (host)}
=\rm 1.54\ 10^{-14} erg\ cm^{-2} s^{-1} \AA^{-1}$ through an
aperture of 5"$\times$7.5" (Peterson \& Cota 1988), and $\rm
10^{-14} erg\ cm^{-2} sec^{-1} \AA^{-1}$ through an aperture of 3"$\times$10"
(Mal'kov et al. 1997). Bochkarev et al. (1991) determined that
the host galaxy contributed to about 40\% of the total flux of NGC 4151
in the H$\beta$ wavelength band though an aperture 1"$\times$4.0", near the
minimum state (1987). As can be seen from
Table~\ref{tab7}, the minimum flux in the continuum  ($\rm \sim 1.6\ 10^{-14}
erg\ cm^{-2} s^{-1} \AA^{-1}$) obtained with an
aperture 2"$\times$6" was observed from November 29 to December 17, 2000. If
we assume that the host galaxy contribution is about 40\% of the continuum (Bochkarev
et al., 1991), it gives
$F{\rm (host)}\rm >0.6\ 10^{-14} erg\ cm^{-2} s^{-1} \AA^{-1}$ (i.e.
lower limit). Using a linear regression between the continuum
flux and the H$\alpha$ and H$\beta$ broad line fluxes (the
narrow line flux being subtracted) near the low-activity state, and
extrapolating the broad line flux to zero, we estimated
$F$(host)$\approx \rm (0.6\pm 0.3)x10^{-14} erg\ cm^{-2} s^{-1}
\AA^{-1}$. This value is in good agreement with other estimates with
different apertures. {The estimated contribution from the host galaxy
is shown by the horizontal dashed line in Fig.~\ref{fig2} (top).}

{The light curve of the continuum is similar to those of the
 emission lines,  showing
a maximum in 1996 and two minima  in 2001 and 2005.

In Table~\ref{tab10}, we give for the lines and continuum, the mean observed
maximum flux $F$(max) in  the interval JD=2450094-2450402 (1996),
the mean observed  minimum flux $F$(min)
in the intervals JD=2451895-2452043 (December 2000 - May 2001), and
JD=2453416-2453538 (2005), the observed  ratio $F$(max)/$F$(min),
and this ratio for the broad  lines after subtraction of the narrow components
and the contribution of the host galaxy (agn continuum
in Table~\ref{tab10}).}

\begin{table} 
\begin{center}
\begin{minipage}[d]{0.5\textwidth}
\renewcommand{\footnoterule}{}
\caption[]{ Mean observed maximum flux F$^{\rm max}_{\rm obs}$ in the 
intervals
JD=2450094-2450402 (1996), mean observed min flux F$^{\rm min}_{\rm 
obs}$ in
the intervals JD=2451895-2452043 (December 2000 - May 2001)
and 2453416-2453538 (2005),  observed ratio $R_F=$F$^{\rm max}_{\rm 
obs}$/F$^{\rm min}_{\rm obs}$,
and this ratio for the broad  lines ($R_F^{\rm broad}$) after removing the 
narrow components
and the contribution of the host galaxy.}\label{tab10}
\begin{tabular}{|l|l|l|l|l|l|}
\hline
 Lines or  &F$^{\rm max}_{\rm obs}$    &F$^{\rm min}_{\rm obs}$   
&$R_F$  &$R_F^{\rm broad}$\\
 continuum\footnote{Line fluxes are in units $\rm 10^{-12} erg\ cm^{-2}s^{-1} $;
and continuum flux is in units $1\rm 0^{-14} erg \ cm^{-2}s^{-1} \AA^{-1}$}
 &  &  &   &   \\
\hline
 H$\alpha$   &34.860    &15.340   &2.3             & 3.6          \\
 H$\beta$    & 9.170    & 2.640   &3.5             & 5.0          \\
 H$\gamma$  & 4.365    & 1.178   &3.7             &24!(not broad)\\
 HeII$\lambda$4686  & 3.170    & 0.714   &4.4             & 7.9
\\
\hline
cont(5117A)&10.69     &1.9      &5.6             & 7.8          \\
           &          &         &                &agn continuum\\
\hline
\end{tabular}

\end{minipage}
\end{center}
\end{table}

 The maximum amplitude ratios of the  broad component line flux during
the 1996-2006 period were: $\sim3.6$ for H$\alpha$; $\sim 5.0$ for H$\beta$
line; and $\sim 7.8$ for the $\lambda 5117\,\AA$  agn continuum
after subtraction of the host galaxy flux. In the
low-activity state, the broad component of
HeII$\lambda$4686 and H$\gamma$ was almost absent.

\subsection{ Flux variability in the wings and core of the H$\alpha$ and
H$\beta$ emission lines} \label{sec3.2}

\begin{figure*}
\includegraphics[width=19.0cm]{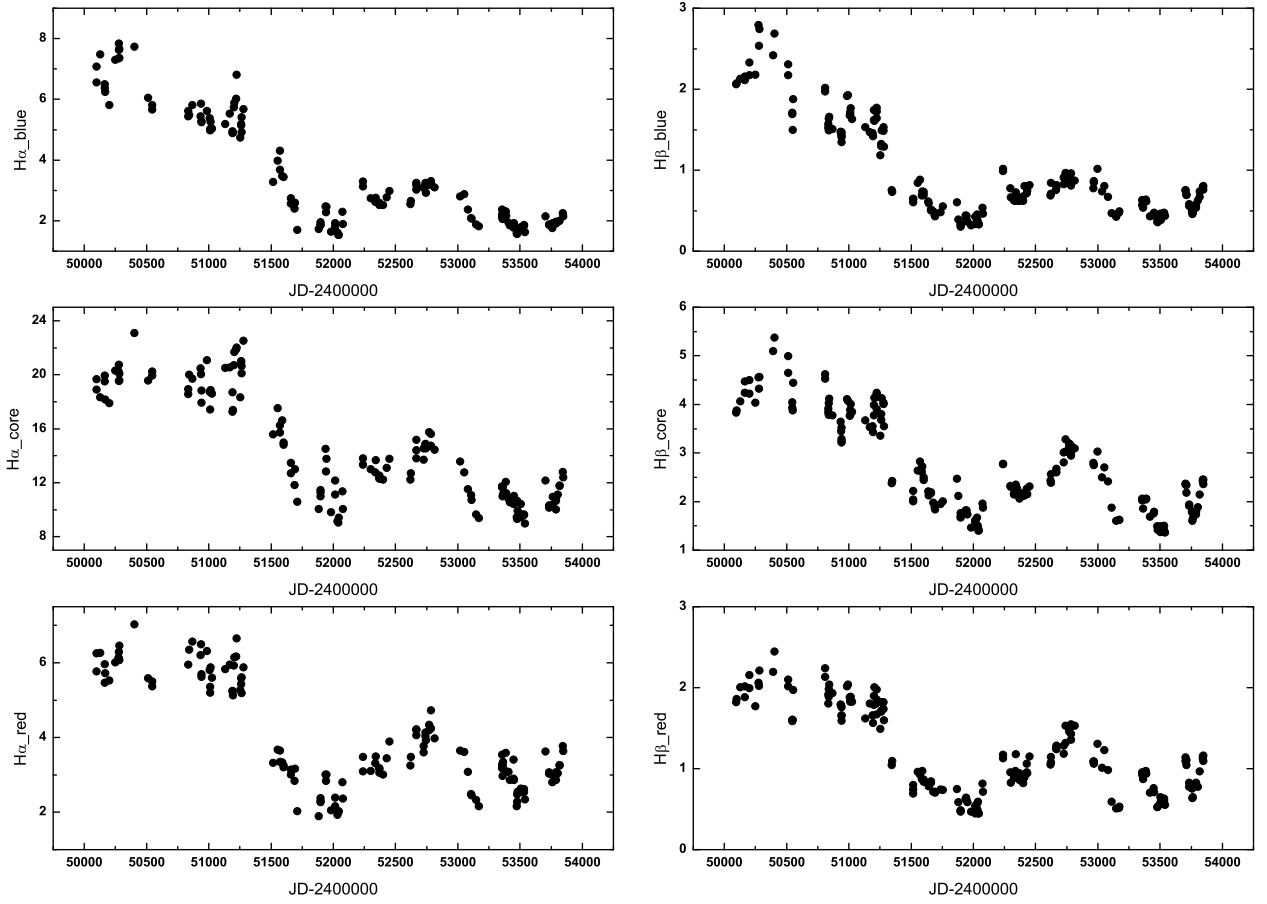}
\caption{ The variation of the wings and core of H$\alpha$
(left) and H$\beta$ (right) from 1996 to 2006. The flux  is
given in units of 10$^{-12} \rm erg\ cm^{-2} s^{-1}$.}\label{fig3}
\end{figure*}

\begin{figure*}
\includegraphics[width=19.0cm]{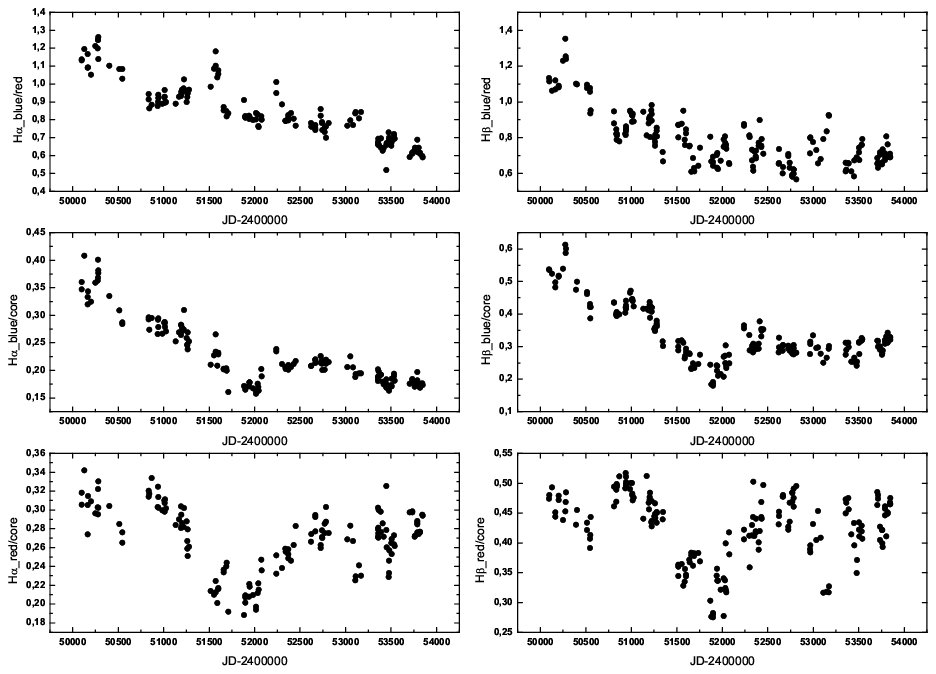}
\caption{ The variation of the ratio of the fluxes in the line wings
and in the core of H$\alpha$ (left) and H$\beta$ (right) (see the text), from
1996 to 2006. }
\label{fig4}
\end{figure*}

{We have divided the H$\alpha$ and H$\beta$ profiles into three
parts: the blue wing, the core, and the red wing,  each part covering
a range of 3000 km/s. Distinct features or peaks observed
at different epochs in the wings were included,
and  the corresponding narrow lines were
included in the core. In Table 11, we give the wavelength intervals used
to measure the flux in the three parts {of each profile}. We also
give the corresponding velocity intervals with respect to the center of the
narrow component.

\begin{table} 
\begin{center}
\caption[]{The wavelength intervals for the
H$\alpha$ and H$\beta$ wings and cores.}\label{tab11}
\begin{tabular}{|l|l|l|}
\hline
Component       &Integration   &Integration\\
                &interval      &interval   \\
                &in \AA\       &in km/s  \\
\hline
 H$\alpha$ blue wing   &6486-6552     &(-4510) - (-1503)\\
 H$\alpha$ core        &6553-6616     &(-1458) - (+1402)\\
 H$\alpha$ red wing    &6617-6684     &(+1458) - (+4510)\\
 H$\beta$ blue wing   &4804-4853     &(-4493) - (-1479)\\
 H$\beta$ core        &4854-4900     &(-1417) - (+1412)\\
 H$\beta$ red wing    &4901-4950     &(+1474) - (+4488)\\
\hline
\end{tabular}
\end{center}
\end{table}

Light curves for the  wings and cores of  H$\alpha$ and H$\beta$ are
presented in Fig. \ref{fig3}. As it can be seen, the flux in the wings and cores
of both lines had a very similar behaviour during the
monitoring period.

In Fig. \ref{fig4} we present {the flux} ratios between the three parts of the
H$\alpha$ and H$\beta$ {profiles}. In both lines,
the blue wing had a stronger flux than the red one in the period of
maximal activity (March 1996 to 1997 {or JD=2450094-2450500) ($F$(blue)/$F$(red)$ >$1, see  Fig. \ref{fig4}, top-left). From 1997 to 2000 (or JD=2450540-2451500), the H$\alpha$
 blue/red flux ratio was very close to unity, while  {the H$\beta$ blue/red} ratio
varied from 0.95 to 0.8 (Fig. \ref{fig4}, top-right). In 2000-2006 (JD=2451550-2453850), the red wing was the strongest
for both lines, and { the H$\alpha$ blue/red ratio} decreased
almost monotonically  from 0.8 to 0.6
(Fig. \ref{fig4}, top-left), while it
varied from 0.6 to 0.8 for  H$\beta$ (Fig. \ref{fig4},
top-right)}.

The $F$(blue)/$F$(core) ratio of both lines was decreasing nearly
monotonically from 1996 to 2001 {(JD=2450094-2452000), and} after
2002 {(JD$>$2452300)}, the  H$\beta$ ratio remained nearly
constant ($\approx$ 0.3), while the H$\alpha$ ratio decreased very  slightly
{(Fig. \ref{fig4}, middle).} On the other hand, the
$F$(red)/$F$(core) ratio of both lines showed approximaly the same rapid
changes in the monitoring period.

\subsection{ Mean and Root-Mean-Square Spectra}
\label{sec3.3}

\begin{figure}  
\includegraphics[width=4.350cm]{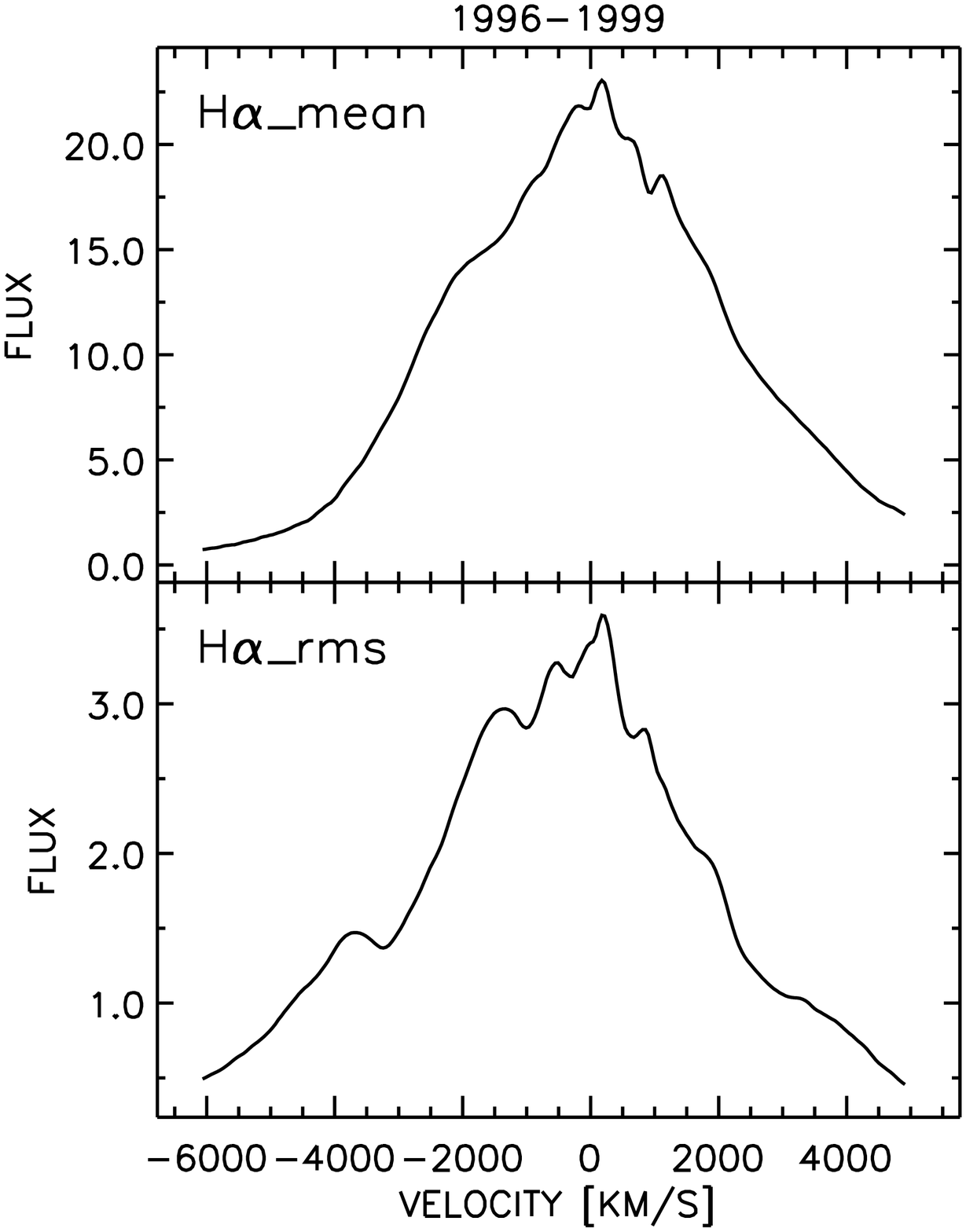}
\includegraphics[width=4.350cm]{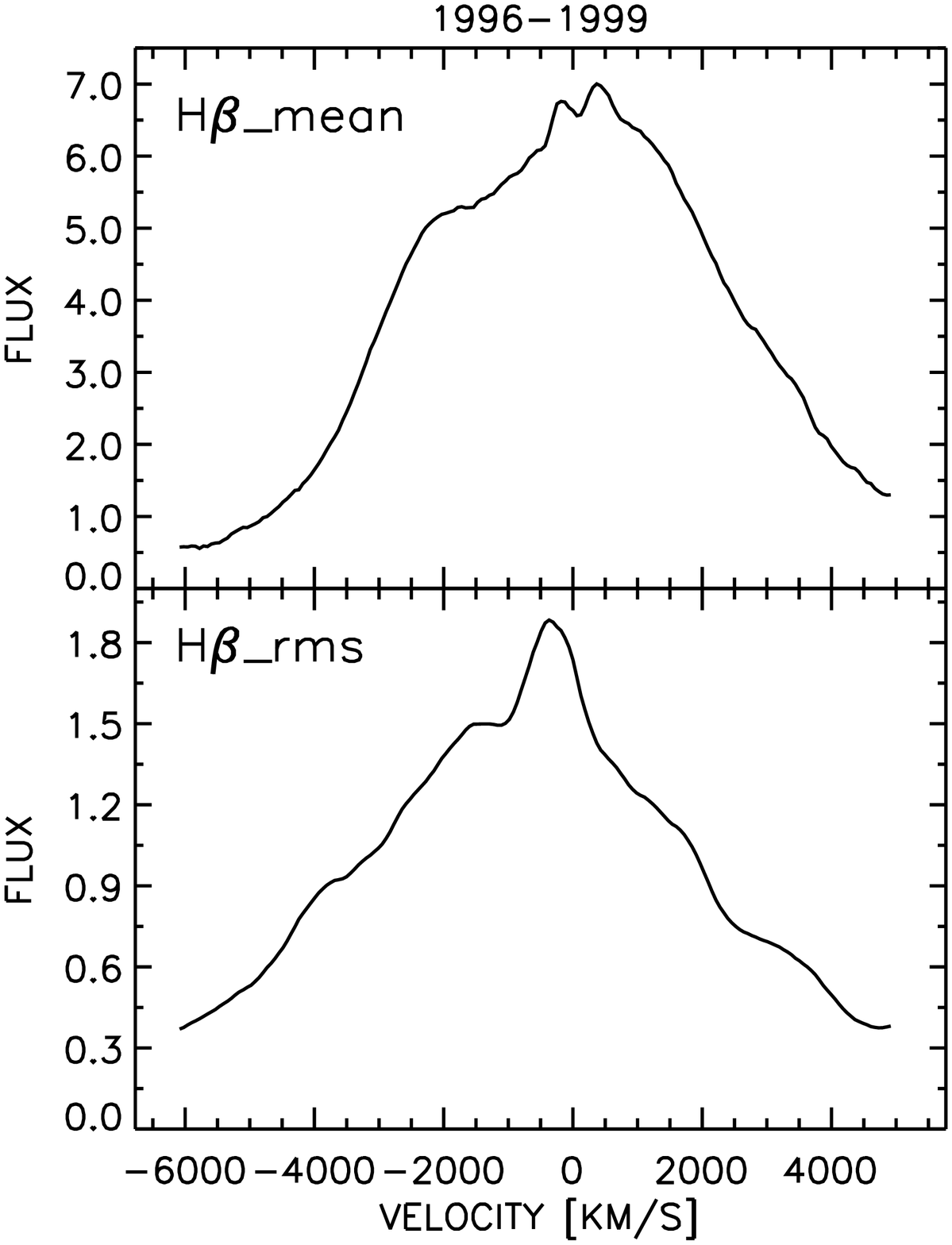}
\caption{The averaged and rms profiles of \textbf{the broad} H$\alpha$ (
left) and
H$\beta$ (right) \textbf{lines} for  the first
period.}
\label{fig5}
\end{figure}

\begin{figure}  
\includegraphics[width=4.350cm]{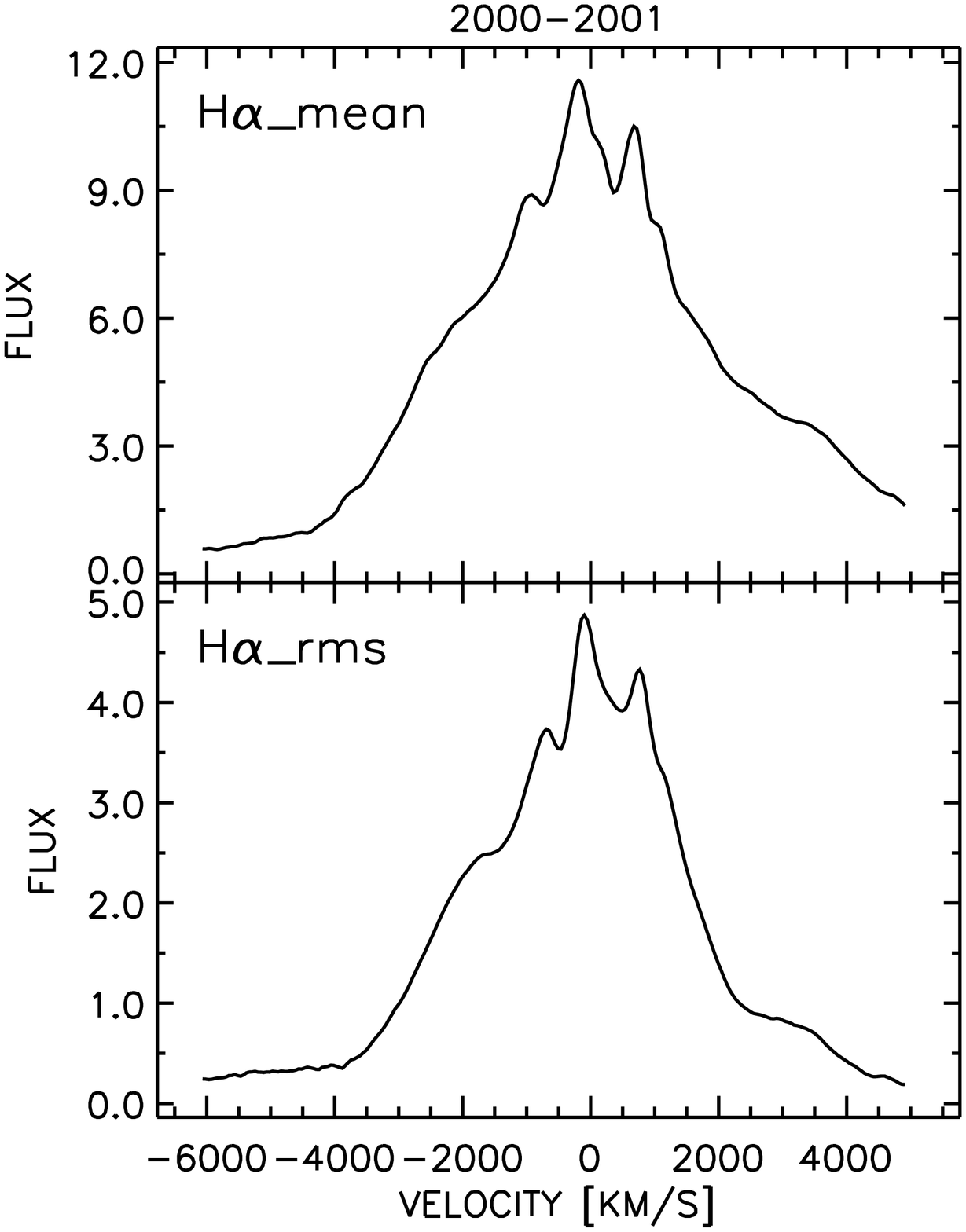}
\includegraphics[width=4.350cm]{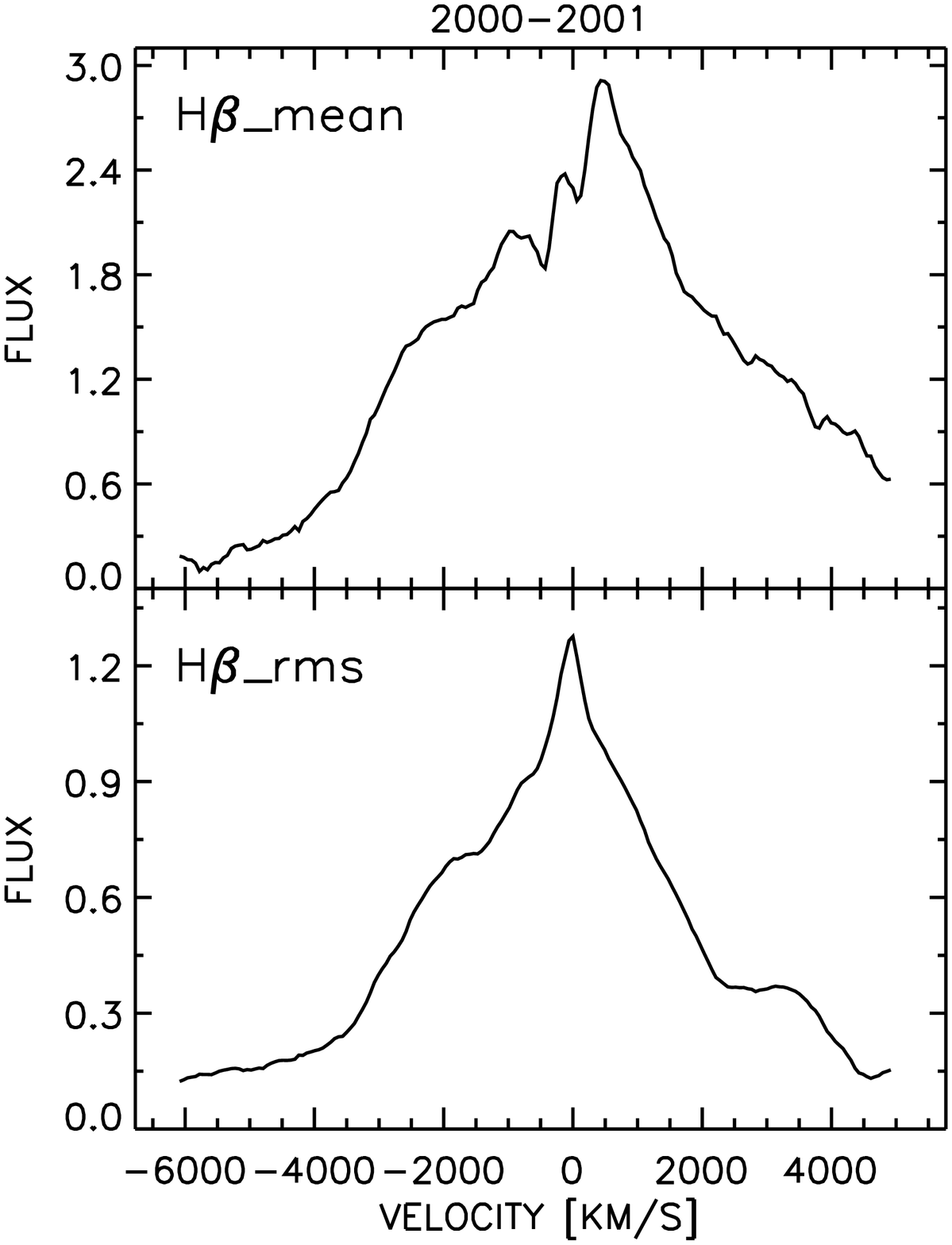}
\caption{The averaged and rms profiles of \textbf{the broad}  H$\alpha$
{(left) and
H$\beta$ (right) \textbf{lines}} for the second
period.}
\label{fig6}
\end{figure}

 The comparison between the averaged and the root-mean-square (rms)
 spectra allows to investigate the line profile variability. We first made
an inspection of the H$\alpha$ and H$\beta$
profiles for different periods, {using spectra with a
resolution of 8 \AA}. With a criteria
based on the similarity of line profiles, we found three
characteristic profiles during the  period 1996-2006. In the first
period (1996-1999, JD=(2450094.466-2451515.583), where the lines were
very intense, a red asymmetry and a shoulder in the blue wing
were present. In the second period (2000-2001,
JD=2451552.607-2452238.000), the broad lines were weaker and the
shoulder in the blue wing {is smaller and a shoulder in the red part is
present}. From 2002 to 2006 (third period,
JD=2452299.374-2453846.403),  the lines showed a blue asymmetry,
and a shoulder in the red part \textbf{was dominant in the line profiles}
(peak at $\lambda\approx$ 4915\AA\ or at
 $\approx$ 2000 km/s relative to the narrow component, see Figs. 5-7, {top}).

Averaged and rms profiles of H$\beta$ and H$\alpha$ for
each of  these three periods and for the whole monitoring period
 (1996-2006) were calculated after removing the continuum. They are shown
in Figs. 5-8.

\textbf{ We measured
the Full
Width at Half Maximum  (FWHM) in the rms and averaged broad line profiles,  and we defined the asymmetry $A$ as the
ratio of the red/blue Half Width at Half Maximum (HWHM), i.e. $A=HWHM_{\rm
red}/HWHM_{\rm blue}$. The measured values for the broad H$\beta$ and
H$\alpha$
lines and their rms are given in Table 12.}

As can be seen in Fig. \ref{fig5} (bottom), the blue component
was highly variable in the first period, the rms profile of  H$\alpha$ and
  in a lesser extent of  H$\beta$, showing two peaks or shoulders
at $\sim -4000$ km/s and $\sim -2000$ km/s. In the red part
of the rms profile of H$\alpha$, a weak bump at ~+2000 km/s and a
shoulder
at ~+3500 km/s was also detected, while in the rms profile of  H$\beta$,
only weak shoulders were seen at the same
places. \textbf{On the other hand, the line and their rms profiles show a
blue
asymmetry ($A<1$, see Table 12).}
In the second period (see Fig. \ref{fig6},  bottom), the feature
at $\sim -4000$ km/s in the blue part of the rms profile of both lines
disappeared, and only a shoulder at $\sim
-2000$ km/s was present. \textbf{The averaged H$\alpha$ profile has a blue
asymmetry, but the  H$\beta$ one is  almost symmetric.} In
the  red part of the rms profile of both lines,  the
shoulder seen in the first period at {$\sim 3500$} km/s was still
present. This feature, but shifted at  ~2500  km/s, was dominant in
the third period (Fig. \ref{fig7}), not only in  the rms profiles of
H$\alpha$ and H$\beta$,
but also in  their averaged profiles. \textbf{Both lines show a red asymmetry  in this
period, but it is interesting to note that the
H$\alpha$ rms profile shows a significant blue asymmetry  ($A\approx$
0.82), while the H$\beta$ rms profile has a significant red asymmetry
($A\approx$1.17). }
  Averaged and rms profiles
are given in Fig. \ref{fig8} for the whole monitoring period from 1996 to 2006.
As can  be seen, two shoulders dominate the rms profiles: a blue one at
$\sim$-2000 km/s
and a red one at $\sim$1500 km/s. Also, the variation in the blue part
is more important than in the red one, because the line intensities during
the whole monitoring period are dominated by the first period, when the
variation  in the blue part was the most important and at the same time
the lines were the most
intense. \textbf{ Table 12 shows also that for the whole monitoring period, the rms
profiles  of  both lines have a significant blue asymmetry, and that the averaged and rms FWHM of H$\beta$
are larger than those of H$\alpha$ by about 1000 km/s.}

\begin{table*}
\begin{center}
       \caption[]{The Full Widths at Half Maximum (FWHM) and the ratio $A={\rm HWHM_{red}/HWHM_{blue}}$ of the
red/blue of
Half Widths at Half Maximum (HWHM), for the averaged and rms profiles of
H$\alpha$, H$\beta$  during the three
periods. }
\begin{tabular}{|c|c|c|c|c|c|c|c|c|}
\hline

period &FWHM (H$\alpha$)& A$_\alpha$ & FWHM (H$\beta$) &
A$_\beta$\\
\hline
 first & 4780$\pm$350 &0.944$\pm$0.012 & 5980$\pm$550 &0.935$\pm$0.048 \\
second & 4020$\pm$570 &0.872$\pm$0.053 & 5550$\pm$750 &1.086$\pm$0.067 \\
third & 5790$\pm$410 &1.491$\pm$0.075 & 6350$\pm$430 &1.282$\pm$0.112 \\
\hline
mean profile & 4650$\pm$420 &1.000$\pm$0.023 & 6110$\pm$440
&1.056$\pm$0.018 \\
\hline
\hline
period &FWHM (rms H$\alpha$)& A$_\alpha$ - rms & FWHM (rms H$\beta$) &
A$_\beta - rms$\\
\hline
 first & 4790$\pm$350 &0.810$\pm$0.046 & 5430$\pm$530 &0.596$\pm$0.017 \\
second & 3100$\pm$480 &1.061$\pm$0.026 & 3700$\pm$550 &0.811$\pm$0.068 \\
third & 3150$\pm$350 &0.816$\pm$0.038 & 3760$\pm$1100 &1.168$\pm$0.210
\\
\hline
mean rms & 4420$\pm$320 &0.830$\pm$0.011 & 5490$\pm$420 &0.775$\pm$0.023
\\
\hline
\end{tabular}
\end{center}
\end{table*}

 Such a behaviour of the rms profiles during the three periods indicate
that the BLR
of NGC 4151 has a complex structure and that its geometry may change in time.
More informations about  line profiles will be given in a forthcoming
paper.

\begin{figure} 
\includegraphics[width=4.350cm]{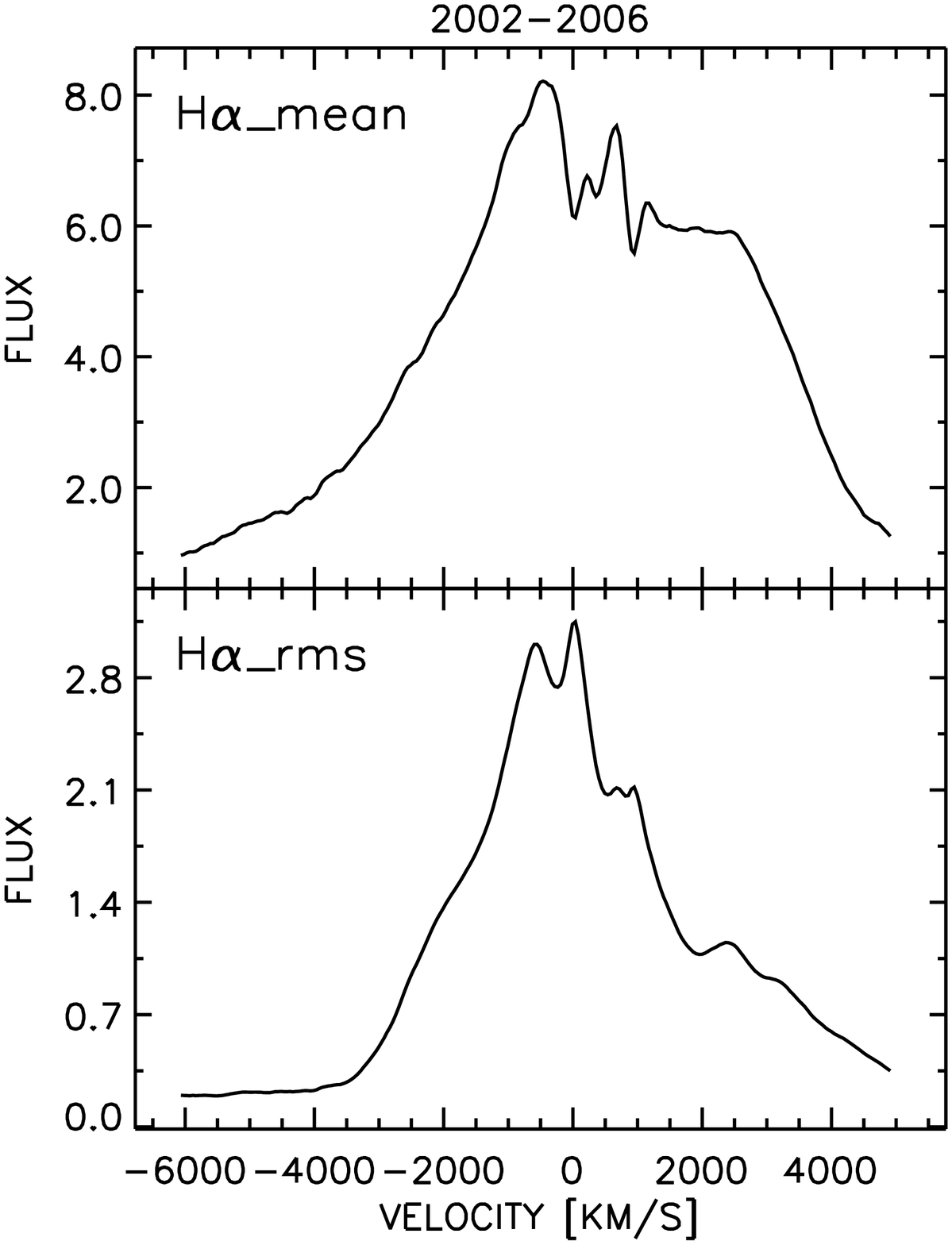}
\includegraphics[width=4.350cm]{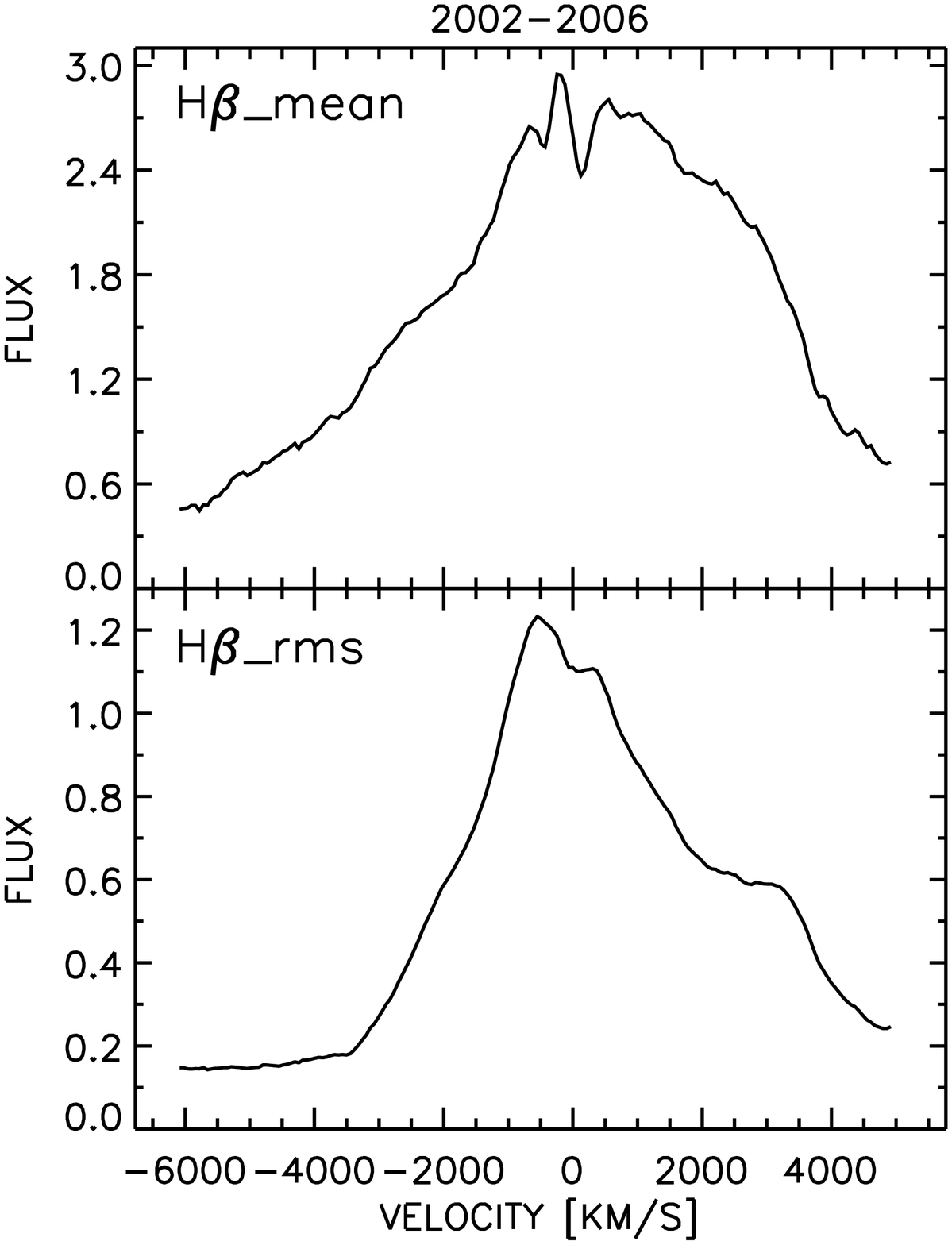}
\caption{The averaged and rms profiles of \textbf{the broad} H$\alpha$
{(left) and
H$\beta$ (right)} \textbf{lines} for the third period.}
\label{fig7}
\end{figure}

\begin{figure}  
\includegraphics[width=4.350cm]{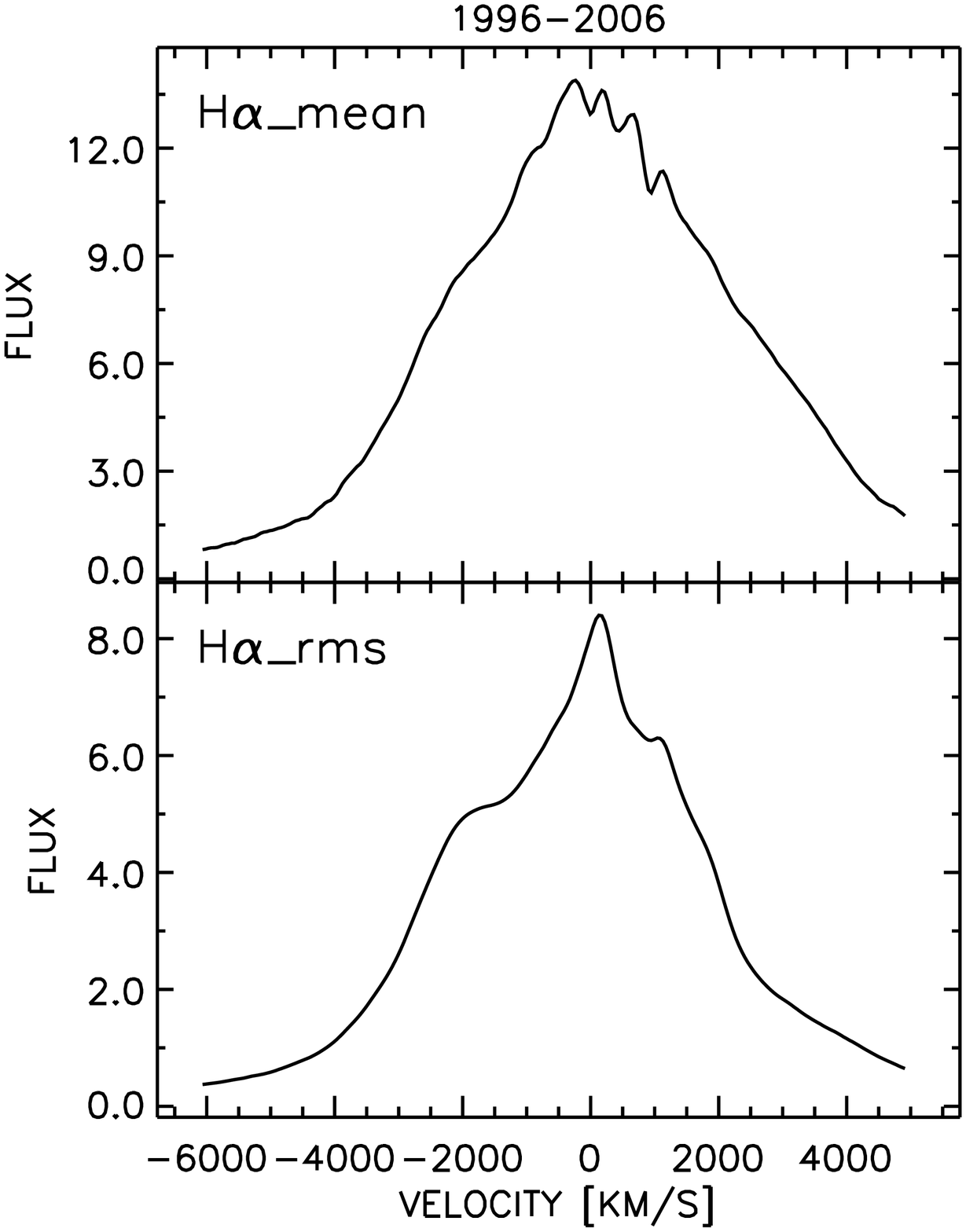}
\includegraphics[width=4.350cm]{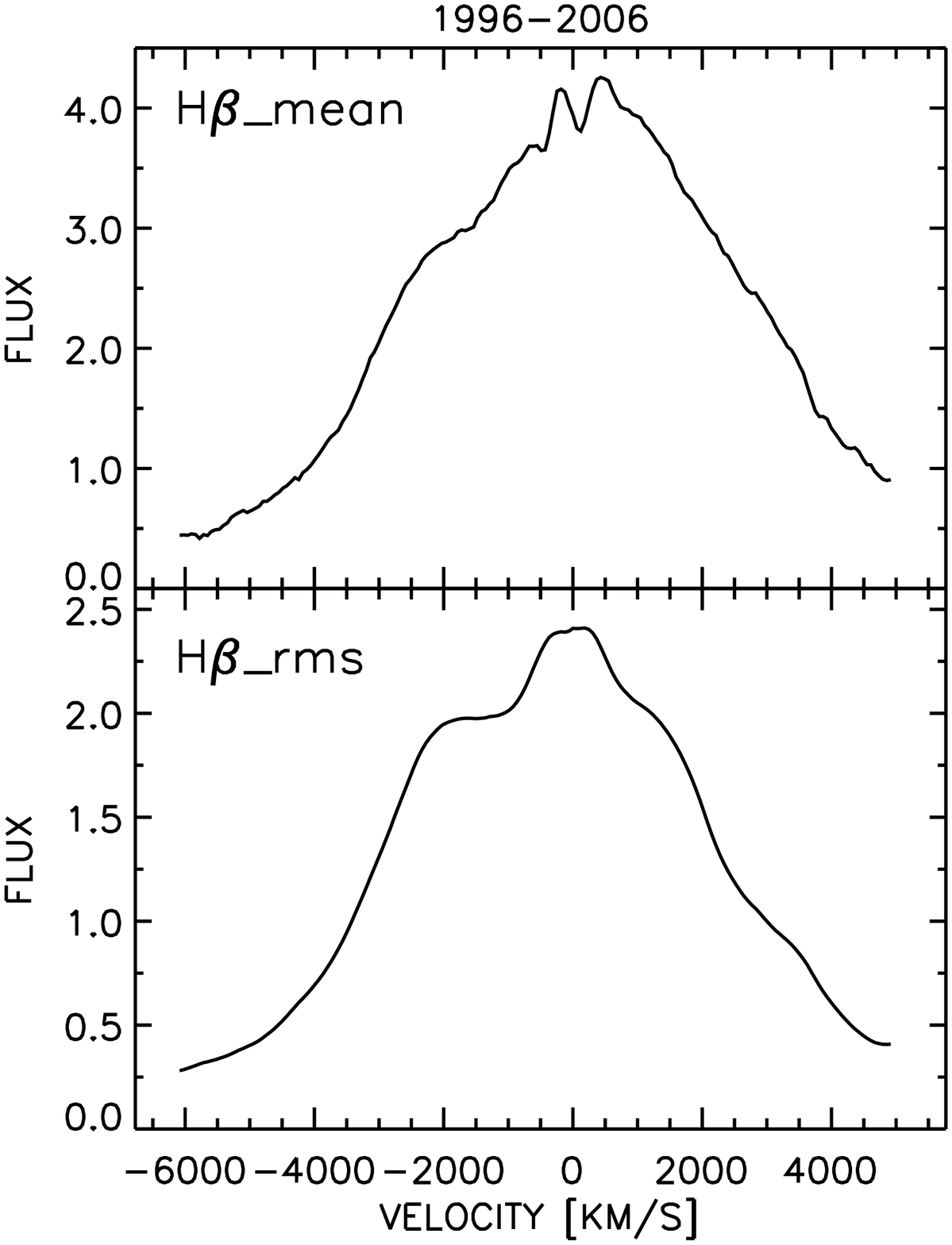}
\caption{The averaged and rms profiles of \textbf{the broad} H$\alpha$
{(left) and
H$\beta$ (right)} \textbf{lines} for the whole
monitoring period.}
\label{fig8}
\end{figure}

\begin{figure}   
\includegraphics[width=9.5cm]{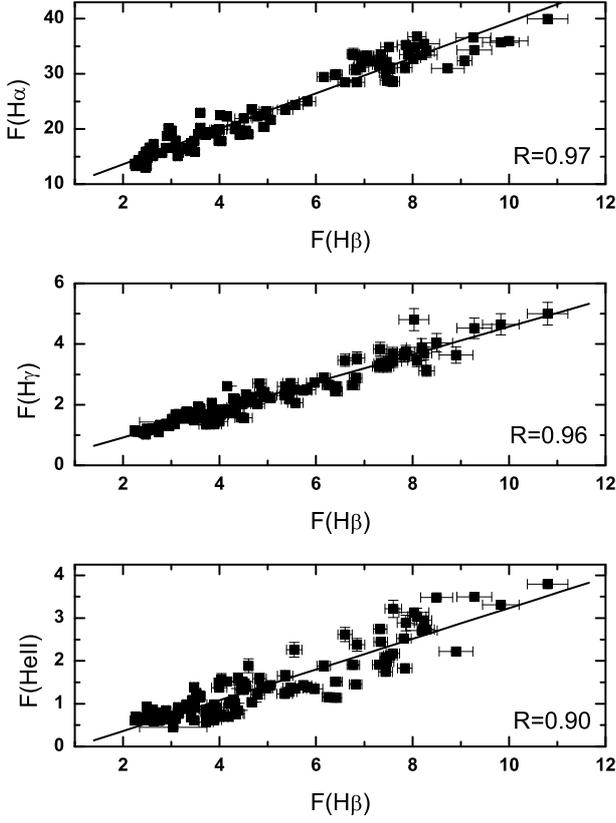}
\caption{ The  H$\alpha$, H$\gamma$, HeII$\lambda$4686 fluxes
versus the H$\beta$ flux.The correlation coefficients $r$ are
given inside the plot. The flux  is given in units of 10$^{-12}
\rm erg\ cm^{-2} s^{-1}$.  }
\label{fig9}
\end{figure}

\begin{figure}  
\includegraphics[width=9.5cm]{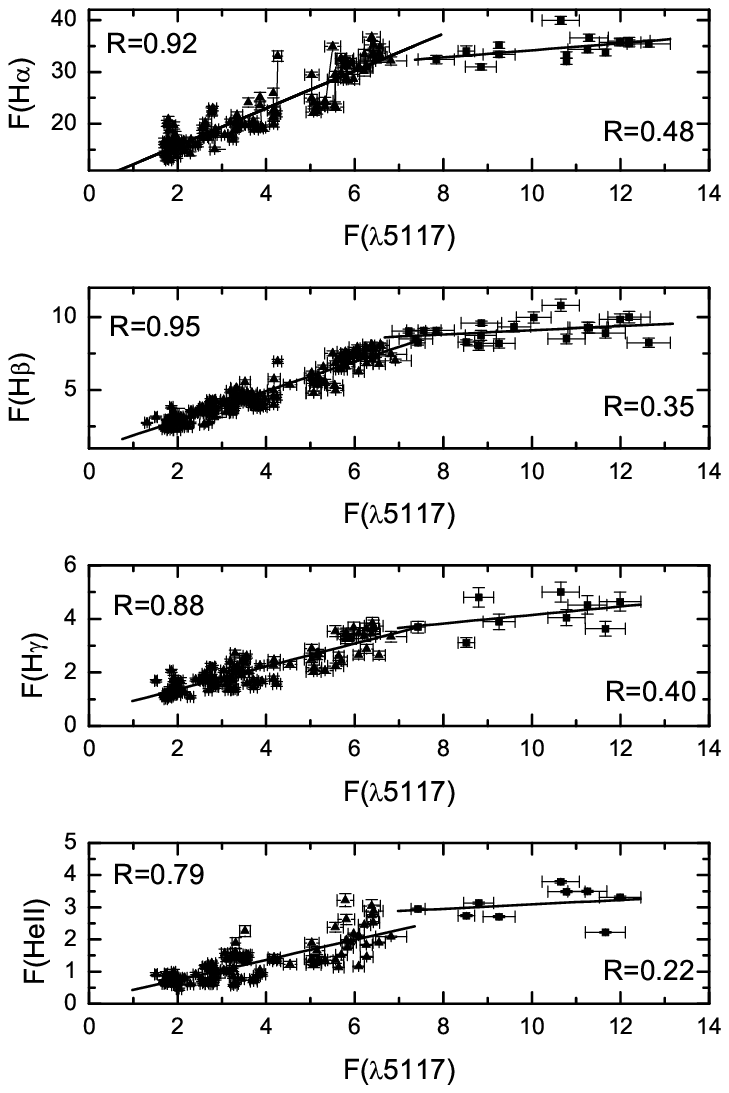}
\caption{ The H$\alpha$, H$\beta$, H$\gamma$, HeII$\lambda$4686
fluxes versus the continuum flux (at $\lambda=$5117 \AA ). {The
correlation coefficients $r$ are given inside the plot. The line fluxes
are given
in units of 10$^{-12} \rm erg\ cm^{-2} s^{-1}$, and  the continuum flux in 
units
of 10$^{-14}\rm erg\ cm^{-2} s^{-1} A^{-1}$.}}
\label{fig10}
\end{figure}

\subsection{The line and continuum flux correlations}
\label{sec3.4}

\subsubsection{The line-line and continuum-line {relationship}}

To find the line-line {flux  relationships
we have plotted the fluxes of  H$\alpha$, H$\gamma$ and
He$\lambda$4686  vs.  the  {  H$\beta$ flux} (Fig. \ref{fig9}).
As  expected, there is a linear relation
{between the {  H$\beta$ flux} and that of the other lines}.  The correlation
coefficients between the H$\beta$
flux and the H$\alpha$, H$\gamma$ and
He$\lambda$4686 fluxes are $r\approx$ 0.97,
0.96 and 0.90, respectively. { The slight differences between
the correlation coefficients may be caused by uncertainty of the measurements
(i.e. very weak H$\gamma$ and He$\lambda$4686 in some periods)}

We have also plotted  the H$\alpha$, H$\beta$, H$\gamma$
and He$\lambda$4686 fluxes against the continuum flux $Fc$ (see Fig. \ref{fig10}).
We find that  the relationship between the line and
continuum fluxes can be divided into two separated sequences.
The first sequence, corresponding
to $ Fc\ ^<_ \sim\  7\cdot 10^{-14} \rm erg \ cm^{-2}s^{-1}\AA^{-1}$,
took place in the period 1998 to
2006; there was a linear relationship between the lines and continuum
fluxes  with a high correlation coefficient (0.88-0.95) for the Balmer lines.
The behaviour of He$\lambda$4686 vs. $Fc$ is not clear, tending to be linear,
but with a high dispersion ($r=0.79$) when $Fc$ was equal to
$ 5.5-7\ 10^{-14}\rm\ erg\ cm^{-2} s^{-1} \AA^{-1}$,
corresponding to the period 1998-1999. The second
sequence corresponds to large values of $Fc$ ($^>_ \sim\
7\
10^{-14} \rm\ erg\ cm^{-2} s^{-1} \AA^{-1}$) and belongs to the period
from 1996 to 1997. Here the line fluxes tended to remain constant,
with a very weak correlation with $Fc$ (see Fig. \ref{fig10}; $r \ \sim
0.2-0.48$ is given in the right-down corner for each line).

\begin{figure*}  
\includegraphics[width=19cm]{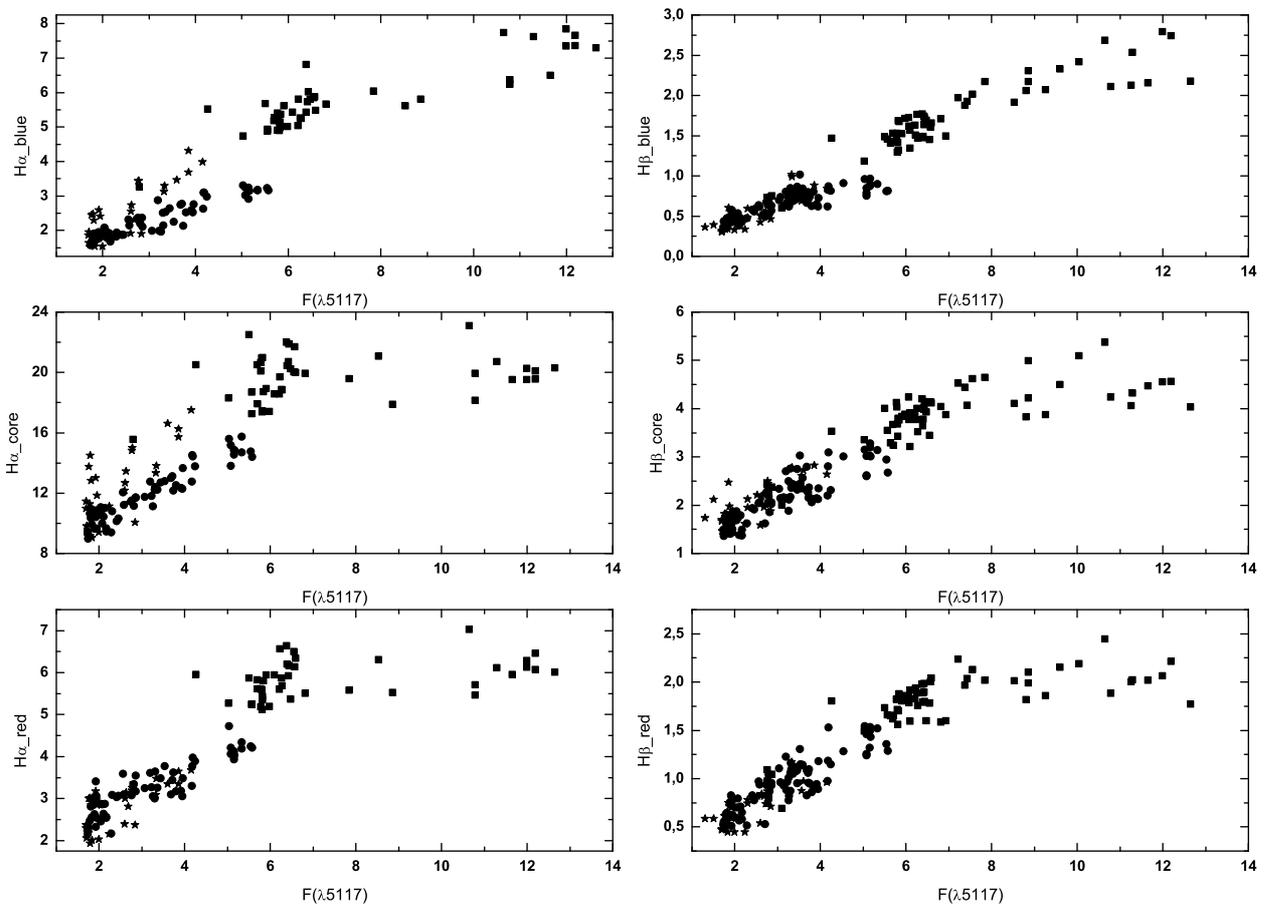}
\caption {The wings and core fluxes as a function of the continuum
flux (at $\lambda=$5117 \AA ) for H$\alpha$ (left) and
H$\beta$ (right). The line and continuum fluxes  are given
in the same units as in Fig. \ref{fig10} . }
\label{fig11}
\end{figure*}

In Fig. \ref{fig11} we show the relation between the continuum flux and
the different parts of the H$\alpha$ and H$\beta$ profiles (the
blue, the core, and the red). As can be seen, the relation for
the core and the blue/red wing is similar to that of the whole line
(see Fig. \ref{fig10}).

To find an explanation of this behaviour, we inspected
{these relationships} within the three periods of observations
mentioned above  (see \S 3.3). We concluded that there was not only
a difference in the line profiles, but also in
 the continuum  vs. line flux relationships.

{ In the first period, when the
lines were the most intense, there was a weak
correlation between the lines and the continuum fluxes.  The
H$\alpha$ flux changed by only $\sim\pm$ 40\%, while
the continuum flux changed by a factor three (see Fig. \ref{fig12}, left).
During the same period, H$\beta$ was also very weakly correlated to
the continuum (Fig. \ref{fig12}, right), exept for five points,
corresponding to observations between June and December 1999: at this time
the continuum and H$\beta$  fluxes decreased  nearly by a factor two, but
the H$\beta$ and H$\alpha$ profiles remained almost identical.}
 { In the second period,
 a large dispersion was observed for
$ Fc\ ^<_\sim\  3.3\cdot 10^{-14} \rm erg \ cm^{-2}s^{-1}\AA^{-1}$,
 and  the lines did not respond to the continuum. Some points (5) with   $Fc\ge
3.3\ 10^{-14}$\,erg\,cm$^{-2}$\,s$^{-1}$\,\AA$^{-1}$ in Fig.12
(middle) correspond to spectra taken in January and February
2000, and their H$\alpha$ and H$\beta$ profiles are similar as in
the first period (i.e. they have a red asymmetry and a shoulder in
the blue wing).} In the third period, the response of the lines
to the continuum was linear.

\begin{figure*}  
\includegraphics[width=19cm]{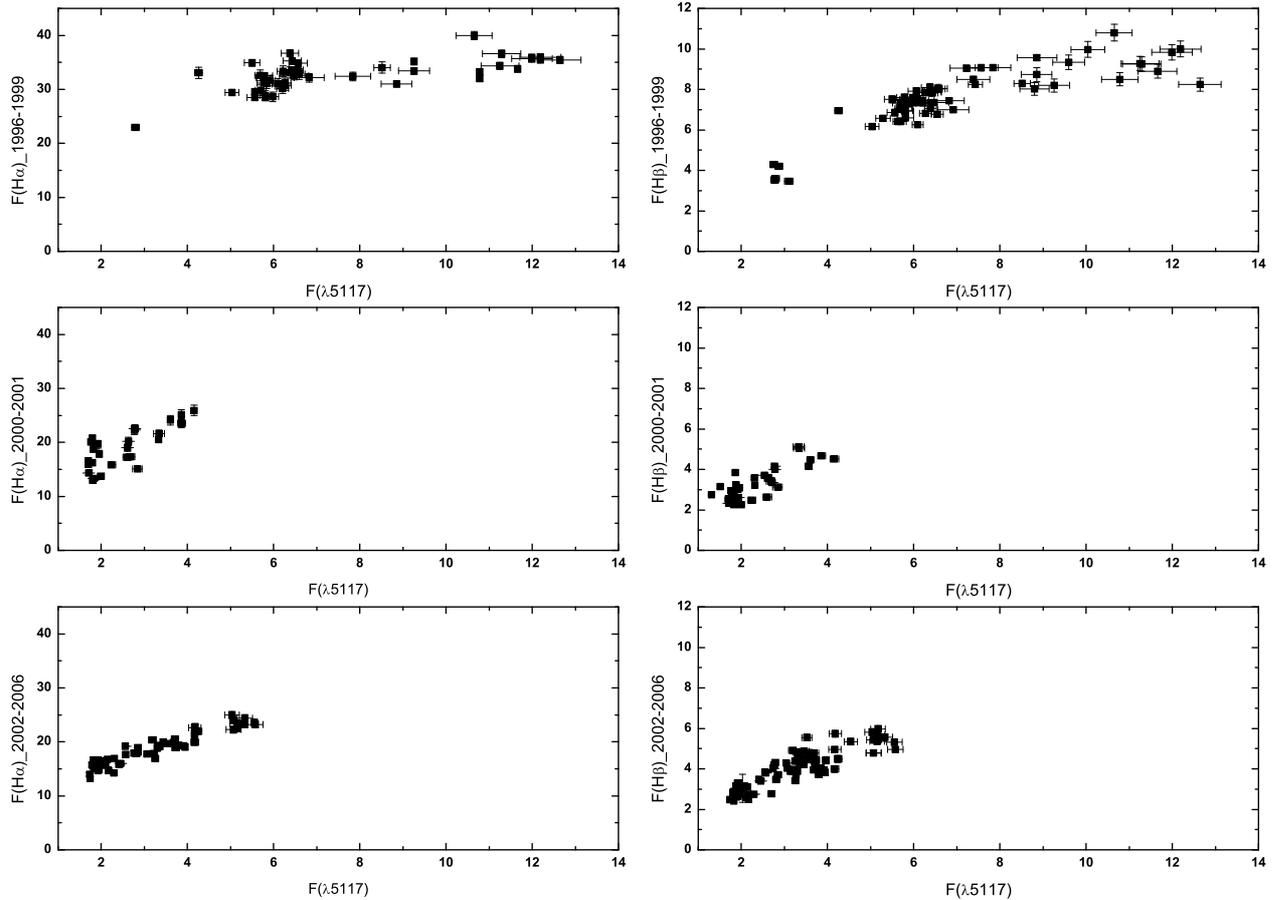}
\caption{H$\alpha$ (left) and H$\beta$ (right) fluxes for the
three periods (first to third from top to down) as a function of the
continuum flux. The line and conntinuum fluxes  are given
in the same units as in Fig. \ref{fig10}. }
\label{fig12}
\end{figure*}

\subsubsection{{ Cross-correlation analysis}}
\begin{figure*}
\includegraphics[width=12 cm]{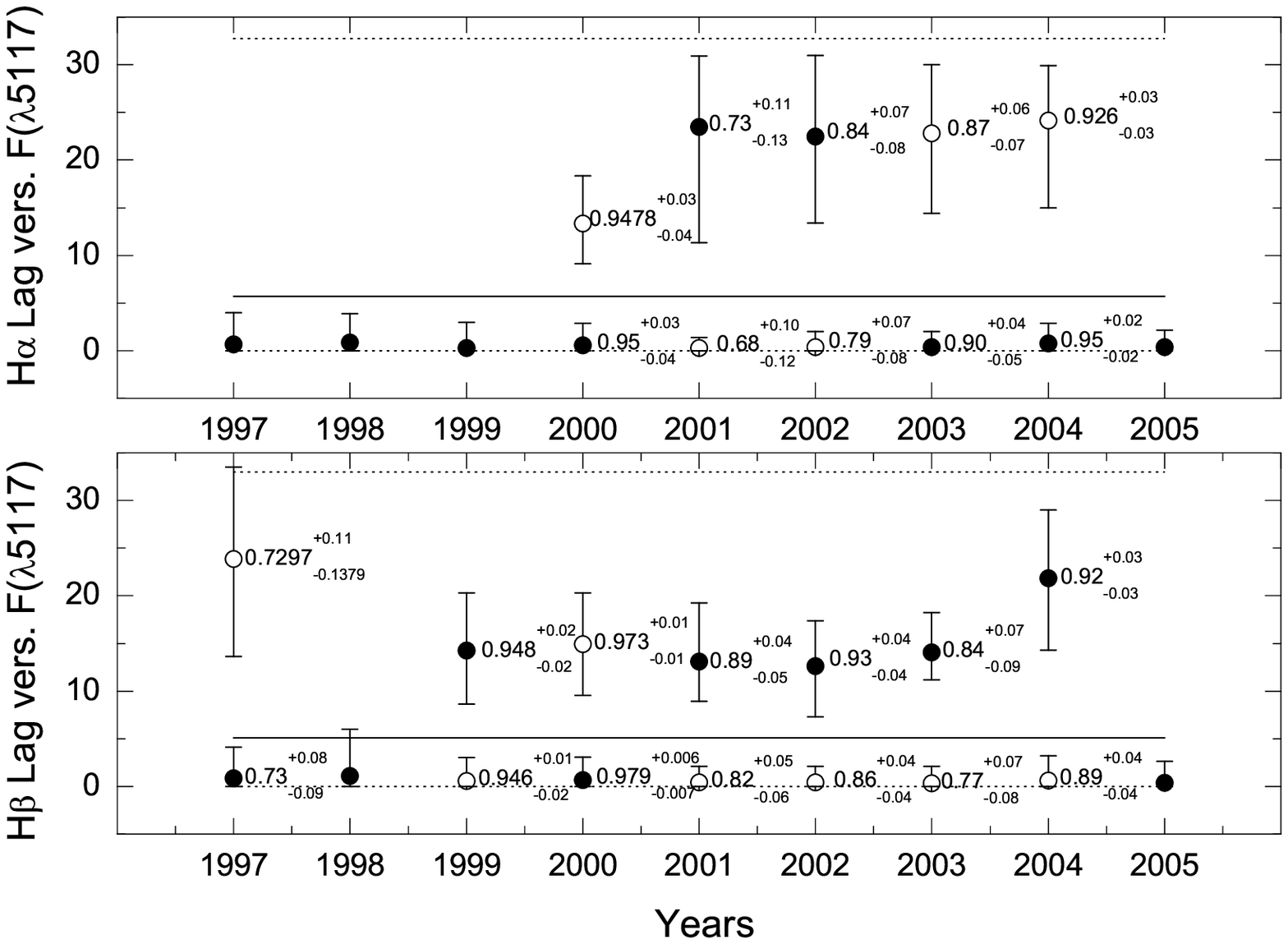}
\caption{\textbf{The time lags obtained by binning the whole period into
time intervals of three years,
starting from 1996 with one year step.  Full circles correspond to the maximal CCF
values, and open circles to  CCF values
which are consistent within the error bars with the maximal one (where there are present). The time
lag obtained in during the whole monitoring
period is shown as a solid horizontal line, and the error-bars are shown as dashed
horizontal lines.} }
 \end{figure*}

In order to
determine the time lag between the optical continuum and the line
variations \textbf{(line lagging continuum)},
we used the \textbf{cross correlation function --} CCF method introduced
by Alexander (1997), the
Z-Transformed Discrete Correlation Function (ZDCF), which contains
the idea  of  the Discrete Correlation Function (DCF) method (Edelson
\& Krolik 1988)  avoiding an interpolation.
The ZDCF approximates the bin distribution by a  bi-normal
distribution. This algorithm differs from  the DCF in that the data points
are equally binned, and it uses the Fisher's
z-transform to stabilize a highly-skewed distribution of the
correlation coefficient. According to Alexander (1997), ZDCF
is much more efficient than DCF in detecting correlations
involving the variability time-scale, and it is more sensitive to
under-sampled light curves than DCF and { Interpolated Cross-Correlation
Function (Gaskell \& Sparke 1986; Gaskell \& Peterson, 1987).}

  The CCF analysis has been carried out for the full data set which
covers the whole monitoring period from 1996 to 2006
and the three periods mentioned above. The
time lags and CCF for H$\beta$ and H$\alpha$ are given in
Table~\ref{tab12}.  As can be seen, the time lags
for H$\alpha$ and H$\beta$  in the whole period are $\sim$5 days,
but they were different in the three periods. In the first and in the third
period, the time lags were much smaller
for both lines (from 0.6 to 1.1 days) than in the second period
(11 and 21 days).

\begin{table*}
\begin{center}
       \caption[]{ Time lags and CCF coefficients for the whole monitoring
 period, \textbf{for the three periods based on the profile shapes, and for the division based on the three 
values of the
 continuum intensity ($F_{\rm con}^1\ge$7$\cdot$
10$^{-14}\rm erg\ cm^{-2} s^{-1} \AA^{-1}$; $F_{\rm con}^1>F_{\rm
con}^2\ge$ 4$\cdot$10$^{-14}\rm erg\ cm^{-2} s^{-1} \AA^{-1}$ and $F_{\rm
con}^3<F_{\rm
con}^2\ge$).} The CCF
coefficients are
calculated between the continuum flux and the H$\alpha$, H$\beta$ fluxes
\textbf{(line lagging contninuum). Positive  time lags mean that the line light curve lags behind the
continuum light curve}.
The time lags are given in days.}\label{tab12}
\begin{tabular}{|c|c|c|c|c|c|c|}

\hline
Period &H$\alpha$&  CCF&H$\alpha$-cent& H$\beta$&  CCF&H$\beta$-cent\\
\hline
&&&&&&\\
 1996-2006 & 5.70$^{+27.05}_{-5.70}$ & 0.87$^{+0.02}_{-0.02}$&80.36$\pm$12.93& 
5.09$^{+27.89}_{-5.09}$& 0.93$^{+0.01}_{-0.01}$& 69.79$\pm$11.91\\
&&&&&&\\
\hline
&&&&&&\\
  First& 0.68$^{+3.30}_{-0.68}$ & 0.67$^{+0.09}_{-0.10}$& 0.62$\pm$1.11&
1.11$^{+4.90}_{-1.11}$ & 0.82$^{+0.04}_{-0.05}$&11.61$\pm$2.87 \\
&&&&&&\\
Second& 21.86$^{+9.03}_{-10.53}$ &0.78$^{+0.11}_{-0.14}$& 66.07$\pm$9.95&
11.15$^{+5.00}_{-4.17}$ &0.88$^{+0.06}_{-0.07}$&8.15$\pm$2.38     \\
&&&&&&\\
Third& 0.81$^{+1.55}_{-0.81}$ &0.94$^{+0.02}_{-0.02}$&3.18$\pm$1.76&
0.81$^{+2.19}_{-0.81}$& 0.86$^{+0.03}_{-0.04}$& 16.17$\pm$3.14\\
&&&&&&\\
\hline
\hline
&&&&&&\\
Continuum &H$\alpha$&  CCF&H$\alpha$-cent& H$\beta$&  CCF&H$\beta$-cent\\
\hline
&&&&&&\\
$F_{\rm con}^1$ & 0.27$^{+0.73}_{-0.27}$ & 0.51$^{+0.22}_{-0.27}$&--&
25.15$^{+13.77}_{-20.16}$ &0.50$^{+0.23}_{-0.28}$&--\\
&&&&&&\\
$F_{\rm con}^2$& 19.36$^{+10.42}_{-5.44}$ & 0.66$^{+0.14}_{-0.17}$&0.56$\pm$6.245& 
0.62$^{+3.49}_{-0.62}$ &0.70$^{+0.07}_{-0.08}$&2.71$\pm$5.22\\
&&&&&&\\
$F_{\rm con}^3$& 1.19$^{+6.01}_{-1.19}$ & 0.67$^{+0.07}_{-0.08}$&13.65$\pm$4.47&
0.70$^{+2.94}_{-0.70}$ &0.82$^{+0.04}_{-0.04}$&12.28$\pm$5.26\\
&&&&&&\\
\hline
\end{tabular}
\end{center}
\end{table*}

\textbf{Moreover, we calculated the CCF and the time lags by dividing the data
set into three groups according to the continuum intensity: (i) $F_{\rm
con}^1\ge$7$\cdot$
10$^{-14}\rm erg\ cm^{-2} s^{-1} \AA^{-1}$; (ii) $F_{\rm con}^1>F_{\rm
con}^2\ge$ 4$\cdot$10$^{-14}\rm erg\ cm^{-2} s^{-1} \AA^{-1}$ and (iii)
$F_{\rm
con}^3<F_{\rm
con}^2$.
As it can be seen in  Fig. 12,  $F_{\rm con}^1$ is only present during
the first period, while $F_{\rm con}^2$ and $F_{\rm con}^3$  are present in
 all three  periods. Table 13 shows that for such a division, the CCF is
small when the continuum is high.
 Also, there is a big
difference in the time lags between these three cases.
 Note that the
highest CCF is obtained for the lowest continuum, but it is still
smaller than the one obtained in the third period based on the line profiles.}

We can summarize  the CCF  analysis as follows. During the whole
monitoring period, the
time lags for the lines were: H$\alpha$ -- 5.70$^{+27.05}_{-5.70}$
days (CCF= 0.87$^{+0.02}_{-0.02} $); H$\beta$ --
5.09$^{+27.89}_{-5.09}$ days   (CCF 0.93$^{+0.01}_{-0.01}$). The large
scatter of these values is due to the fact that
the time lags were very different \textbf{for the three periods based on the profiles and also for periods based on the continuum intensity}.

\textbf{To clarify these differences, we binned the observations using
a time interval of  three years starting from 1996, i.e. we calculated the time
lags and CCFs for
1996-1997-1998, 1997-1998-1999, etc. The results are presented in Fig. 13,
where the time lags (in light days) are given as a function of the central year
 of a three-year interval (e.g. for 1996-1997-1998,
it is 1997). Full circles correspond to  the maximal CCF values, and open circles correspond to CCF values consistent with the maximal values (i.e. within the error bars).  By
inspection of  the CCF values and  analysis of the results in Fig. 13, we found that there
are always small lags corresponding to maximal CCF values or to CCF values consistent with the maximal ones. On  the other hand, there are sometimes also large lags corresponding to maximal CCF values or to CCF marginally consistent with these values.  To summarize:
a) in periods where the CCF peaks indicate larger lags, there are CCFs
which  are marginally weaker (within the error-bars) than the ones corresponding to
larger lags, indicating also shorter lags; b) in all
three-year periods  there are small time lags (between 0.3-07
light days) which are in agreement with the one obtained in the third period, for the division based on the line profiles.
Of course, we cannot exclude that some effects, for instance a contribution to the
line or continuum fluxes from two different regions can cause such time
lags, but the results of CCF analysis indicate at least a compact
component of the BLR (0 - 2
light days) is always present. }

\section{Discussion}

 During the monitoring period, the spectrum of NGC 4151 has shown strong
changes, not only in the
 line and continuum  fluxes, but also in the H$\beta$ and H$\alpha$
 line profiles. Using the line profiles, we characterized three periods (see
 \S3.3, Tables 12 and 13, and Fig. \ref{fig12}). \textbf{We found that the
FWHM of H$\alpha$ and H$\beta$ are different, H$\beta$ being
significantly broader
than H$\alpha$ ($\sim$1000 km/s). This may indicate that H$\beta$ is
formed deeper in the BLR, i.e.  closer to the Black Hole. But on the
other hand, the blue asymmetry in the averaged rms of H$\beta$ and
H$\alpha$ could indicate a contribution of the emitting gas with an approaching
motion, i.e. an outflow. Finally, the presence  of the central spike in the
rms spectrum is hard to explain unless it is produced by a remote component with an axisymmetric distribution and no outward motion. A suggestion is that it comes from a region heated and
ionised by the jet. In conclusion, there seems to be two BLR components, the first being closer and in outward motion, the second being located further away with no outward motions. Only the first component is permanent. Such a structure is also suggested by the time lags and the CCFs.
} We
leave for the following paper a detailed discussion based on the study
 of the line profiles, and we focus here on the global line variations.

We found that the responses of
the H$\beta$ and H$\alpha$  fluxes to the continuum flux  were different
 in the
 three periods, but it can be due to the limited range of  fluxes in periods
2 and 3. For
low values  of the continuum flux ($Fc \le 6\
10^{-14} \rm\ erg\ cm^{-2} s^{-1} \AA^{-1}$), there was a linear relation
between the lines and the continuum  in the  second and
third periods
(see Fig. 12).  The dispersion of the points is larger  in the second than in
the third period, so we think that the results concerning  the last period
are more reliable.  Note that $Fc$ is smaller than   $4\
10^{-14} \rm\ erg\ cm^{-2} s^{-1} \AA^{-1}$ in period 2 and smaller than $6\
10^{-14} \rm\ erg\ cm^{-2} s^{-1} \AA^{-1}$ in period 3,  and that the linear
relation seems to flatten for H$\beta$ between 4 and 6$\  10^{-14}\ \rm erg\
cm^{-2} s^{-1}$ in period 3. In the first period, there are  only a few
points for $Fc\le 4\ 10^{-14}\  \rm erg\ cm^{-2} s^{-1}$ and  they all
correspond almost to the same value of $Fc$, but  if one  interpolates
between these points and those at $Fc\sim 6\ 10^{-14}\  \rm erg\ cm^{-2}
s^{-1}$, about the same relation as in periods 2 and 3  is obtained. Still in
the first period, when the continuum  flux was more intense, i.e. $
Fc \ge 7\
10^{-14} \rm\ erg\ cm^{-2} s^{-1} \AA^{-1}$,  the linear relation
between the
line and continuum fluxes disappeared,
and the line fluxes saturated at values   $\sim 30\ 10^{-12} \rm erg\ cm^{-2}
s^{-1}$ for
H$\alpha$ and $\sim 8\ 10^{-12} \rm erg\ cm^{-2} s^{-1}$ for
H$\beta $ (see Fig. 12).
Note finally that if one extrapolates
linearly the
continuum flux to zero, e.g. in the third period,  it seems that the line
flux is
still larger than zero, being on the order of 5$\cdot$10$^{-12}$ erg
cm$^{-2}$ s$^{-1}$ for H$\alpha$.
 It appears thus that the relation between the line and the continuum is
not
 linear in the whole continuum flux range (2 to 12 $10^{-14} \rm erg\ cm^{-2}
s^{-1}$), but it steepens at low fluxes and saturates at high fluxes.
This could occur when the ionizing incident flux is intense, so the
medium reprocesses the irradiating flux  into continua (Balmer, Pashen...)
and not into lines (cf. for instance  Collin-Souffrin \& Lasota 1988).

 Let us try to model roughly the response of the lines to a given continuum
flux. For the  highest value of the continuum flux,
one  gets  an optical luminosity  $L_{\rm opt}=\lambda L_{\lambda}\sim
10^{43}$ erg s$^{-1}$. One can  then compute the ionizing flux incident on
the BLR, assuming it to be on  the order of the optical flux, $F_{\rm
ion}\sim L_{\rm opt}/(4\pi R^2)$,  where $R$ is the radius of the BLR (of
course it is a very rough approximation;  it will be refined in the
following paper). Assuming an average  energy of the ionizing photons $<h\nu_{\rm ion}>$ equal
to 2 Rydbergs, and using the grids of  models computed with Cloudy published
by Korista et al. (1997) for a typical  AGN spectrum (AGN3 in their list),
one finds  the results shown in Fig.  \ref{fig-Hb-cloudy}. On the left
panel, the H$\beta$ fluxes at Earth are  given as a function of the density,
 for a covering factor equal to unity, and $R=1$ and 3 light days. On the
 right panel, the computed H$\beta$ flux at the source is given as a function
  of the ionizing photon flux \textbf{$\phi=L_{\rm ion}/(4\pi R^2<h\nu_{\rm ion}>$}, whose minimum and maximum \textbf{recorded} values are
marked as
 the two vertical lines. For this computation, the size of the BLR is 3 light days, and $L_{\rm ion}= 10^{43}$ ergs $cm^{-3}$.
 \textbf{From the left panel of  Fig.  \ref{fig-Hb-cloudy}}, it is clear that the observed line flux is always
much larger than the computed one, even for the highest density
 10$^{14}$ cm$^{-3}$.}
 \textbf{(We recall that the observed fluxes of the broad component of
H$\beta$ and H$\alpha$ are: $F_{H\alpha}$=(5.1 -- 31.1)
10$^{-12}\rm erg\ cm^{-2}s^{-1}$ and $F_{H\beta}$=(2.3 -- 9.8)
10$^{-12}\rm erg\ cm^{-2}s^{-1}$.)}
\textbf{From the right panel of  Fig.  \ref{fig-Hb-cloudy}, we see that} no saturation effect appears in the computed
fluxes, except for the low density 10$^{10}$ cm$^{-2}$, for which the
 line flux is much too small (see the left panel).

There are two possibilities to account for this behavior:

 \noindent (i) The ionizing flux is considerably underestimated by our
assumption of equality with the
optical flux.  This could be linked with  the fact that, in general, UV
variations in Seyfert nuclei are stronger  than optical ones, corresponding
 to a ``flattening" of the spectrum when the  object brightens.  If the
 ionizing flux is underestimated,  a saturation effect could appear if the
 BLR consists in a mixture of gas with different densities ranging from
10$^{10}$ to 10$^{14}$ cm$^{-3}$ (cf. Fig. \ref{fig-Hb-cloudy}, right).

 \noindent (ii) A non-photoionized region is contributing  to the Balmer
lines. Such a
 ``mechanically heated"  region was invoked by Dumont et al.  (1998) to
account for
the strong intensities of the Balmer lines in NGC 5548.  This region could be
associated with the radio-jet. Indeed the radio image of  NGC 4151 reveals a
0.2-pc
two-sided base to the well-known arc-second radio jet  (Ulvestad et al.
2005)\footnote{Note that Arshakian et al. (2006) found a correlation
between  the optical continuum and  the radio-jet emission  variability in
the case of 3C 390.3}.
Thus the BLR would be made of two-components: the usual one,  ionized by the
radiation of the accretion disc and its corona, and another  component,
possibly associated with a rotating outflow   surrounding the
jet (Murray \& Chiang 1997). In this second component, ionization  and
heating
could be due either to relativistic particles or to a shock at  the basis of
the jet, and they could not be directly correlated to the ionizing continuum. \textbf{ This interpretation would be in agreement with the results of the study of the time lags and CCF, as well as with the behavior of the line profiles.}

\begin{figure}
\begin{center}
\includegraphics[width=8cm]{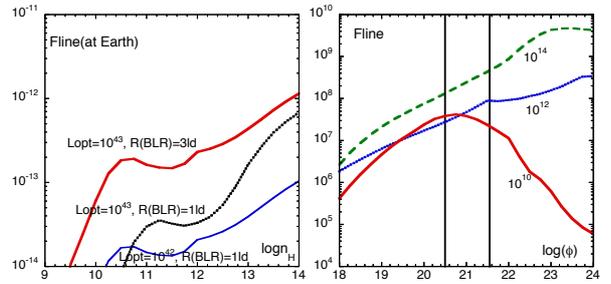}
 \caption{Computed H$\beta$ flux for the conditions of NGC 4151:  left, at
 Earth, and as a function of the density in cm$^{-3}$; right, at the
source,
 and as a function of the photon flux $\phi$ in cm$^{-2}$ s$^{-1}$.  The
two
 vertical lines mark the lowest and highest values of the photon flux as defined in the text, deduced from the observations. On the right panel, the
 curves are labeled by the density in cm$^{-3}$.  See the text for more
 explanations.}
\label{fig-Hb-cloudy}
\end{center}
\end{figure}

It is worthwhile recalling here the results obtained from spectro-polarimetry of
H$\alpha$ by Martel (1998). He presented evidence that
the scattering axes of different parts of the H$\alpha$
profile correlate with the major morphological axes of the host
galaxy of NGC 4151.
 If scattering is the dominant
polarization mechanism in the BLR, then there are multiple lines
of sight towards large-scale structures in
the host galaxy, specifically the central bar, the radio jet, and
the dynamical axis.  Martel (1998) suggested that
the line-emitting and scattering regions are cospatial, possibly
in bulk flows moving along preferential axes defined
by large-scale structures, such as streaming along
the bar and inflow/outflow along the radio jet. In this case, the
observed variability of the H$\alpha$ profile and flux could
be caused, at least partly, by dynamical effects, and not
 by a time-variable continuum source, as it is usually
assumed in reverberation mapping studies.

The time lags between the continuum and the lines were $\sim$ 5  days for
H$\beta$ and
H$\alpha$ during the whole monitoring period, but they differ  strongly in
the different periods \textbf(see Fig. 13)}. They were small in the first
and third period (0 to
2 days), and
very large in the second period (10 to 20 days). The time lags of   H$\alpha$
and
H$\beta$ seem different in the first and second periods,  but they are
 compatible within the precision of the results (cf. table 12).
The time lags
for the first
 period seem to be the most realistic, since the relation between the
line
 and continuum flux is linear and well-defined, in particular for H$\alpha$.
\textbf{Moreover, there were always short time lags for the three year periods (Fig. 13) indicating the presence of a permanent small component of the BLR. During some periods, we suspect the presence of a second, much larger component.} Having in
mind the two possibilities mentioned above \textbf{(larger and shorter
time lags)}, it is possible that an additional
emission in the
lines (and perhaps partly in the continuum), or a weak correlation between
 the optical and ionizing flux, give misleading time lags in the first and in
the second period \textbf{(as well as in the whole monitoring period)}.  All
this is
difficult to understand unless the BLR is complex with a structure  changing
with time.  If it would stay unchanged, one would expect the time  lags to be
 small when the continuum is the lowest, i.e. during the second period. On
the other
hand,
one should keep in mind that the
H$\beta$ and H$\alpha$ line profiles were different and respond
differently to the continuum in the three periods. This is another  proof that
the BLR structure changed. So, measuring the time delays  separately
in the different periods is
more realistic than during the whole monitoring period.

The structure of the BLR will be discussed in the following paper,  in
relation with the line profiles and with their variations.

\section{Conclusion}

We have presented the results of a 11-year (1996 - 2006) spectral
monitoring of the NGC 4151 nucleus. We have investigated the continuum and
line variations during this period. We have reached
the following conclusions:

(a) the nucleus of NGC 4151 showed big variations of the line and
continuum fluxes during the monitoring period (1996-2006). The
maximum of activity took place in 1996, and there were two minima between
December 2000 and May 2001 and between February 2005 and June 2005. The continuum
flux changed by a factor  $\sim$ 6, the broad H$\alpha$ and H$\beta$  changed
respectively
by factors $\sim$3.6 and $\sim$5.  The H$\alpha$, H$\gamma$ and He$\lambda$4686
fluxes were well correlated with the H$\beta$ flux (Fig. 9).

(b) There was a good linear relationship between the
 emission line and continuum flux variations when the continuum
flux   $Fc$ was $\ ^<_\sim 7\cdot 10^{-14} \rm\ erg \ cm^{-2}s^{-1}\AA^{-1}$.
When $Fc$ was large ($ Fc\ ^>_\sim 7\cdot
10^{-14} \rm \ erg\ cm^{-2} s^{-1} \AA^{-1}$) the line fluxes either were
weakly correlated, or simply did not correlate at all with the
continuum flux (Figs. 10 and 12).

(c) In the minimum state, the line wings were very weak and were observed only in
H$\beta$ and H$\alpha$. Thus the spectral type of the object
in the monitoring period (1996-2000) changed from Sy1.8 during the  minimum
of activity, to Sy1.5
during the maximum (Fig. 1).

(d) The flux ratios of the
blue/red wings and blue (or red) wings/core varied differently for  H$\alpha$
and H$\beta$
(Fig. 4).
 We found three characteristic periods (1996-1999, 2000-2001 and 2002-2006)
in the profile variability. The behaviour of the rms profiles in the three
periods
indicates that the BLR has a complex structure.

(e) From the CCF analysis, we found a time lag
of about  5 days  for the whole monitoring period, but it was smaller  (0-2
days) in the first and the third periods, and larger (10-20 days) in  the
second period, implying that the dimension of the BLR varied with time.
\textbf{An analysis of the time lags and CCFs for  a period of three years
starting from 1996 shows that short time lags are present in all
periods. Therefore,}  we propose that the time lags during  the
third period is more
realistic, since line fluxes were srtongly correlated with the
continuum flux as expected if the optical  is proportional to
the  ionizing continuum. \textbf{In summary we suspect that the BLR contains a permanent small component, and sometimes an additional component of much larger dimensions.}

(f) The
lines and the continuum variations behave differently during the three
periods. In the first period, when the continuum was strong, the line fluxes
saturate, meaning that  the optical continuum was not proportional to
the  ionizing continuum  (Fig. 12). More generally, we found an excess of
line
emission  with respect to a pure photoionization model during the whole
monitoring  period. This result could imply, either the presence of a
non-radiatively  heated region, or an ionizing to optical flux ratio larger
than expected for a typical AGN spectrum.

A discussion of the line profiles and of the structure of the BLR will be
given in a fortcoming paper (paper II).}

  \section*{Acknowledgments}

This work was supported by INTAS (grant N96-0328),
RFBR (grants N97-02-17625 N00-02-16272, N03-02-17123 and 06-02-16843),
State program 'Astronomy' (Russia), CONACYT research grant 39560-F
and 54480 (M\'exico) and the Ministry of Science of Republic of
Serbia through the project Astrophysical Spectroscopy of
Extragalactic Objects (146002). L. \v C. P. is supported by Alexander
von Humboldt foundation through Fritz Thyssen Special Programme.
\textbf{We would
like to thank Tal Alexander for useful discussions concerning the time lags,
and Ian McHardy for his comments and suggestions which contributed to improve the paper.}

 \Online

\onllongtab{2}{
\begin{longtable}{rllllrrl}
\caption{\label{tab2} Log of the spectroscopic observations:
Columns: 1 - UT date; 2 - Julian date (JD); 3- code according to
Table~\ref{tab1}; 4 - projected spectrograph entrance apertures;
5 - wavelength range covered; 6 - spectral resolution; 7 - slit
 position angle (PA) in degrees;  8 - mean seeing in arcsec.}\\
\hline

UT-date  &   JD     &Code  &Apertura &Sp.range &Res.&PA  &Seeing \\
        &(2400000+)&      &(arcsec) &  (\AA-\AA)   &(\AA)&(deg)&  (")  \\
\hline
   1         &2      &3       &4         &5      &6   &7   &8   \\
\hline
\endfirsthead
\caption{Continued.}\\
\hline
\hline
UT-date  &   JD     &Code  &Apertura &Sp.range &Res.&PA  &Seeing \\
         &(2400000+)&      &(arcsec) &  (\AA-\AA)   &(\AA)&(deg)&  (")  \\
\hline
   1         &2      &3       &4         &5      &6   &7   &8   \\
\hline
\endhead
\hline
\endfoot
\hline
\endlastfoot
Jan. 11, 1996&  +50094.5&  L1(G)&  4.0$\times$12.4&  3840-5600&   8  &
33&   3   \\
Jan. 15, 1996&  +50097.6&  L1(G)&  4.0$\times$13.8&  3640-7140&  10  &
77&   3   \\
Jan. 16, 1996&  +50098.6&  L1(G)&  4.0$\times$15  &  3640-7140&  10  &
90&   4.5 \\
Feb. 14, 1996&  +50128.0&  L(N) &  1.5$\times$6.0 &  3400-5440&  10  &
211&   2.4 \\
Mar. 19, 1996&  +50162.4&  L(N) &  2.0$\times$6.0 &  3650-5540&   9  &
235&   4.8 \\
Mar. 20, 1996&  +50163.3&  L(N) &  2.0$\times$6.0 &  4750-7340&  13  &
&   3.5 \\
Mar. 21, 1996&  +50164.4&  L(N) &  2.0$\times$6.0 &  3650-5540&   8  &
40&   1.6 \\
Mar. 22, 1996&  +50165.4&  L1(G)&  4.0$\times$19.8&  5700-7500&   8  &
&   4   \\
Mar. 23, 1996&  +50166.3&  L(N) &  2.0$\times$6.0 &  4750-7340&  13  &
&       \\
Apr. 26, 1996&  +50200.3&  L(U) &  2.0$\times$6.0 &  4440-5200&   5  &
353&   2.4 \\
Apr. 27, 1996&  +50201.3&  L(U) &  2.0$\times$6.0 &  4440-5200&   5  &
351&   1.6 \\
Jun. 14, 1996&  +50249.3&  L(U) &  2.0$\times$6.0 &  4440-5240&   5  &
131&   2.4 \\
Jul. 10, 1996&  +50275.3&  L(U) &  2.0$\times$6.0 &  3700-5340&  10  &
97&   1.2 \\
Jul. 11, 1996&  +50276.3&  L(U) &  2.0$\times$6.0 &  6140-6950&   6  &
131&   1.4 \\
Jul. 12, 1996&  +50277.3&  L(U) &  2.0$\times$6.0 &  4450-5250&   5  &
117&   1.6 \\
Jul. 15, 1996&  +50280.3&  L(U) &  2.0$\times$6.0 &  4450-5250&   5  &
123&   1.6 \\
Jul. 16, 1996&  +50281.3&  L(U) &  2.0$\times$6.0 &  6140-6950&   6  &
&   1.3 \\
Nov. 05, 1996&  +50392.6&  L(U) &  2.0$\times$6.0 &  4450-5250&   5  &
27&   3.2 \\
Nov. 15, 1996&  +50402.6&  L(U) &  2.0$\times$6.0 &  3700-5400&  11  &
349&   2   \\
Mar. 02, 1997&  +50510.4&  L(U) &  2.0$\times$6.0 &  4390-5200&   5  &
333&   2.4 \\
Mar. 03, 1997&  +50511.4&  L(U) &  2.0$\times$6.0 &  4440-5250&   5  &
9&   2.4 \\
Apr. 05, 1997&  +50543.6&  L1(G)&  4.2$\times$19.8&  4140-5850&   8  &
90&   4.4 \\
Apr. 06, 1997&  +50544.4&  L1(G)&  4.2$\times$19.8&  4140-5800&   8  &
90&   4.4 \\
Apr. 08, 1997&  +50547.3&  L(U) &  2.0$\times$6.0 &  4640-5450&   5  &
353&   3.2 \\
Apr. 13, 1997&  +50552.3&  L(U) &  2.0$\times$6.0 &  4440-5250&   5  &
337&   2   \\
Dec. 27, 1997&  +50809.7&  L(U) &  2.0$\times$6.0 &  4540-5340&   5  &
153&   1.6 \\
Dec. 28, 1997&  +50810.7&  L(U) &  2.0$\times$6.0 &  3840-7240&  15  &
149&   4   \\
Jan. 20, 1998&  +50833.6&  L(N) &  2.0$\times$6.0 &  3840-6150&   8  &
141&   1.2 \\
Jan. 21, 1998&  +50834.6&  L(N) &  2.0$\times$6.0 &  3840-6150&   8  &
142&   1.2 \\
Jan. 23, 1998&  +50836.7&  L(N) &  2.0$\times$6.0 &  4240-5500&   5  &
131&   1.2 \\
Jan. 23, 1998&  +50836.7&  L(N) &  2.0$\times$6.0 &  3840-6150&   8  &
130&   1.2 \\
Jan. 28, 1998&  +50842.4&  L(U) &  2.0$\times$6.0 &  4540-5350&   5  &
357&   1.6 \\
Feb. 22, 1998&  +50867.4&  L(N) &  2.0$\times$6.0 &  3840-6150&   8  &
352&   1.2 \\
Apr. 30, 1998&  +50934.5&  L1(G)&  8.0$\times$19.8&  4100-5750&   8  &
147&   4.4 \\
May. 04, 1998&  +50938.3&  L(N) &  2.0$\times$6.0 &  3740-6150&   8  &
140&   2.8 \\
May. 07, 1998&  +50940.5&  L(N) &  2.0$\times$6.0 &  3740-6150&   8  &
90&   3.6 \\
May. 08, 1998&  +50941.5&  L(N) &  2.0$\times$6.0 &  3740-6150&   8  &
92&   2   \\
May. 08, 1998&  +50942.5&  L(N) &  2.0$\times$6.0 &  3740-6150&   8  &
119&   1.6 \\
May. 08, 1998&  +50942.5&  L(N) &  2.0$\times$6.0 &  4140-5400&   5  &
115&   1.6 \\
Jun. 20, 1998&  +50985.3&  L1(G)&  8.0$\times$19.8&  4120-5740&   8  &
0&   4.4 \\
Jun. 26, 1998&  +50991.3&  L(N) &  2.0$\times$6.0 &  3640-6040&   8  &
&   2   \\
Jul. 14, 1998&  +51008.7&  GH   &  2.5$\times$6.0 &  3970-7210&  15  &
90&   2   \\
Jul. 16, 1998&  +51010.7&  GH   &  2.5$\times$6.0 &  3630-6890&  15  &
90&   2   \\
Jul. 17, 1998&  +51011.6&  GH   &  2.5$\times$6.0 &  4210-7470&  16  &
90&   2   \\
Jul. 21, 1998&  +51015.7&  GH   &  2.5$\times$6.0 &  4040-7280&  12  &
90&   2.7 \\
Jul. 30, 1998&  +51025.3&  L(U) &  2.0$\times$6.0 &  4540-5350&   5  &
120&   3.6 \\
Nov. 13, 1998&  +51130.6&  L1(G)&  8.0$\times$19.8&  4140-5810&   8  &
90&   4.4 \\
Dec. 19, 1998&  +51166.6&  L1(G)&  4.0$\times$19.8&  4140-5760&   8  &
90&   6.6 \\
Jan. 12, 1999&  +51190.7&  L(U) &  2.0$\times$6.0 &  6250-7100&   9  &
&   2.5 \\
Jan. 13, 1999&  +51191.9&  GH   &  2.5$\times$6.0 &  4140-7420&  15  &
90&   1.5 \\
Jan. 14, 1999&  +51193.0&  GH   &  2.5$\times$6.0 &  4150-7430&  15  &
90&   1.3 \\
Jan. 22, 1999&  +51200.5&  L1(G)&  4.2$\times$19.8&  4100-5740&   8  &
90&   2.2 \\
Jan. 23, 1999&  +51201.6&  L1(G)&  4.2$\times$19.8&  4100-5740&   8  &
0&   4.4 \\
Jan. 25, 1999&  +51203.6&  L1(G)&  4.2$\times$19.8&  4100-5750&   9  &
0&   6.6 \\
Feb. 09, 1999&  +51218.6&  L(U) &  2.0$\times$6.0 &  3640-8040&  14  &
156&   4   \\
Feb. 12, 1999&  +51221.7&  L(U) &  2.0$\times$6.0 &  3640-7940&  14  &
156&   1.6 \\
Feb. 13, 1999&  +51222.6&  L(U) &  2.0$\times$6.0 &  4290-5500&   5  &
181&   5.2 \\
Feb. 14, 1999&  +51223.6&  L(U) &  3.0$\times$6.0 &  4290-5500&   5  &
143&   2.4 \\
Mar. 15, 1999&  +51252.9&  GH   &  2.5$\times$6.0 &  4200-7490&  17  &
90&   2.0 \\
Mar. 20, 1999&  +51258.5&  L1(G)&  4.2$\times$19.8&  4090-5740&   8  &
0&   2.2 \\
Mar. 23, 1999&  +51261.5&  L1(G)&  4.2$\times$19.8&  4090-5740&   8  &
0&   8.8 \\
Mar. 24, 1999&  +51262.3&  L1(G)&  4.2$\times$19.8&  5590-7300&   9  &
&   4   \\
Mar. 24, 1999&  +51262.5&  L(U) &  2.0$\times$6.0 &  4290-5450&   5  &
62&   4   \\
Apr. 09, 1999&  +51277.6&  L1(G)&  4.2$\times$19.8&  4090-5750&   8  &
0&   4.4 \\
Apr. 11, 1999&  +51279.5&  L1(G)&  4.2$\times$19.8&  4040-5740&   8  &
0&   6.6 \\
Apr. 15, 1999&  +51283.5&  L1(G)&  3.0$\times$19.8&  4090-5770&   8  &
50&   4.4 \\
Jun. 14, 1999&  +51344.3&  L1(G)&  4.2$\times$19.8&  4090-5790&   8  &
90&   4.4 \\
Jun. 16, 1999&  +51346.4&  L1(G)&  4.2$\times$19.8&  4090-5790&   8  &
90&   4.4 \\
Dec. 02, 1999&  +51514.6&  L1(G)&  4.2$\times$19.8&  4090-5700&   8  &
90&   4.4 \\
Dec. 03, 1999&  +51515.6&  L1(G)&  4.2$\times$19.8&  4140-5750&   8  &
90&   4.4 \\
Dec. 05, 1999&  +51517.6&  L1(G)&  4.2$\times$19.8&  4090-5750&   8  &
90&   4.4 \\
Jan. 09, 2000&  +51552.6&  L1(G)&  4.2$\times$19.8&  4090-5790&   8  &
90&   2.2 \\
Jan. 10, 2000&  +51553.6&  L1(G)&  4.2$\times$19.8&  5640-7310&   8  &
90&   3.0 \\
Jan. 27, 2000&  +51570.9&  GH   &  2.5$\times$6.0 &  4070-7340&  13  &
90&   3.0 \\
Jan. 28, 2000&  +51571.9&  GH   &  2.1$\times$6.0 &  4070-7340&  12  &
90&   2.5 \\
Feb. 11, 2000&  +51585.5&  L1(G)&  4.2$\times$19.8&  4040-5740&   8  &
90&   4.4 \\
Feb. 14, 2000&  +51588.5&  L1(G)&  4.2$\times$19.8&  4040-5740&   8  &
90&   6.6 \\
Feb. 14, 2000&  +51589.5&  L1(G)&  4.2$\times$19.8&  5590-7300&   8  &
90&   4   \\
Feb. 26, 2000&  +51600.8&  GH   &  2.1$\times$6.0 &  4560-7590&  12  &
90&   2   \\
Feb. 27, 2000&  +51601.8&  GH   &  2.1$\times$6.0 &  4300-7580&  12  &
90&   3   \\
Apr. 03, 2000&  +51638.5&  L1(G)&  4.2$\times$19.8&  4090-5790&   8  &
90&   4.4 \\
Apr. 05, 2000&  +51640.5&  L1(G)&  4.2$\times$19.8&  4040-5740&   8  &
90&   6.6 \\
Apr. 25, 2000&  +51659.8&  GH   &  2.1$\times$6.0 &  4210-7490&  12  &
90&   2.5 \\
Apr. 26, 2000&  +51660.8&  GH   &  2.1$\times$6.0 &  4210-7490&  15  &
90&   2.5 \\
May. 11, 2000&  +51676.4&  L1(G)&  4.2$\times$19.8&  4090-5790&  10  &
90&   4.4 \\
May. 25, 2000&  +51689.7&  GH   &  2.5$\times$6.0 &  4140-7390&  15  &
90&   2.5 \\
May. 26, 2000&  +51690.7&  GH   &  2.5$\times$6.0 &  4140-7390&  15  &
90&   3.5 \\
Jul. 10, 2000&  +51736.4&  L1(G)&  4.2$\times$19.8&  4060-5750&   8  &
90&   6.6 \\
Jul. 29, 2000&  +51755.3&  L1(G)&  4.2$\times$19.8&  4060-5750&   8  &
90&   4.4 \\
Nov. 21, 2000&  +51869.6&  L(U) &  2.0$\times$6.0 &  4242-5398&   5  &
90&   1.2 \\
Nov. 30, 2000&  +51878.6&  L1(G)&  4.2$\times$19.8&  4090-5740&   8  &
90&   4.4 \\
Dec. 17, 2000&  +51895.9&  GH   &  2.5$\times$6.0 &  4010-7310&  13  &
90&   2   \\
Dec. 18, 2000&  +51897.0&  GH   &  2.5$\times$6.0 &  4010-7340&  13  &
90&   2   \\
Dec. 19, 2000&  +51898.0&  GH   &  2.5$\times$6.0 &  4000-7270&  15  &
90&   4   \\
Jan. 26, 2001&  +51936.5&  L1(G)&  4.2$\times$19.8&  5640-7300&   8  &
90&   2   \\
Jan. 28, 2001&  +51937.5&  L1(G)&  4.2$\times$19.8&  4090-5750&   8  &
90&   4.4 \\
Jan. 31, 2001&  +51940.6&  L1(G)&  4.2$\times$19.8&  4040-5700&   8  &
90&   4.4 \\
Feb. 02, 2001&  +51943.5&  L1(G)&  4.2$\times$19.8&  4140-5750&   8  &
90&   6.6 \\
Feb. 11, 2001&  +51952.4&  L1(G)&  4.2$\times$19.8&  4090-5750&   8  &
0&   4.4 \\
Mar. 13, 2001&  +51981.7&  GH   &  2.5$\times$6.0 &  4130-7430&  14  &
90&   2   \\
Apr. 13, 2001&  +52013.3&  L1(G)&  4.2$\times$19.8&  4090-5750&   8  &
90&   3.0 \\
Apr. 16, 2001&  +52016.4&  L1(G)&  4.2$\times$19.8&  4090-5750&   8  &
90&   4.4 \\
Apr. 29, 2001&  +52029.5&  L1(G)&  4.2$\times$19.8&  4090-5800&   8  &
90&   6.6 \\
May. 05, 2001&  +52034.6&  GH   &  2.5$\times$6.0 &  3600-6800&  12  &
90&   3.5 \\
May. 06, 2001&  +52036.4&  L1(G)&  4.2$\times$19.8&  4040-5750&   8  &
90&   4.4 \\
May. 12, 2001&  +52041.8&  GH   &  2.5$\times$6.0 &  3600-6880&  15  &
90&   1.5 \\
May. 14, 2001&  +52043.7&  GH   &  2.5$\times$6.0 &  4030-7340&  13  &
90&   2   \\
Jun. 14, 2001&  +52074.7&  GH   &  2.5$\times$6.0 &  4020-7310&  13  &
90&   3.5 \\
Jun. 15, 2001&  +52075.7&  GH   &  2.5$\times$6.0 &  4020-7310&  17  &
90&   2.5 \\
Nov. 24, 2001&  +52237.6&  L(U) &  2.0$\times$6.0 &  3640-6000&  10  &
91&   3.6 \\
Nov. 24, 2001&  +52238.0&  GH   &  2.5$\times$6.0 &  4230-5900&   8  &
90&   2.5 \\
Nov. 25, 2001&  +52239.0&  GH   &  2.5$\times$6.0 &  5680-7380&   7.5&
90&   2.5 \\
Jan. 23, 2002&  +52297.6&  L1(G)&  4.2$\times$19.8&  4040-5750&   9  &
90&   4.4 \\
Jan. 24, 2002&  +52299.4&  L1(G)&  4.2$\times$19.8&  4040-5750&   8  &
90&   2.2 \\
Feb. 21, 2002&  +52327.4&  L1(G)&  4.2$\times$19.8&  4040-5750&   8  &
90&   6.6 \\
Feb. 22, 2002&  +52327.5&  L(U) &  2.0$\times$6.0 &  3500-5840&   9  &
90&   3.2 \\
Mar. 05, 2002&  +52338.7&  GH   &  2.5$\times$6.0 &  3900-7190&  13  &
90&   2   \\
Mar. 06, 2002&  +52339.7&  GH   &  2.5$\times$6.0 &  5690-7390&   7.5&
90&   2   \\
Mar. 07, 2001&  +52340.8&  GH   &  2.5$\times$6.0 &  4260-5940&   8  &
90&   2   \\
Mar. 17, 2002&  +52350.8&  GH   &  2.5$\times$6.0 &  4260-5940&   8  &
90&   3   \\
Apr. 03, 2002&  +52367.7&  GH   &  2.5$\times$6.0 &  4280-5960&   8  &
90&   2   \\
Apr. 04, 2002&  +52368.7&  GH   &  2.5$\times$6.0 &  5740-7440&   7.5&
90&   2   \\
Apr. 05, 2002&  +52369.7&  GH   &  2.5$\times$6.0 &  3820-7130&  12  &
90&   2   \\
Apr. 06, 2002&  +52370.7&  GH   &  2.5$\times$6.0 &  3820-7130&  12  &
90&   1.5 \\
May. 03, 2002&  +52397.7&  GH   &  2.5$\times$6.0 &  4260-5940&   8  &
90&   2   \\
May. 04, 2002&  +52398.7&  GH   &  2.5$\times$6.0 &  5680-7370&   7.5&
90&   2   \\
May. 05, 2002&  +52399.7&  GH   &  2.5$\times$6.0 &  4200-5880&   8  &
90&   2   \\
May. 16, 2002&  +52411.4&  L1(G)&  4.2$\times$19.8&  4090-5790&   8  &
90&   4.4 \\
Jun. 02, 2002&  +52427.7&  GH   &  2.5$\times$6.0 &  4150-5820&   8  &
90&   2.5 \\
Jun. 04, 2002&  +52429.7&  GH   &  2.5$\times$6.0 &  3990-7290&  12  &
90&   3.5 \\
Jun. 05, 2002&  +52430.7&  GH   &  2.5$\times$6.0 &  4240-5920&   7.5&
90&   3.0 \\
Jun. 24, 2002&  +52450.4&  L(U) &  2.0$\times$6.0 &  3500-5880&   8  &
90&   2   \\
Dec. 11, 2002&  +52620.0&  GH   &  2.5$\times$6.0 &  4230-6070&   7.5&
90&   1.5 \\
Dec. 12, 2002&  +52621.0&  GH   &  2.5$\times$6.0 &  5750-7430&   8  &
90&   1.8 \\
Dec. 13, 2002&  +52622.0&  GH   &  2.5$\times$6.0 &  3740-7380&  14  &
90&   1.8 \\
Dec. 14, 2002&  +52623.0&  GH   &  2.5$\times$6.0 &  4240-6080&   8  &
90&   1.8 \\
Jan. 25, 2003&  +52665.0&  GH   &  2.5$\times$6.0 &  4300-5960&   7.5&
90&   1.5 \\
Jan. 26, 2003&  +52665.9&  GH   &  2.5$\times$6.0 &  5670-7360&   7.5&
90&   4   \\
Jan. 27, 2003&  +52666.9&  GH   &  2.5$\times$6.0 &  3920-7240&  15  &
90&   1.5 \\
Jan. 28, 2003&  +52667.9&  GH   &  2.5$\times$6.0 &  3980-7300&  12  &
90&   2.5 \\
Mar. 25, 2003&  +52723.8&  GH   &  2.5$\times$6.0 &  4240-6070&   7.5&
90&   3.5 \\
Mar. 26, 2003&  +52724.8&  GH   &  2.5$\times$6.0 &  3747-7385&  12  &
90&   4.5 \\
Mar. 27, 2003&  +52725.8&  GH   &  2.5$\times$6.0 &  5600-7460&  7.5 &
90&   2.5 \\
Apr. 10, 2003&  +52739.7&  GH   &  2.5$\times$6.0 &  5640-7500&   8  &
90&   4   \\
Apr. 11, 2003&  +52740.8&  GH   &  2.5$\times$6.0 &  4130-5960&  13  &
90&   5.4 \\
Apr. 12, 2003&  +52741.8&  GH   &  2.5$\times$6.0 &  3700-7340&  12  &
90&   4.0 \\
Apr. 13, 2003&  +52743.3&  L1(G)&  4.2$\times$19.8&  5640-7330&   9  &
90&   2   \\
May. 08, 2003&  +52768.4&  L1(U)&  4$\times$20.25 &  3750-6047&   8  &
90&   2.0 \\
May. 08, 2003&  +52768.3&  L(U) &  2.0$\times$6.0 &  3690-6044&   9  &
90&   1.6 \\
May. 9 , 2003&  +52769.3&  L(U) &  2.0$\times$6.0 &  3690-6044&   9  &
90&   1.5 \\
May. 10, 2003&  +52770.3&  L(U) &  2.0$\times$6.0 &  5740-8096&   8  &
90&   1.5 \\
May. 23, 2003&  +52782.7&  GH   &  2.5$\times$6.0 &  3540-7188&  12  &
90&   3.1 \\
May. 24, 2003&  +52783.8&  GH   &  2.5$\times$6.0 &  4240-6070&  7.5 &
90&   3.5 \\
May. 25, 2003&  +52784.7&  GH   &  2.5$\times$6.0 &  5582-7450&   8  &
90&   3.7 \\
May. 26, 2003&  +52785.7&  GH   &  2.5$\times$6.0 &  4230-6075&  7.5 &
90&   2.7 \\
Jun. 22, 2003&  +52812.8&  GH   &  2.5$\times$6.0 &  4270-6970&  7.5 &
90&   1.8 \\
Jun. 23, 2003&  +52813.7&  GH   &  2.5$\times$6.0 &  5620-7330&  7.5 &
90&   1.8 \\
Nov. 22, 2003&  +52965.6&  L1(G)&  4.2$\times$19.8&  4089-5798&   9  &
90&   2.0 \\
Nov. 23, 2003&  +52966.6&  L1(G)&  4.2$\times$19.8&  4090-5748&   9  &
90&   1.5 \\
Nov. 24, 2003&  +52967.9&  L1(G)&  4.2$\times$19.8&  4090-5748&   9  &
90&   1.5 \\
Dec. 22, 2003&  +52995.6&  L1(U)&  4.0$\times$20.2&  3750-6950&  10  &
90&   3.5 \\
Jan. 28, 2004&  +53032.9&  GH   &  2.5$\times$6.0 &  4238-5950&  12  &
90&   2   \\
Feb. 17, 2004&  +53053.0&  GH   &  2.5$\times$6.0 &  3736-7120&  17  &
90&   1.6 \\
Mar. 17, 2004&  +53081.8&  GH   &  2.5$\times$6.0 &  4205-5920&  12  &
90&   2   \\
Mar. 18, 2004&  +53082.7&  GH   &  2.5$\times$6.0 &  5664-7400&  14  &
90&       \\
Apr. 12, 2004&  +53107.6&  Z2K  &  4.0$\times$9.45&  3784-7170&   7.5&
90&   4.0 \\
Apr. 13, 2004&  +53108.7&  GH   &  2.5$\times$6.0 &  4216-5930&  12  &
90&   2.5 \\
Apr. 14, 2004&  +53109.7&  GH   &  2.5$\times$6.0 &  5622-7320&  14  &
90&       \\
May. 20, 2004&  +53145.7&  GH   &  2.5$\times$6.0 &  4193-5910&  12  &
90&   2.22\\
May. 21, 2004&  +53146.7&  GH   &  2.5$\times$6.0 &  5664-7390&   7.5&
90&       \\
Jun. 12, 2004&  +53168.7&  GH   &  2.5$\times$6.0 &  4208-5920&  10  &
90&   2.73\\
Jun. 13, 2004&  +53169.7&  GH   &  2.5$\times$6.0 &  5592-7320&   8  &
90&       \\
Jun. 17, 2004&  +53173.7&  GH   &  2.5$\times$6.0 &  4213-5920&  11  &
90&   2.50\\
Dec. 15, 2004&  +53355.0&  GH   &  2.5$\times$6.0 &  4185-5870&   7.5&
90&   2.96\\
Dec. 16, 2004&  +53356.0&  GH   &  2.5$\times$6.0 &  5739-7440&   7.5&
90&       \\
Dec. 18, 2004&  +53357.6&  L(S) &  1.0$\times$6.0 &  3900-7537&  12. &
270&   1.6 \\
Dec. 22, 2004&  +53361.5&  L(S) &  1.0$\times$6.0 &  3900-7537&  14  &
270&   2.5 \\
Jan. 16, 2005&  +53386.9&  GH   &  2.5$\times$6.0 &  3707-7096&  12  &
90&   3.05\\
Jan. 17, 2005&  +53388.0&  GH   &  2.5$\times$6.0 &  4183-5900&   9  &
90&   2.4 \\
Jan. 18, 2005&  +53388.9&  GH   &  2.5$\times$6.0 &  5577-7320&  12  &
90&       \\
Feb. 07, 2005&  +53408.8&  S-P  &  2.5$\times$6.0 &  5722-7590&   6.5&
90&   2.3 \\
Feb. 15, 2005&  +53416.9&  S-P  &  2.5$\times$6.0 &  3708-5807&   7  &
90&   2.5 \\
Feb. 15, 2005&  +53417.5&  L1(U)&  4.0$\times$9.45&  3750-7400&   8  &
90&   4.0 \\
Mar. 17, 2005&  +53446.8&  GH   &  2.5$\times$6.0 &  5557-7300&  13  &
90&       \\
Mar. 18, 2005&  +53447.8&  GH   &  2.5$\times$6.0 &  3689-7090&  13  &
90&   3.0 \\
Mar. 21, 2005&  +53451.3&  L1(U)&  4.0$\times$9.45&  3750-7400&   9  &
90&   8   \\
Apr. 13, 2005&  +53474.4&  L1(U)&  4.0$\times$9.45&  3750-7400&   8  &
90&   2.5 \\
Apr. 15, 2005&  +53475.8&  GH   &  2.5$\times$6.0 &  4245-5960&   9  &
90&   2.82\\
Apr. 16, 2005&  +53476.8&  GH   &  2.5$\times$6.0 &  5521-7256&  10  &
90&       \\
Apr. 16, 2005&  +53477.3&  L1(U)&  4.0$\times$9.45&  3750-7400&   8  &
90&   5.5 \\
Apr. 18, 2005&  +53478.7&  GH   &  2.5$\times$6.0 &  3745-7190&  13  &
90&       \\
May. 12, 2005&  +53503.3&  L1(U)&  4.0$\times$9.45&  3750-7400&   8  &
90&   2.0 \\
May. 13, 2005&  +53503.7&  GH   &  2.5$\times$6.0 &  4216-5910&   9  &
90&   3.0 \\
May. 14, 2005&  +53504.7&  GH   &  2.5$\times$6.0 &  5583-7300&   7  &
90&       \\
May. 16, 2005&  +53507.4&  L1(U)&  4.0$\times$9.45&  3750-7400&   8  &
90&   3.2 \\
Jun. 09, 2005&  +53530.7&  GH   &  2.5$\times$6.0 &  3714-7070&  12  &
90&   1.94\\
Jun. 10, 2005&  +53531.6&  GH   &  2.5$\times$6.0 &  4274-5970&   9  &
90&   3.37\\
Jun. 11, 2005&  +53531.6&  GH   &  2.5$\times$6.0 &  5676-7395&   7.5&
90&       \\
Jun. 16, 2005&  +53538.3&  L1(U)&  4.0$\times$9.45&  3740-7350&   9  &
90&   2.5 \\
Nov. 28, 2005&  +53703.0&  GH   &  2.5$\times$6.0 &  3590-6900&  15  &
90&   3.0 \\
Nov. 29, 2005&  +53704.0&  GH   &  2.5$\times$6.0 &  4230-5910&   7  &
90&   2.8 \\
Dec. 06, 2005&  +53711.0&  S-P  &  2.5$\times$6.0 &  3690-5780&   7  &
90&   2.5 \\
Dec. 07, 2005&  +53712.0&  S-P  &  2.5$\times$6.0 &  3690-5780&   7  &
90&   2.5 \\
Dec. 27, 2005&  +53732.0&  GH   &  2.5$\times$6.0 &  3890-7270&  17  &
90&   3   \\
Dec. 28, 2005&  +53733.0&  GH   &  2.5$\times$6.0 &  3880-7260&  17  &
90&   2.4 \\
Jan. 21, 2006&  +53756.9&  GH   &  2.5$\times$6.0 &  4330-6040&   9  &
90&   2.7 \\
Jan. 22, 2006&  +53757.9&  GH   &  2.5$\times$6.0 &  4330-6040&   9  &
90&   3   \\
Jan. 24, 2006&  +53760.5&  L1(U)&  4.0$\times$9.45&  3740-7400&   9  &
90&   3.5 \\
Jan. 25, 2006&  +53761.5&  L1(U)&  4.0$\times$9.45&  3740-7400&   9  &
90&   2.5 \\
Feb. 20, 2006&  +53786.9&  GH   &  2.5$\times$6.0 &  3740-7120&  17  &
90&   2.8 \\
Feb. 21, 2006&  +53787.5&  L1(U)&  4.0$\times$9.45&  3740-7400&   8  &
90&   2   \\
Feb. 22, 2006&  +53788.5&  L1(U)&  4.0$\times$9.45&  3740-7400&   8  &
90&   2   \\
Feb. 23, 2006&  +53789.5&  GH   &  2.5$\times$6.0 &  3740-7400&  15  &
90&   2.5 \\
Mar. 09, 2006&  +53803.8&  GH   &  2.5$\times$6.0 &  3730-7100&  14  &
90&   5.1 \\
Mar. 21, 2006&  +53816.4&  L1(U)&  4.0$\times$9.45&  3740-7400&   8  &
90&   5   \\
Mar. 22, 2006&  +53817.4&  L1(U)&  4.0$\times$9.45&  3740-7400&   8  &
90&   4   \\
Apr. 18, 2006&  +53843.7&  GH   &  2.5$\times$6.0 &  3720-7090&   9  &
90&   2.7 \\
Apr. 19, 2006&  +53844.8&  GH   &  2.5$\times$6.0 &  4240-5940&   7  &
90&   2.5 \\
Apr. 20, 2006&  +53845.7&  GH   &  2.5$\times$6.0 &  4240-5940&   7  &
90&   2.1 \\
Apr. 20, 2006&  +53846.4&  L1(U)&  4.0$\times$9.45&  3740-7400&   8  &
90&   2.5 \\
\hline
\end{longtable}
}
 \onllongtab{7}{
\begin{longtable}{lrllll}
\caption{\label{tab7} Observed continuum, H$\alpha$, H$\beta$,
H$\gamma$ and HeII$\lambda\,4686$ fluxes, reduced to the 6 m telescope
aperture $2''\times 6''$. The fluxes are corrected for  position
angle (to PA=90 degree), seeing and aperture effects. Columns: 1
- Julian date; 2 - $F$(cont), the continuum flux at 5117\,\AA\,
(in units of $10^{-14}$\, erg\,s$^{-1}$\,cm$^{-2}$\,\AA$^{-1}$);
and $\varepsilon_{cont}$, the estimated  continuum flux error; 3 -
$F(H\alpha)$, the H$\alpha$ total flux (in units of
$10^{-12}$\,erg\,s$^{-1}$\,cm$^{-2}$), and  $\varepsilon_{\rm
H\alpha}$, the H$\alpha$ flux error; 4 - $F(H\beta)$, the
H$\beta$ total flux (in units of
10$^{-12}$\,erg\,s$^{-1}$\,cm$^{-2}$), and $\varepsilon_{\rm
H\beta}$, the H$\beta$ flux error; 5 - $F(H\gamma)$, the
H$\gamma$ total flux (in units of
10$^{-12}$\,erg\,s$^{-1}$\,cm$^{-2}$), and $\varepsilon_{\rm
H\gamma}$, the H$\gamma$ flux error; 6 - $F(HeII)$, the
HeII$\lambda$4686 total flux (in units of
10$^{-12}$\,erg\,s$^{-1}$\,cm$^{-2}$), and $\varepsilon_{\rm
HeII}$, the HeII$\lambda$4686 flux error.}\\
 \hline
 JD     &
$F{\rm (cont)} \pm\varepsilon_{cont}$&
$F(H\alpha)\pm\varepsilon_{\rm H\alpha}$&
$F(H\beta)\pm\varepsilon_{\rm H\beta}$&
$F(H\gamma)\pm\varepsilon_{\rm H\gamma}$&
$F(HeII)\pm\varepsilon_{\rm HeII}$ \\
\hline
\endfirsthead
\caption{Continued.}\\
\hline
 JD     &
$F{\rm (cont)}\pm\varepsilon_{cont}$&
$F(H\alpha)\pm\varepsilon_{\rm H\alpha}$&
$F(H\beta)\pm\varepsilon_{\rm H\beta}$&
$F(H\gamma)\pm\varepsilon_{\rm H\gamma}$&
$F(HeII)\pm\varepsilon_{\rm HeII}$ \\
\hline
\endhead
\hline
\endfoot
\hline
\endlastfoot
50094.5&   8.8    $\pm$0.34 &--        --  & 8.03$\pm$0.31&  4.807$\pm$0.361&  3.136$\pm$ 0.031\\
50097.6&   9.26   $\pm$0.36 &33.41$\pm$0.63& 8.19$\pm$0.32&  3.893$\pm$0.292&  2.707$\pm$ 0.027\\
50098.6&   9.26   $\pm$0.36:&35.16$\pm$0.63& --       --  &  --        --   &  --         --   \\
50128  &   11.25  $\pm$0.44 &34.34$\pm$0.65& 9.28$\pm$0.36&  4.524$\pm$0.339&  3.495$\pm$ 0.035\\
50162.4&   11.66  $\pm$0.45 &--        --  & 8.9 $\pm$0.35&  3.638$\pm$0.273&  2.22 $\pm$ 0.022\\
50163.3&   11.66  $\pm$0.45:&33.79$\pm$0.64& --       --  &  --        --   &  --         --   \\
50164.4&   10.78  $\pm$0.42 &--        --  & 8.5 $\pm$0.33&  4.044$\pm$0.303&  3.485$\pm$ 0.035\\
50165.4&   10.78  $\pm$0.42:&33.24$\pm$0.63& --  --       &  --        --   &  --         --   \\
50166.3&   10.78  $\pm$0.42:&32.04$\pm$0.61& --  --       &  --        --   &  --         --   \\
50200.3&   9.6    $\pm$0.37 &--        --  & 9.34$\pm$0.36&  --        --   &  --         --   \\
50201.3&   8.85   $\pm$0.35 &30.97$\pm$0.59& 8.73$\pm$0.34&  --        --   &  --         --   \\
50249.3&   12.64  $\pm$0.49 &35.45$\pm$0.67& 8.24$\pm$0.32&  --        --   &  --         --   \\
50275.3&   11.99  $\pm$0.47 &35.72$\pm$0.68& 9.83$\pm$0.38&  4.649$\pm$0.349&  3.313$\pm$ 0.033\\
50276.3&   11.99  $\pm$0.47:&35.88$\pm$0.68& --       --  &  --        --   &  --         --   \\
50277.3&   11.29  $\pm$0.44 &36.6 $\pm$0.70& 9.26$\pm$0.36&  --        --   &  --         --   \\
50280.3&   12.19  $\pm$0.48 &35.95$\pm$0.68& 10  $\pm$0.39&  --        --   &  --         --   \\
50281.3&   12.19  $\pm$0.48:&35.53$\pm$0.68& --       --  &  --        --   &  --         --   \\
50392.6&   10.04  $\pm$0.39 &--        --  & 9.97$\pm$0.39&  --        --   &  --         --   \\
50402.6&   10.65  $\pm$0.42 &39.95$\pm$0.76& 10.8$\pm$0.42&  5.003$\pm$0.375&  3.792$\pm$ 0.038\\
50510.4&   7.84   $\pm$0.41 &32.4 $\pm$0.81& 9.08$\pm$0.15&  --        --   &  --         --   \\
50511.4&   8.86   $\pm$0.46 &--        --  & 9.57$\pm$0.15&  --        --   &  --         --   \\
50543.6&   6.82   $\pm$0.35 &32.15$\pm$0.80& 7.44$\pm$0.12&  3.336$\pm$0.200&  2.088$\pm$ 0.021\\
50544.4&   6.47   $\pm$0.34 &32.54$\pm$0.81& 7.35$\pm$0.12&  --        --   &  --         --   \\
50547.3&   6.92   $\pm$0.36 &--        --  & 7   $\pm$0.11&  --        --   &  --         --   \\
50552.3&   7.38   $\pm$0.38 &--        --  & 8.5 $\pm$0.14&  --        --   &  --         --   \\
50809.7&   7.56   $\pm$0.39 &--        --  & 9.07$\pm$0.15&  --        --   &  --         --   \\
50810.7&   7.22   $\pm$0.38 &--        --  & 9.04$\pm$0.14&  --        --   &  --         --   \\
50833.6&   5.91   $\pm$0.13 &31.65$\pm$0.98& 7.31$\pm$0.14&  3.256$\pm$0.189&  1.912$\pm$ 0.019\\
50834.6&   6.09   $\pm$0.13 &31.03$\pm$0.96& 7.54$\pm$0.14&  3.466$\pm$0.201&  2.12 $\pm$ 0.021\\
50836.7&   6.4    $\pm$0.14 &--        --  & 7.33$\pm$0.14&  3.837$\pm$0.223&  2.745$\pm$ 0.027\\
50836.7&   6.42   $\pm$0.14 &--        --  & 7.82$\pm$0.15&  3.619$\pm$0.210&  2.524$\pm$ 0.025\\
50842.4&   6.59   $\pm$0.14 &32.87$\pm$1.02& 8.02$\pm$0.15&  --        --   &  --         --   \\
50867.4&   6.22   $\pm$0.14 &33.34$\pm$1.03& 7.34$\pm$0.14&  3.363$\pm$0.195&  2.445$\pm$ 0.024\\
50934.5&   6.39   $\pm$0.14 &33.14$\pm$1.03& 7.06$\pm$0.13&  --        --   &  --         --   \\
50938.3&   6.55   $\pm$0.14 &33.6 $\pm$1.04& 6.77$\pm$0.13&  2.643$\pm$0.153&  1.902$\pm$ 0.019\\
50940.5&   6.27   $\pm$0.14 &30.78$\pm$0.95& 6.82$\pm$0.13&  2.878$\pm$0.167&  1.457$\pm$ 0.015\\
50941.5&   5.64   $\pm$0.12 &--        --  & 6.41$\pm$0.12&  2.608$\pm$0.151&  1.135$\pm$ 0.011\\
50942.5&   5.7    $\pm$0.13 &29.87$\pm$0.93& 6.42$\pm$0.12&  2.432$\pm$0.141&  1.516$\pm$ 0.015\\
50942.5&   6.09   $\pm$0.13 &--        --  & 6.25$\pm$0.12&  2.653$\pm$0.154&  1.157$\pm$ 0.012\\
50985.3&   8.52   $\pm$0.19 &34.03$\pm$1.05& 8.29$\pm$0.16&  3.118$\pm$0.181&  2.738$\pm$ 0.027\\
50991.3&   7.43   $\pm$0.16 &--        --  & 8.24$\pm$0.16&  3.692$\pm$0.214&  2.946$\pm$ 0.029\\
51008.7&   5.84   $\pm$0.13 &31.09$\pm$0.96& 7.44$\pm$0.14&  3.355$\pm$0.195&  1.743$\pm$ 0.017\\
51010.7&   5.98   $\pm$0.13 &28.62$\pm$0.89& 7.59$\pm$0.14&  3.677$\pm$0.213&  2.173$\pm$ 0.022\\
51011.6&   5.81   $\pm$0.13 &28.82$\pm$0.89& 7.47$\pm$0.14&  3.258$\pm$0.189&  1.974$\pm$ 0.020\\
51015.7&   6.26   $\pm$0.14 &31.15$\pm$0.97& 7.84$\pm$0.15&  3.637$\pm$0.211&  1.829$\pm$ 0.018\\
51025.3&   6.21   $\pm$0.14 &30.15$\pm$0.93& 7.45$\pm$0.14&  --        --   &  --         --   \\
51130.6&   5.69   $\pm$0.13 &32.59$\pm$1.01& 7.05$\pm$0.13&  --        --   &  --         --   \\
51166.6&   4.26   $\pm$0.09 &33.07$\pm$1.03& 6.96$\pm$0.13&  --        --   &  --         --   \\
51190.7&   5.56   $\pm$0.18:&29.56$\pm$0.53& --       --  &  --        --   &  --         --   \\
51191.9&   5.56   $\pm$0.18 &28.47$\pm$0.51& 6.85$\pm$0.16&  3.54 $\pm$0.202&  2.377$\pm$ 0.150\\
51193  &   5.81   $\pm$0.19 &28.47$\pm$0.51& 6.6 $\pm$0.15&  3.465$\pm$0.198&  2.62 $\pm$ 0.165\\
51200.5&   6.07   $\pm$0.19 &--        --  & 7.36$\pm$0.17&  --        --   &  --         --   \\
51201.6&   6.42   $\pm$0.21 &33.51$\pm$0.60& 7.91$\pm$0.18&  --        --   &  --         --   \\
51203.6&   6.57   $\pm$0.21 &34.83$\pm$0.63& 8.05$\pm$0.19&  --        --   &  --         --   \\
51218.6&   6.43   $\pm$0.21 &35.24$\pm$0.63& 7.86$\pm$0.18&  3.74 $\pm$0.213&  2.888$\pm$ 0.182\\
51221.7&   6.38   $\pm$0.20 &36.73$\pm$0.66& 8.09$\pm$0.19&  3.452$\pm$0.197&  3.043$\pm$ 0.192\\
51222.6&   6.06   $\pm$0.19 &--        --  & 7.92$\pm$0.18&  --        --   &  --         --   \\
51223.6&   5.69   $\pm$0.18 &--        --  & 7.34$\pm$0.17&  --        --   &  --         --   \\
51252.9&   5.03   $\pm$0.16 &29.45$\pm$0.53& 6.16$\pm$0.14&  2.886$\pm$0.165&  1.883$\pm$ 0.119\\
51258.5&   5.79   $\pm$0.19 &32.55$\pm$0.59& 6.95$\pm$0.16&  --        --   &  --         --   \\
51261.5&   5.81   $\pm$0.19 &31.19$\pm$0.56& 6.93$\pm$0.16&  --        --   &  --         --   \\
51262.3&   5.77   $\pm$0.18:&31.47$\pm$0.57& --       --  &  --        --   &  --         --   \\
51262.5&   5.77   $\pm$0.18 &32.2 $\pm$0.58& 7.12$\pm$0.16&  --        --   &  --         --   \\
51277.6&   5.79   $\pm$0.19 &--        --  & 7.6 $\pm$0.17&  3.425$\pm$0.195&  3.217$\pm$ 0.203\\
51279.5&   5.5    $\pm$0.18 &34.91$\pm$0.63& 7.51$\pm$0.17&  --        --   &  --         --   \\
51283.5&   5.29   $\pm$0.17 &--        --  & 6.56$\pm$0.15&  --        --   &  --         --   \\
51344.3&   2.87   $\pm$0.09 &--        --  & 4.2 $\pm$0.10&  1.716$\pm$0.098&  0.998$\pm$ 0.063\\
51346.4&   2.75   $\pm$0.09 &--        --  & 4.29$\pm$0.10&  2.212$\pm$0.126&  1.092$\pm$ 0.069\\
51514.6&   2.77   $\pm$0.09 &--        --  & 3.52$\pm$0.08&  1.668$\pm$0.095&  1.252$\pm$ 0.079\\
51515.6&   2.79   $\pm$0.09 &22.94$\pm$0.41& 3.6 $\pm$0.08&  1.58 $\pm$0.090&  1.193$\pm$ 0.075\\
51517.6&   3.1    $\pm$0.10 &--        --  & 3.47$\pm$0.08&  1.574$\pm$0.090&  1.391$\pm$ 0.088\\
51552.6&   4.16   $\pm$0.10 &--        --  & 4.51$\pm$0.09&  --        --   &  --         --   \\
51553.6&   4.16   $\pm$0.10:&25.93$\pm$0.93& --       --  &  --        --   &  --         --   \\
51570.9&   3.86   $\pm$0.09:&25.13$\pm$0.90& --       --  &  --        --   &  --         --   \\
51571.9&   3.86   $\pm$0.09 &23.57$\pm$0.85& 4.67$\pm$0.10&  2.257$\pm$0.065&  1.035$\pm$ 0.049\\
51585.5&   3.56   $\pm$0.08 &--        --  & 4.16$\pm$0.09&  2.615$\pm$0.076&  1.52 $\pm$ 0.071\\
51588.5&   3.6    $\pm$0.08 &--        --  & 4.48$\pm$0.09&  2.021$\pm$0.059&  1.517$\pm$ 0.071\\
51589.5&   3.6    $\pm$0.08:&24.12$\pm$0.87& --       --  &  --        --   &  --         --   \\
51600.8&   2.78   $\pm$0.06 &22.51$\pm$0.81& 4.01$\pm$0.08&  --        --   &  --         --   \\
51601.8&   2.77   $\pm$0.06 &22.28$\pm$0.80& 4.15$\pm$0.09&  --        --   &  --         --   \\
51638.5&   2.53   $\pm$0.06 &--        --  & 3.7 $\pm$0.08&  --        --   &  --         --   \\
51640.5&   2.3    $\pm$0.05 &--        --  & 3.57$\pm$0.07&  --        --   &  --         --   \\
51659.8&   2.63   $\pm$0.06 &20.19$\pm$0.73& 3.6 $\pm$0.08&  1.88 $\pm$0.055&  1.156$\pm$ 0.054\\
51660.8&   2.61   $\pm$0.06 &19.04$\pm$0.69& 3.56$\pm$0.07&  1.943$\pm$0.056&  1.144$\pm$ 0.054\\
51676.4&   1.88   $\pm$0.04 &--        --  & 3.23$\pm$0.07&  --        --   &  --         --   \\
51689.7&   1.95   $\pm$0.04 &17.89$\pm$0.64& 3.09$\pm$0.06&  1.678$\pm$0.049&  0.649$\pm$ 0.031\\
51690.7&   1.92   $\pm$0.04 &19.68$\pm$0.71& 3.01$\pm$0.06&  1.528$\pm$0.044&  0.711$\pm$ 0.033\\
51736.4&   2.31   $\pm$0.05 &--        --  & 3.22$\pm$0.07&  --        --   &  --         --   \\
51755.3&   2.72   $\pm$0.06 &--        --  & 3.35$\pm$0.07&  --        --   &  --         --   \\
51869.6&   1.86   $\pm$0.04 &--        --  & 3.84$\pm$0.08&  2.061$\pm$0.060&  0.844$\pm$ 0.040\\
51878.6&   1.51   $\pm$0.03 &--        --  & 3.14$\pm$0.07&  1.687$\pm$0.049&  0.918$\pm$ 0.043\\
51895.9&   1.7    $\pm$0.04 &15.92$\pm$0.57& 2.49$\pm$0.05&  1.203$\pm$0.035&  0.71 $\pm$ 0.033\\
51897  &   1.7    $\pm$0.04 &16.61$\pm$0.60& 2.55$\pm$0.05&  1.18 $\pm$0.034&  0.698$\pm$ 0.033\\
51898  &   1.79   $\pm$0.04 &16.22$\pm$0.58& 2.64$\pm$0.06&  1.2  $\pm$0.035&  0.806$\pm$ 0.038\\
51936.5&   1.79   $\pm$0.07:&20.81$\pm$0.60& --       --  &  --        --   &  --         --   \\
51937.5&   1.79   $\pm$0.07 &--        --  & 2.96$\pm$0.05&  --        --   &  --         --   \\
51940.6&   1.81   $\pm$0.07 &18.73$\pm$0.54& 2.91$\pm$0.05&  --        --   &  --         --   \\
51943.5&   1.76   $\pm$0.07 &20.09$\pm$0.58& 2.94$\pm$0.05&  --        --   &  --         --   \\
51952.4&   1.31   $\pm$0.05 &--        --  & 2.75$\pm$0.04&  --        --   &  --         --   \\
51981.7&   1.71   $\pm$0.06 &14.31$\pm$0.41& 2.32$\pm$0.04&  1.11 $\pm$0.047&  0.625$\pm$ 0.008\\
52013.3&   2.24   $\pm$0.09 &15.86$\pm$0.46& 2.47$\pm$0.04&  --        --   &  --         --   \\
52016.4&   2.6    $\pm$0.10 &17.27$\pm$0.50& 2.63$\pm$0.04&  --        --   &  --         --   \\
52029.5&   1.78   $\pm$0.07 &--        --  & 2.79$\pm$0.04&  --        --   &  --         --   \\
52034.6&   1.8    $\pm$0.07 &12.97$\pm$0.38& 2.46$\pm$0.04&  1.127$\pm$0.047&  0.684$\pm$ 0.009\\
52036.4&   1.93   $\pm$0.07 &--        --  & 2.61$\pm$0.04&  1.217$\pm$0.051&  0.86 $\pm$ 0.011\\
52041.8&   1.83   $\pm$0.07 &13.31$\pm$0.39& 2.26$\pm$0.04&  1.107$\pm$0.046&  0.696$\pm$ 0.009\\
52043.7&   1.99   $\pm$0.08 &13.71$\pm$0.40& 2.25$\pm$0.04&  1.155$\pm$0.049&  0.608$\pm$ 0.008\\
52074.7&   2.68   $\pm$0.10 &17.38$\pm$0.50& 3.41$\pm$0.05&  1.667$\pm$0.070&  0.687$\pm$ 0.009\\
52075.7&   2.84   $\pm$0.11 &15.08$\pm$0.44& 3.13$\pm$0.05&  1.539$\pm$0.065&  0.798$\pm$ 0.010\\
52237.6&   3.34   $\pm$0.13 &21.66$\pm$0.63& 5.06$\pm$0.08&  2.221$\pm$0.093&  1.432$\pm$ 0.019\\
52238  &   3.33   $\pm$0.13 &--        --  & 5.12$\pm$0.08&  --        --   &  --         --   \\
52239  &   3.33   $\pm$0.13:&20.55$\pm$0.60& --       --  &  --        --   &  --         --   \\
52297.6&   3.45   $\pm$0.07 &--        --  & 4.23$\pm$0.16&  --        --   &  --         --   \\
52299.4&   3.68   $\pm$0.08 &19.87$\pm$0.54& 3.95$\pm$0.15&  --        --   &  --         --   \\
52327.4&   3.21   $\pm$0.07 &--        --  & 3.86$\pm$0.15&  --        --   &  --         --   \\
52327.5&   3.14   $\pm$0.07 &--        --  & 3.89$\pm$0.15&  1.35 $\pm$0.073&  0.988$\pm$ 0.055\\
52338.7&   4.17   $\pm$0.09 &19.93$\pm$0.54& 3.99$\pm$0.16&  1.773$\pm$0.096&  1.378$\pm$ 0.077\\
52339.7&   4.17   $\pm$0.09:&21.14$\pm$0.57& --       --  &  --        --   &  --         --   \\
52340.8&   3.96   $\pm$0.08 &--        --  & 4.42$\pm$0.17&  --        --   &  --         --   \\
52350.8&   2.74   $\pm$0.06 &--        --  & 4.03$\pm$0.16&  1.666$\pm$0.090&  0.755$\pm$ 0.042\\
52367.7&   3.94   $\pm$0.08 &--        --  & 3.83$\pm$0.15&  --        --   &  --         --   \\
52368.7&   3.94   $\pm$0.08:&19.06$\pm$0.51& --       --  &  --        --   &  --         --   \\
52369.7&   3.79   $\pm$0.08 &19.39$\pm$0.52& 3.73$\pm$0.15&  1.605$\pm$0.087&  0.863$\pm$ 0.048\\
52370.7&   3.91   $\pm$0.08 &19.29$\pm$0.52& 3.93$\pm$0.15&  1.671$\pm$0.090&  0.911$\pm$ 0.051\\
52397.7&   3.3    $\pm$0.07 &--        --  & 3.89$\pm$0.15&  1.368$\pm$0.074&  0.622$\pm$ 0.035\\
52398.7&   3.3    $\pm$0.07:&18.82$\pm$0.51& --       --  &  --        --   &  --         --   \\
52399.7&   3.27   $\pm$0.07 &--        --  & 3.72$\pm$0.15&  1.347$\pm$0.073&  0.568$\pm$ 0.032\\
52411.4&   3.84   $\pm$0.08 &--        --  & 4.04$\pm$0.16&  --        --   &  --         --   \\
52427.7&   3.72   $\pm$0.08 &--        --  & 4   $\pm$0.17&  1.447$\pm$0.078&  0.693$\pm$ 0.039\\
52429.7&   3.71   $\pm$0.08 &20.55$\pm$0.55& 4.34$\pm$0.17&  1.718$\pm$0.093&  0.762$\pm$ 0.043\\
52430.7&   3.75   $\pm$0.08 &--        --  & 4.04$\pm$0.16&  --        --   &  --         --   \\
52450.4&   4.24   $\pm$0.09 &21.93$\pm$0.59& 4.5 $\pm$0.18&  1.571$\pm$0.085&  1.325$\pm$ 0.074\\
52620  &   3.36   $\pm$0.07 &--        --  & 4.29$\pm$0.17&  --        --   &  --         --   \\
52621  &   3.36   $\pm$0.07:&19.15$\pm$0.52& --       --  &  --        --   &  --         --   \\
52622  &   3.44   $\pm$0.07 &19.95$\pm$0.54& 4.32$\pm$0.17&  1.914$\pm$0.103&  0.769$\pm$ 0.043\\
52623  &   3.58   $\pm$0.08 &--        --  & 4.76$\pm$0.19&  --        --   &  --         --   \\
52665  &   5.07   $\pm$0.17 &--        --  & 4.79$\pm$0.14&  --        --   &  --         --   \\
52665.9&   5.07   $\pm$0.17:&24.08$\pm$0.67& --       --  &  --        --   &  --         --   \\
52666.9&   5.07   $\pm$0.17 &22.26$\pm$0.62& 4.79$\pm$0.14&  2.027$\pm$0.097&  1.21 $\pm$ 0.092\\
52667.9&   5.57   $\pm$0.18 &23.24$\pm$0.65& 4.96$\pm$0.14&  2.246$\pm$0.108&  1.364$\pm$ 0.104\\
52723.8&   4.54   $\pm$0.15 &--        --  & 5.35$\pm$0.16&  2.304$\pm$0.111&  1.227$\pm$ 0.093\\
52724.8&   4.17   $\pm$0.14 &22.59$\pm$0.63& 4.96$\pm$0.14&  2.277$\pm$0.109&  1.369$\pm$ 0.104\\
52725.8&   4.17   $\pm$0.14:&21.54$\pm$0.60& --       --  &  --        --   &  --         --   \\
52739.7&   5.17   $\pm$0.17:&23.31$\pm$0.65& --       --  &  --        --   &  --         --   \\
52740.8&   5.17   $\pm$0.17 &--        --  & 5.97$\pm$0.17&  2.729$\pm$0.131&  1.347$\pm$ 0.102\\
52741.8&   5.15   $\pm$0.17 &23.5 $\pm$0.66& 5.36$\pm$0.16&  2.603$\pm$0.125&  1.65 $\pm$ 0.125\\
52743.3&   5.15   $\pm$0.17:&22.47$\pm$0.63& --       --  &  --        --   &  --         --   \\
52768.4&   5.16   $\pm$0.17 &--        --  & 5.67$\pm$0.16&  --        --   &  --         --   \\
52768.3&   5.08   $\pm$0.17 &--        --  & 5.44$\pm$0.16&  2.173$\pm$0.104&  1.277$\pm$ 0.097\\
52769.3&   5.33   $\pm$0.18 &24.39$\pm$0.68& 5.57$\pm$0.16&  2.064$\pm$0.099&  1.353$\pm$ 0.103\\
52770.3&   5.33   $\pm$0.18:&23.29$\pm$0.65& --       --  &  --        --   &  --         --   \\
52782.7&   5.03   $\pm$0.17 &24.98$\pm$0.70& 5.82$\pm$0.17&  2.488$\pm$0.119&  1.387$\pm$ 0.105\\
52783.8&   5.17   $\pm$0.17 &--        --  & 5.47$\pm$0.16&  2.714$\pm$0.130&  1.374$\pm$ 0.104\\
52784.7&   5.55   $\pm$0.17:&23.56$\pm$0.66& --       --  &  --        --   &  --         --   \\
52785.7&   5.55   $\pm$0.18 &--        --  & 5.31$\pm$0.15&  --        --   &  --         --   \\
52812.8&   4.18   $\pm$0.14 &--        --  & 5.74$\pm$0.17&  2.467$\pm$0.118&  1.429$\pm$ 0.109\\
52813.7&   4.18   $\pm$0.14:&22.76$\pm$0.64& --       --  &  --        --   &  --         --   \\
52965.6&   3.68   $\pm$0.12 &--        --  & 4.79$\pm$0.14&  --        --   &  --         --   \\
52966.6&   3.46   $\pm$0.11 &--        --  & 4.87$\pm$0.14&  2.428$\pm$0.117&  1.371$\pm$ 0.104\\
52967.9&   3.3    $\pm$0.11 &--        --  & 4.83$\pm$0.14&  2.709$\pm$0.130&  1.588$\pm$ 0.121\\
52995.6&   3.52   $\pm$0.12 &--        --  & 5.55$\pm$0.16&  2.504$\pm$0.120&  2.258$\pm$ 0.172\\
53032.9&   3.26   $\pm$0.09 &--        --  & 4.38$\pm$0.10&  2.111$\pm$0.082&  0.849$\pm$ 0.062\\
53053  &   3.19   $\pm$0.09 &20.35$\pm$0.53& 4.92$\pm$0.11&  2.405$\pm$0.094&  1.4  $\pm$ 0.102\\
53081.8&   2.76   $\pm$0.08 &--        --  & 4.18$\pm$0.10&  1.861$\pm$0.073&  0.695$\pm$ 0.051\\
53082.7&   2.76   $\pm$0.08:&17.91$\pm$0.47& --       --  &  --        --   &  --         --   \\
53107.6&   2.03   $\pm$0.06:&16.54$\pm$0.43& --       --  &  --        --   &  --         --   \\
53108.7&   2.03   $\pm$0.06 &--        --  & 3.04$\pm$0.70&  1.432$\pm$0.056&  0.452$\pm$ 0.033\\
53109.7&   2.03   $\pm$0.06:&16.14$\pm$0.42& --       --  &  --        --   &  --         --   \\
53145.7&   1.92   $\pm$0.05 &--        --  & 2.64$\pm$0.06&  1.18 $\pm$0.046&  0.585$\pm$ 0.043\\
53146.7&   1.92   $\pm$0.05:&14.7 $\pm$0.38& --       --  &  --        --   &  --         --   \\
53168.7&   2.29   $\pm$0.07 &--        --  & 2.74$\pm$0.06&  1.088$\pm$0.042&  0.601$\pm$ 0.044\\
53169.7&   2.29   $\pm$0.07:&14.21$\pm$0.37& --       --  &  --        --   &  --         --   \\
53173.7&   2.7    $\pm$0.08 &--        --  & 2.76$\pm$0.06&  1.338$\pm$0.052&  0.607$\pm$ 0.044\\
53355  &   2.85   $\pm$0.08 &--        --  & 3.72$\pm$0.09&  1.618$\pm$0.063&  0.857$\pm$ 0.061\\
53356  &   2.85   $\pm$0.08:&18.13$\pm$0.47& --       --  &  --        --   &  --         --   \\
53357.6&   2.85   $\pm$0.08 &18.91$\pm$0.49& 3.71$\pm$0.09&  1.466$\pm$0.057&  0.799$\pm$ 0.058\\
53361.5&   2.81   $\pm$0.08 &17.85$\pm$0.46& 3.48$\pm$0.08&  1.481$\pm$0.058&  0.616$\pm$ 0.045\\
53386.9&   2.56   $\pm$0.06 &19.21$\pm$0.52& 3.84$\pm$0.08&  1.736$\pm$0.057&  0.756$\pm$ 0.056\\
53388  &   2.57   $\pm$0.06 &--        --  & 3.79$\pm$0.08&  1.692$\pm$0.056&  0.625$\pm$ 0.046\\
53388.9&   2.57   $\pm$0.06:&17.64$\pm$0.48& --       --  &  --        --   &  --         --   \\
53408.8&   2.3    $\pm$0.06:&16.94$\pm$0.46& --       --  &  --        --   &  --         --   \\
53416.9&   2.04   $\pm$0.05 &--        --  & 2.96$\pm$0.06&  1.294$\pm$0.043&  0.639$\pm$ 0.047\\
53417.5&   2.04   $\pm$0.05:&16.36$\pm$0.44& --       --  &  --        --   &  --         --   \\
53446.8&   1.92   $\pm$0.05:&16.16$\pm$0.44& --       --  &  --        --   &  --         --   \\
53447.8&   1.92   $\pm$0.05 &16.62$\pm$0.45& 3.07$\pm$0.06&  1.376$\pm$0.045&  0.716$\pm$ 0.053\\
53451.3&   2.14   $\pm$0.05 &16.8 $\pm$0.45& 3.1 $\pm$0.06&  --        --   &  --         --   \\
53474.4&   1.73   $\pm$0.05:&14.01$\pm$0.38& --       --  &  --        --   &  --         --   \\
53475.8&   1.73   $\pm$0.04 &--        --  & 2.47$\pm$0.05&  1.014$\pm$0.033&  0.62 $\pm$ 0.046\\
53476.8&   1.73   $\pm$0.04:&13.97$\pm$0.38& --       --  &  --        --   &  --         --   \\
53477.3&   1.78   $\pm$0.04:&15.77$\pm$0.43& --       --  &  --        --   &  --         --   \\
53478.7&   1.83   $\pm$0.04 &15.08$\pm$0.41& 2.42$\pm$0.05&  1.071$\pm$0.035&  0.684$\pm$ 0.051\\
53503.3&   2.16   $\pm$0.05 &14.79$\pm$0.40& 2.51$\pm$0.05&  --        --   &  --         --   \\
53503.7&   2.16   $\pm$0.05 &--        --  & 2.74$\pm$0.05&  1.241$\pm$0.041&  0.796$\pm$ 0.059\\
53504.7&   2.16   $\pm$0.05:&14.65$\pm$0.40& --       --  &  --        --   &  --         --   \\
53507.4&   2.1    $\pm$0.05 &15.76$\pm$0.43& 2.6 $\pm$0.05&  --        --   &  --         --   \\
53530.7&   1.89   $\pm$0.05 &15.07$\pm$0.41& 2.62$\pm$0.05&  1.168$\pm$0.039&  0.681$\pm$ 0.050\\
53531.6&   1.94   $\pm$0.05 &--        --  & 2.74$\pm$0.05&  --        --   &  --         --   \\
53531.6&   1.94   $\pm$0.05:&14.92$\pm$0.40& --       --  &  --        --   &  --         --   \\
53538.3&   1.74   $\pm$0.04 &13.25$\pm$0.36& 2.49$\pm$0.05&  1.207$\pm$0.040&  0.93 $\pm$ 0.069\\
53703  &   3.73   $\pm$0.09 &18.92$\pm$0.51& 4.42$\pm$0.09&  --        --   &  --         --   \\
53704  &   3.57   $\pm$0.09 &--        --  & 4.44$\pm$0.09&  1.577$\pm$0.052&  1.357$\pm$ 0.100\\
53711  &   3.04   $\pm$0.07 &--        --  & 4.3 $\pm$0.09&  2.071$\pm$0.068&  1.036$\pm$ 0.077\\
53712  &   2.79   $\pm$0.07 &--        --  & 4.32$\pm$0.09&  2.197$\pm$0.073&  1.01 $\pm$ 0.075\\
53732  &   2.41   $\pm$0.06 &15.84$\pm$0.43& 3.49$\pm$0.07&  1.733$\pm$0.057&  0.904$\pm$ 0.067\\
53733  &   2.45   $\pm$0.06 &16.01$\pm$0.43& 3.39$\pm$0.07&  --        --   &  --         --   \\
53756.9&   1.88   $\pm$0.05 &--        --  & 3.14$\pm$0.05&  --        --   &  --         --   \\
53757.9&   1.92   $\pm$0.05 &--        --  & 3.16$\pm$0.05&  --        --   &  --         --   \\
53760.5&   1.8    $\pm$0.04 &15.62$\pm$0.33& 2.81$\pm$0.04&  --        --   &  --         --   \\
53761.5&   1.82   $\pm$0.04 &16.59$\pm$0.35& 2.89$\pm$0.05&  1.372$\pm$0.091&  0.845$\pm$ 0.073\\
53786.9&   2.07   $\pm$0.05 &15.64$\pm$0.33& 3.16$\pm$0.05&  1.69 $\pm$0.112&  0.767$\pm$ 0.066\\
53787.5&   1.91   $\pm$0.05 &16.57$\pm$0.35& 3.31$\pm$0.05&  1.775$\pm$0.117&  0.982$\pm$ 0.084\\
53788.5&   1.94   $\pm$0.05 &16.39$\pm$0.34& 3.31$\pm$0.05&  1.544$\pm$0.102&  0.876$\pm$ 0.075\\
53789.5&   1.94   $\pm$0.05:&16.55$\pm$0.35& --       --  &  1.724$\pm$0.114&  1.027$\pm$ 0.088\\
53803.8&   3.25   $\pm$0.08 &16.9 $\pm$0.35& 3.43$\pm$0.05&  1.684$\pm$0.111&  1.087$\pm$ 0.093\\
53816.4&   3.22   $\pm$0.08 &17.83$\pm$0.37& 3.98$\pm$0.06&  1.843$\pm$0.122&  1.471$\pm$ 0.127\\
53817.4&   3.06   $\pm$0.07 &17.79$\pm$0.37& 4.04$\pm$0.06&  1.755$\pm$0.116&  1.571$\pm$ 0.135\\
53843.7&   3.53   $\pm$0.08 &19.68$\pm$0.41& 4.55$\pm$0.07&  2.34 $\pm$0.154&  1.407$\pm$ 0.121\\
53844.8&   3.41   $\pm$0.08 &--        --  & 4.52$\pm$0.07&  2.071$\pm$0.137&  1.458$\pm$ 0.125\\
53845.7&   3.32   $\pm$0.08 &--        --  & 4.38$\pm$0.07&  2.054$\pm$0.136&  1.58 $\pm$ 0.136\\
53846.4&   3.31   $\pm$0.08 &19.01$\pm$0.40& 4.6 $\pm$0.07&  2.083$\pm$0.137&  1.887$\pm$ 0.162\\
\hline
\end{longtable}
}

\end{document}